\pdfoutput=1
\documentclass[shortcut,eqnsubsec,bbams]{amtp-pp}
\usepackage[utf8]{inputenc}
\usepackage[a4paper,margin=2.25cm,bottom=3.25cm]{geometry}
\allowdisplaybreaks

\usepackage{cleveref}
\crefformat{equation}{(#2#1#3)}
\crefmultiformat{equation}{(#2#1#3)}{, (#2#1#3)}{, (#2#1#3)}{, (#2#1#3)}
\crefrangeformat{equation}{(#3#1#4)--(#5\crefstripprefix{#1}{#2}#6)}
\renewcommand{\eqref}{\cref}

\usepackage{booktabs}

\usepackage{float}
\usepackage[labelformat=simple]{subcaption}

\usepackage{amsthm}

\newcommand{\eqend}[1]{\,\mathrm{#1}}

\newcommand{\subline}[1]{\smash[b]{\substack{#1}}}

\newcommand{\expect}[1]{{\left\langle{#1}\right\rangle}}

\newcommand{\dalembert}{\mathop{}\!\square}

\newcommand{\tr}{\operatorname{tr}}
\newcommand{\op}{\mathcal{O}}

\newcommand{\st}{\mathop{}\!\hat{\mathsf{s}}\hskip 0.05em\relax}
\newcommand{\stq}{\mathop{}\!\hat{\mathsf{q}}\hskip 0.05em\relax}
\newcommand{\stQ}{\mathcal{Q}}
\newcommand{\bvq}{\mathcal{B}}
\newcommand{\opint}{\mathcal{O}_\mathcal{I}}

\renewcommand{\varXi}{\vbox{\hrule height 0.15ex\kern0.4ex\hbox{\kern-0.05em$\Xi$\kern-0.05em}}}

\usepackage{xspace}

\newcommand{\ie}{\textit{i.\,e.}\xspace}
\newcommand{\eg}{\textit{e.\,g.}\xspace}
\newcommand{\etc}{\textit{etc.}\xspace}

\newcommand{\vv}{\textit{vice versa}\xspace}

\newtheorem{theorem}{Theorem}
\newtheorem{proposition}[theorem]{Proposition}
\newtheorem{corollary}[theorem]{Corollary}
\newtheorem{definition}{Definition}
\newtheorem{lemma}{Lemma}
\newtheorem*{lemma*}{Lemma}

\begin{document}

\title{All-order existence of and recursion relations for the operator product expansion in Yang-Mills theory}

\author{Markus B. Fr\"ob}
\email{mfroeb@itp.uni-leipzig.de}

\author{Jan Holland}
\email{holland@itp.uni-leipzig.de}

\affiliation{Institut f\"ur Theoretische Physik, Universit\"at Leipzig, Br\"uderstra\ss e 16, 04103 Leipzig, Germany}

\begin{abstract}
We prove the existence of the operator product expansion (OPE) in Euclidean Yang-Mills theories as a short-distance expansion, to all orders in perturbation theory. We furthermore show that the Ward identities of the underlying gauge theory are reflected in the OPE; especially, the OPE of an arbitrary number of gauge-invariant composite operators only involves gauge-invariant composite operators. Moreover, we derive recursion relations which allow to construct the OPE coefficients, the quantum BRST differential and the quantum antibracket order by order in perturbation theory, starting from the known free-theory objects. These relations are completely finite from the start, and do not need any further renormalisation as is usually the case in other approaches. Our results underline the importance of the OPE as a general structure underlying quantum field theories. The proofs are obtained within the framework of the Wilson-Wegner-Polchinski-Wetterich renormalisation group flow equations, and generalise similar results recently obtained for scalar field theories [J.~Holland and S.~Hollands, \href{http://dx.doi.org/10.1007/s00220-014-2274-8}{Commun.~Math.~Phys. \textbf{336} (2015) 1555}; J.~Holland and S.~Hollands, \href{http://dx.doi.org/10.1063/1.4937811}{J.~Math.~Phys. \textbf{56} (2015) 122303}]. Combining their results with our recursion formula, we also obtain associativity of the OPE coefficients.
\end{abstract}

\maketitle

\section{Introduction}

Ever since its inception by Wilson, the operator product expansion (OPE) has found a multitude of applications in high-energy physics, conformal field theory and condensed matter systems, among others. In its original form, it is the statement that one can approximate an expectation value of some composite operators at different points by a sum of products of certain coefficients and expectation values of a single composite operator~\cite{wilson1969,wilsonzimmermann1972,zimmermann1973}
\begin{equation}
\label{ope_general}
\expect{ \op_{A_1}(x_1) \cdots \op_{A_s}(x_s) }_\Psi \approx \sum_{B\colon [\op_B] \leq D} \mathcal{C}^B_{A_1 \cdots A_s}(x_1,\ldots,x_s) \expect{ \op_B(x_s) }_\Psi \eqend{,}
\end{equation}
where $A_1, \ldots, A_s, B, \ldots$ label the composite operators in the theory, and the expectation value $\expect{ \cdot }_\Psi$ is taken in some state $\ket{\Psi}$. This approximation should hold in an arbitrary (well-behaved) state $\ket{\Psi}$, and the sum over $B$ includes all composite operators up to a fixed dimension $[\op_B] \leq D$ with the same quantum numbers as the product of the composite operators on the left-hand side. The OPE coefficients $\mathcal{C}^B_{A_1 \cdots A_s}$ are state-independent distributions which are singular if one or more points $x_l$ coincide, and their singular behaviour as the points come close to each other (reflecting the singular behaviour of the expectation value on the left-hand side) is bounded by $\min_{k,l} \abs{x_k-x_l}^{[\op_B]-[\op_\vec{A}]}$, where $[\op_\vec{A}] \equiv \sum_{k=1}^s [\op_{A_k}]$ is the sum of the dimensions of the composite operators on the left-hand side. Thus, including more and more composite operators in the sum over $B$, the remainder becomes less and less singular, and vanishes as a $D$-dependent power of the minimum distance if all points scale together -- the OPE should be understood as an asymptotic expansion for small distances between the points $x_k$.

However, in the context of both massive and massless perturbative scalar field theory Holland, Hollands and Kopper have recently shown that the expansion~\eqref{ope_general} is actually much better behaved than expected. Namely, the sum on the right-hand side, and thus the OPE itself, is convergent if we let $D \to \infty$ for an arbitrary finite separation of the points $x_k$~\cite{hollandskopper2012,hollandetal2014}, at least to all orders in perturbation theory. The result holds for a class of states which are obtained by the application of smeared field operators on the (interacting) vacuum, which in Minkowski space corresponds to states of finite energy, and which in the massless case must be further restricted to avoid well-known infrared (IR) divergences for exceptional momenta. Together with further associativity conditions on the OPE coefficients, which under the same assumptions were also proven recently~\cite{hollandhollands2013,hollandhollands2015b}, one can then view the OPE as a defining property of a quantum field theory which encodes all the algebraic content, while the state $\ket{\Psi}$ is specified via its one-point functions $\expect{ \op_B }_\Psi$. If it is possible to directly construct the OPE coefficients, \eg, as is done in the case of two-dimensional conformal field theories (CFTs) where conformal symmetry and associativity fixes the OPE coefficients up to so-called ``conformal blocks'', which are then obtained in the ``bootstrap approach''~\cite{ferraragrillogatto1973,polyakov1974,luescher1976,mack1976,belavinetal1984,fredenhagenjoerss1994,elshowketal2012}, not even a Lagrangian is needed, which is very much in the original spirit of Wilson as evidenced by the title of his original paper~\cite{wilson1969}.

Although these conditions on the OPE coefficients can be nicely formulated using vertex algebras~\cite{borcherds1983,kac1996,hollandsolbermann2009}, and a large class of models of two-dimensional interacting CFTs have been constructed (see, \eg,~\cite{belavinetal1984,verlinde1988,lechner2008,bostelmanncadamuro2012}), it seems that in four dimensions and without conformal symmetry they are not enough to allow the direct construction of the OPE coefficients, and more input is necessary. Using associativity and imposing that the interacting field equation should be reflected in the OPE coefficients, \ie,
\begin{equation}
\mathcal{C}^B_{(\dalembert \phi) A_1 \cdots A_s} = \dalembert \mathcal{C}^B_{\phi A_1 \cdots A_s} = \lambda \mathcal{C}^B_{\phi^3 A_1 \cdots A_s}
\end{equation}
for a massless scalar $\lambda \phi^4$ interaction, an explicit recursive construction of the OPE coefficients as a power series in $\lambda$ was given by Holland~\cite{holland2009}, which involves an inversion of the d'Alembertian at each step. More recently, an explicit formula for the $\lambda$ derivative of the OPE coefficients was derived both for massive and massless scalar field theory~\cite{hollandhollands2015a,hollandhollands2015b} in the context of perturbation theory. This formula involves an integration over the OPE coefficient with an additional insertion of the interaction operator $\phi^4$, which is what one would expect naively from a path integral formulation, but in addition contains products of lower-order OPE coefficients which act as ``counterterms'' to cancel UV and IR divergences. Thus, the final formula is completely finite without any additional renormalisation, and moreover makes no explicit reference to perturbation theory. Since this formula is a first-order differential equation in the coupling constant, it could be possible to show that there exists a unique solution, given the free-theory coefficients as ``boundary conditions'', which then would completely determine the OPE coefficients; promising results in that direction have been obtained for the two-dimensional Gross-Neveu model~\cite{hollandhollands2016}.

The aim of this article is to extend these results to the case of Yang-Mills theories, with are both technically more challenging and more interesting from a physical point of view, providing the foundation for the standard model of particle physics. In comparison with the scalar case, there are two major obstacles which must be overcome:
\begin{enumerate}
\item The theory involves massless fields. As is well-known, this leads to IR divergences for exceptional momenta, which must be treated with care since such exceptional configurations occur in loop integrals in intermediate steps even if the final momentum configuration is not exceptional. While for scalar theories the massless case can be obtained as a limit from the massive one after a suitable redefinition of composite operators~\cite{hollandhollands2015b}, this is not possible for gauge theories where adding a mass by hand breaks gauge invariance, while spontaneous symmetry breaking which is compatible with gauge invariance leaves some of the fields massless in the general case (the exception being U(1) and SU(2) gauge theories~\cite{kellerkopper1991,koppermueller2000a,koppermueller2000b,mueller2003,koppermueller2009}).
\item Most regularisations, and especially the renormalisation group flow equations which were used for the scalar case break gauge invariance at intermediate steps. To restore it in the renormalised theory, and especially to obtain an appropriate set of Ward identities, stringent consistency conditions must be derived on possible anomalies. This is most effectively done using the BRST formalism~\cite{becchietal1975,henneauxteitelboim1992,weinberg_v2,barnichetal2000,piguetsorella} in the (extended) formulation of Batalin and Vilkovisky~\cite{batalinvilkovisky1981,batalinvilkovisky1983,batalinvilkovisky1984}, which involves a nilpotent fermionic operator, the BRST/Slavnov-Taylor differential. Gauge-invariant observables are elements of the cohomology of this operator, and a proper extension to the quantum theory must be found.
\end{enumerate}
Fortunately, for both of these obstacles a solution was found, proving mathematically rigorously the existence of correlation functions of arbitrary composite operators in (Euclidean) Yang-Mills theories, to all orders in perturbation theory, in such a way that the result after removing the UV and IR cutoffs is fully gauge invariant in the physically correct sense (\ie, with suitable Ward identities)~\cite{froebhollandhollands2015}. As for the results explained above, this proof was obtained within the renormalisation group flow equation approach to quantum field theory~\cite{polchinski1984,wetterich1993,kopper1998,mueller2003,kopper2007}.

\subsection{Results}

We build on the work of Fr{\"o}b, Holland and Hollands~\cite{froebhollandhollands2015}, who considered pure Euclidean Yang-Mills theory based on some semi-simple Lie group. However, our results only depend very weakly on the concrete form of the theory, such that we can keep the discussion quite general. We thus consider correlation functions of arbitrary composite operators $\op_{A_k}$, whose existence to all orders in perturbation theory was proven in Ref.~\cite{froebhollandhollands2015}. In fact, our first result is even valid for general massless theories, and is given by
\begin{theorem}
\label{thm1}
For an arbitrary massless, superficially renormalisable theory and up to an arbitrary, but fixed perturbation order, one can choose renormalisation conditions such that the correlation functions with insertions admit an expansion of the form
\begin{equation}
\label{thm1_ope}
\expect{ \op_{A_1}(x_1) \cdots \op_{A_s}(x_s) } \sim \sum_B \mathcal{C}^B_{A_1 \cdots A_s}(\vec{x}) \expect{ \op_B(x_s) }
\end{equation}
for a suitable definition of the OPE coefficients $\mathcal{C}^B_{A_1 \cdots A_s}$, which are well-defined distributions with singularities whenever two points coincide, and where $\vec{x} = (x_1, \ldots, x_s)$. The expansion is asymptotic in the sense that if one restricts the sum to operators with engineering dimension below a threshold $D$, the remainder vanishes as
\begin{equation}
\label{thm1_ope_asymptotic}
\lim_{\tau \to 0} \tau^{[\op_\vec{A}]-D + \delta} \left[ \expect{ \op_{A_1}(\tau x_1) \cdots \op_{A_s}(\tau x_s) } - \sum_{B\colon [\op_B] < D} \mathcal{C}^B_{A_1 \cdots A_s}( \tau \vec{x} ) \expect{ \op_B(\tau x_s) } \right] = 0
\end{equation}
for all $\delta > 0$, where $[\op_\vec{A}] \equiv \sum_{k=1}^s [\op_{A_k}]$ is the sum of the engineering dimensions of the composite operators in the correlation function. The OPE coefficients themselves have bounded scaling degree, \ie, they satisfy
\begin{equation}
\label{thm1_ope_scaling}
\lim_{\tau \to 0} \tau^{[\op_\vec{A}]-[\op_B] + \delta} \, \mathcal{C}^B_{A_1 \cdots A_s}(\tau \vec{x}) = 0
\end{equation}
for all $\delta > 0$. Furthermore, the OPE~\eqref{thm1_ope} holds not only in the vacuum $\expect{ \cdot }$, but in all states $\expect{ \cdot }_\Psi$ which can be obtained by applying (smeared) field operators on the vacuum, as long as the smearing does not involve exceptional momenta. Especially, this includes all states of finite total energy if we perform a Wick rotation to pass to the Minkowski theory.
\end{theorem}
This is the expected result, which was already proven quite early for a number of examples~\cite{zimmermann1971}; our result is in this sense not new but extends the validity of the OPE to \emph{all} massless theories. It also encompasses massless scalar field theory, for which even convergence of the asymptotic series~\eqref{thm1_ope} was proven~\cite{hollandetal2014}. In principle, there should be no obstacle to adopting this proof also to the case of other massless theories, but it is technically quite involved and would need a refinement of the bounds on correlation functions derived in Ref.~\cite{froebhollandhollands2015}.

For gauge theories, one expects that the OPE also respects gauge invariance in a suitable way. That this is indeed so is our second result, given by
\begin{theorem}
\label{thm2}
For an arbitrary massless, superficially renormalisable gauge theory and up to an arbitrary, but fixed perturbation order, one can choose renormalisation conditions such that the OPE coefficients fulfil the Ward identity
\begin{splitequation}
\label{thm2_ward}
0 &= \sum_{k=1}^s \sum_{C\colon [\op_C] \leq [\op_{A_k}]+1} \stQ_{A_k}{}^C \mathcal{C}^B_{A_1 \cdots A_{k-1} C A_{k+1} \cdots A_s}(\vec{x}) - \sum_C \stQ_C{}^B \mathcal{C}^C_{A_1 \cdots A_s}(\vec{x}) \\
&\quad- \hbar \sum_{1 \leq k < l \leq s} \sum_{\subline{E\colon [\op_E] \leq [\op_{A_k}]+[\op_{A_l}]-3 \\ w\colon \abs{w} = [\op_{A_k}]+[\op_{A_l}]-[\op_E]-3}} \mathcal{C}^B_{A_1 \cdots A_{k-1} E A_{k+1} \cdots A_{l-1} A_{l+1} \cdots A_s}(\vec{x}) \bvq^{E,w}_{A_k A_l} \partial^w_{x_k} \delta^4(x_k-x_l)
\end{splitequation}
if the equivariant classical cohomology of the classical Slavnov-Taylor differential $\st$ at form degree $4$ and ghost number $1$ is empty, $H^{1,4}_{\mathrm{E}(4)}(\st\vert\total) = 0$. In this equation, the nilpotent quantum Slavnov-Taylor differential $\stq = \st + \bigo{\hbar}$, which differs in higher orders in $\hbar$ from the classical Slavnov-Taylor differential $\st$, is given by its expansion coefficients defined via
\begin{equation}
\label{stq_matrix_def}
\stq \op_A \equiv \sum_{B\colon [\op_B] \leq [\op_A]+1} \stQ_A{}^B \op_B \eqend{.}
\end{equation}
The terms in the last line are contact terms supported on some diagonal $x_k = x_l$, which come from an expansion of the associative quantum antibracket $\left( \op_A(x), \op_B(y) \right)_\hbar = \left( \op_A(x), \op_B(y) \right) + \bigo{\hbar}$, which again differs in higher orders in $\hbar$ from the classical antibracket $( \cdot, \cdot )$. The expansion then reads
\begin{equation}
\label{bvq_matrix_def}
\left( \op_A(x), \op_B(y) \right)_\hbar \equiv \sum_{C\colon [\op_C] \leq [\op_A]+[\op_B]-3} \op_C(x) \sum_{w\colon \abs{w} = [\op_A]+[\op_B]-[\op_C]-3} \bvq^{C,w}_{AB} \partial^w_x \delta^4(x-y) \eqend{.}
\end{equation}
Note that while the sum over $C$ in the Ward identity~\eqref{thm2_ward} is not restricted by dimensional arguments, the theorem implies that it is convergent; the proof actually shows that at each perturbation order only a finite number of terms contribute.
\end{theorem}
The content of this theorem is best explained using an example in quantum electrodynamics. Assume that all the operators $\op_{A_k}$ are gauge-invariant, such that $\stq \op_{A_k} = 0$ and thus $\stQ_{A_k}{}^C = 0$ for all $C$. For the free theory, where $\hbar = g = 0$, the quantum BRST differential reduces to the free classical one $\st_0$, whose action on fields is given by
\begin{equation}
\label{stq_qed_free}
\st_0 A_\mu = \partial_\mu c \eqend{,} \quad \st_0 c = 0 \eqend{,} \quad \st_0 \bar{c} = B \eqend{,} \quad \st_0 B = 0 \eqend{.}
\end{equation}
The Ward identity~\eqref{thm2_ward} then reduces to
\begin{equation}
\sum_C \stQ_C{}^B \mathcal{C}^C_{A_1 \cdots A_s}(x_1, \ldots, x_s) = 0 \eqend{,}
\end{equation}
and taking, \eg, $\op_B = \partial_\mu \partial_\nu c$, we thus need to search for operators $\op_C$ such that $\op_B$ appears on the right-hand side of the expansion~\eqref{stq_matrix_def} of $\stq \op_C$. For the later results, it will be important that all composite operators are monomials (although of course a redefinition of composite operators can be done if the OPE coefficients, the quantum BRST differential and the quantum antibracket are simultaneously redefined), and we will assume this now as well. The only two monomials $\op_C$ for which $\stQ_C{}^B$ does not vanish are $\op_{C,1} = \partial_\mu A_\nu$ and $\op_{C,2} = \partial_\nu A_\mu$, for which we have
\begin{equation}
\stQ_{C,1}{}^B = \stQ_{\partial_\mu A_\nu}{}^{\partial_\mu \partial_\nu c} = 1 = \stQ_{\partial_\nu A_\mu}{}^{\partial_\mu \partial_\nu c} = \stQ_{C,2}{}^B
\end{equation}
since partial derivatives commute, and we thus obtain
\begin{equation}
\mathcal{C}^{\partial_\mu A_\nu}_{A_1 \cdots A_s}(x_1, \ldots, x_s) + \mathcal{C}^{\partial_\nu A_\mu}_{A_1 \cdots A_s}(x_1, \ldots, x_s) = 0 \eqend{.}
\end{equation}
The OPE~\eqref{thm1_ope} thus reads
\begin{splitequation}
\expect{ \op_{A_1}(x_1) \cdots \op_{A_s}(x_s) } &\sim \cdots + \mathcal{C}^{\partial_\mu A_\nu}_{A_1 \cdots A_s}(x_1, \ldots, x_s) \expect{ \partial_\mu A_\nu(x_s) } + \mathcal{C}^{\partial_\nu A_\mu}_{A_1 \cdots A_s}(x_1, \ldots, x_s) \expect{ \partial_\nu A_\mu(x_s) } + \cdots \\
&= \cdots + \mathcal{C}^{\partial_\mu A_\nu}_{A_1 \cdots A_s}(x_1, \ldots, x_s) \expect{ F_{\mu\nu}(x_s) } + \cdots \eqend{,} \raisetag{\baselineskip}
\end{splitequation}
where the free field strength tensor $F_{\mu\nu} = \partial_\mu A_\nu - \partial_\nu A_\mu$ is gauge invariant in the free theory, $\st_0 F_{\mu\nu} = 0$, as can be seen immediately from the action of the BRST differential~\eqref{stq_qed_free}. We thus see that the Ward identity~\eqref{thm2_ward} especially implies that in the OPE of gauge-invariant operators also only gauge-invariant operators appear on the right-hand side. Note that if we would have shown that the OPE~\eqref{thm1_ope} is a convergent instead of only an asymptotic expansion, as was done for scalar fields~\cite{hollandskopper2012,hollandetal2014}, the Ward identity~\eqref{thm2_ward} would have followed as a simple corollary by acting with $\stq$ on both sides and using the Ward identity for correlation functions derived in Ref.~\cite{froebhollandhollands2015}.

The one condition that needs to be fulfilled for the theorem to hold is the vanishing of the equivariant cohomology class $H^{1,4}_{\mathrm{E}(4)}(\st\vert\total)$ of $\st$ modulo exact terms (\ie, modulo $\total$) at form degree $4$ and ghost number $1$. This class is the subset of the full cohomology class $H^{1,4}(\st\vert\total)$ which is invariant under the action of the Euclidean group $\mathrm{E}(4)$, including parity inversion and time reversal: $\mathrm{E}(4) = \mathrm{O}(4) \rtimes \mathbb{R}^4$. For Yang-Mills theories based on a semisimple Lie group, these have been calculated quite some time ago~\cite{barnichetal1995a,barnichetal1995b,barnichetal2000}, and while the full cohomology class is not empty, all its elements are odd under parity inversion (the well-known axial anomaly). The equivariant cohomology class is thus empty, and the above Ward identity holds for pure Yang-Mills theories -- but not, for instance, in the presence of chiral fermions in non-real representations, which are only $\mathrm{O}(4)$- but not $\mathrm{E}(4)$-invariant. For a more thorough discussion of these condition, we refer the reader to the aforementioned work~\cite{froebhollandhollands2015}.

Lastly, one may wonder about what happens if one performs a partial OPE, \ie, only expands a subset of the operators in a correlation function, and afterwards reexpands the remaining operators again. If the resulting expansions were convergent, this would put stringent consistency conditions on the OPE coefficients, \eg,
\begin{equation}
\label{ope_consistency_condition}
\mathcal{C}^B_{A_1 A_2 A_3}(x_1, x_2, x_3) = \sum_C \mathcal{C}^C_{A_1 A_2}(x_1, x_2) \mathcal{C}^B_{C A_3}(x_2, x_3) = \sum_C \mathcal{C}^C_{A_2 A_3}(x_2, x_3) \mathcal{C}^B_{A_1 C}(x_1, x_3)
\end{equation}
should hold if one expands the operators in a three-point correlation function in different ways. Especially, all OPE coefficients should be completely determined in terms of the simplest ones, which result from the OPE of two operators. This is indeed so for suitable insertion points, and we have
\begin{theorem}
\label{thm6}
For an arbitrary massless, superficially renormalisable theory and up to an arbitrary, but fixed perturbation order the OPE coefficients are associative, \ie,
\begin{equation}
\label{ope_associative}
\mathcal{C}^B_{A_1 \cdots A_s}(\vec{x}) = \sum_C \mathcal{C}^C_{A_1 \cdots A_k}(x_1, \cdots, x_k) \mathcal{C}^B_{C A_{k+1} \cdots A_s}(x_k, \cdots, x_s)
\end{equation}
holds if the insertion points fulfil
\begin{equation}
\max_{1 \leq i < k} \abs{x_i-x_k} < \min_{k < j \leq s} \abs{x_j - x_k} \eqend{.}
\end{equation}
Especially, the sum over $C$ is convergent in this case.
\end{theorem}
Thus, the OPE is associative as long as we perform first an expansion of the operators which are closest together. In Ref.~\cite{hollands2009}, it was shown that consistency conditions of the form~\eqref{ope_consistency_condition} arising from the associativity condition~\eqref{ope_associative} on a suitable domain, together with an assumption of analyticity away from coincidence $x_i = x_j$, does then determine the $n$-point OPE coefficients with $n > 2$ completely in terms of the two-point coefficients.

\subsection{Recursive constructions}

While the previous theorems~\ref{thm1} and~\ref{thm2} assert that an OPE exists and fulfils the appropriate Ward identities, it is not clear how to actually construct the OPE coefficients themselves. Similar to the scalar case~\cite{hollandhollands2015a,hollandhollands2015b}, we have succeeded to derive a recursive formula for the OPE coefficients which makes it possible to calculate them order by order in perturbation theory, starting from the free theory coefficients which are easy to obtain. The concrete result again applies to general massless theories, and is given by
\begin{theorem}
\label{thm3}
For an arbitrary massless, superficially renormalisable theory and up to an arbitrary, but fixed perturbation order, one can choose renormalisation conditions such that the derivative of the OPE coefficients with respect to the coupling constant $g$ can be expressed as
\begin{splitequation}
\label{thm3_ope_recursive}
\hbar \partial_g \mathcal{C}^B_{A_1 \cdots A_s}(\vec{x}) = \int \sum_{E\colon 1 \leq [\op_E] \leq 4} \mathcal{I}^E &\Bigg[ - \mathcal{C}^B_{E A_1 \cdots A_s}(y, \vec{x}) + \sum_{C\colon [\op_C] < [\op_B]} \mathcal{C}^C_{A_1 \cdots A_s}(\vec{x}) \mathcal{C}^B_{E C}(y, x_s) \\
&\quad+ \sum_{k=1}^s \sum_{C\colon [\op_C] \leq [\op_{A_k}]} \mathcal{C}^C_{E A_k}(y,x_k) \mathcal{C}^B_{A_1 \cdots A_{k-1} C A_{k+1} \cdots A_s}(\vec{x}) \Bigg] \total^4 y \eqend{.}
\end{splitequation}
In this equation, all composite operators are $g$-independent monomials, while the $\mathcal{I}^E$ are possibly $g$-dependent coefficients which depend on the concrete form of the interaction Lagrangian. Note that this equation is to be understood in the sense of formal power series in the coupling constant $g$ as it arises in perturbation theory.
\end{theorem}
By expanding both sides in a formal power series in $g$, one obtains a recursion formula which allows to calculate higher-order coefficients in terms of lower-order ones, if the starting point, the coefficients in the free theory, are known. In turn, these can be obtained from a sort of Wick expansion in the free theory, as has been shown for scalar field theory by Holland~\cite{holland2013}. For gauge theories, or in general theories where fields have additional indices or internal quantum numbers such as charge or colour, only some almost trivial modifications to the proof are necessary, and since in contrast to scalar field theories the simplest example in gauge theories (the OPE for two operators $\tr F^2$ with the non-Abelian field strength $F$, checking also gauge invariance) is already lengthy and highly non-trivial, we delegate explicit formulas and examples to subsequent work. For massive scalar field theory, it is possible to choose the renormalisation conditions (namely BPHZ conditions) such that
\begin{equation}
\label{massive_scalar_ie}
\mathcal{I}^E = \begin{cases} 1 & \op_E = \phi^4 \\ 0 & \op_E \neq \phi^4 \eqend{,} \end{cases}
\end{equation}
\ie, the ``interaction operator''
\begin{equation}
\label{op_g_def}
\opint \equiv \sum_{E\colon 1 \leq [\op_E] \leq 4} \mathcal{I}^E \op_E
\end{equation}
is equal to $\phi^4$~\cite{hollandhollands2015a}. The same can be achieved for massless scalar field theory by performing a suitable redefinition of composite operators, OPE coefficients and the coupling constant $g$~\cite{hollandhollands2015b}.\footnote{We note that while the end result given in this reference is mostly correct, a part of the proof is missing.} For gauge theories, however, it may be necessary to change renormalisation conditions from the usual BPHZ ones in order to remove potential anomalies (see Ref.~\cite{froebhollandhollands2015} for details), and while the process is fully under control, obtaining explicit results for the resulting renormalisation conditions engenders considerable difficulties, and we refer the reader to the work~\cite{efremovguidakopper2015} for SU(2) Yang-Mills theory and~\cite{chankowskietal2016}, where a coupled Yang--Mills-scalar-fermion theory was studied at two loops. Nonetheless, the interaction operator is not fully arbitrary, but instead fulfils the conditions of
\begin{theorem}
\label{thm4}
The interaction operator $\opint$ defined by~\eqref{op_g_def} and appearing in Theorem~\ref{thm3} is of the form
\begin{equation}
\label{thm4_opint}
\opint = \partial_g L \rvert_{g = 0} + \bigo{g} + \bigo{\hbar} \eqend{,}
\end{equation}
where $L$ is the classical interaction Lagrangian, including gauge-fixing and ghost terms as well as antifields for the case of Yang-Mills theories. It is only defined up to a total derivative, and fulfils
\begin{equation}
\label{thm4_stq_opint}
\stq \opint = \total \op
\end{equation}
for some composite operator $\op$.
\end{theorem}
At lowest order in the coupling constant, setting $g = 0$, no loops can appear in the correlation functions and thus also $\hbar = 0$. Since possible anomalies are at least of order $\hbar$ as shown in Ref.~\cite{froebhollandhollands2015}, we can choose BPHZ conditions for the lowest order, which leads to equation~\eqref{thm4_opint}. For massive scalar theory, this is already the correct result~\eqref{massive_scalar_ie}, but for gauge theories one must use the other conditions to constrain the form of $\opint$, order by order in perturbation theory. Note also that total derivatives appearing in $\opint$ do not contribute to the integral in the recursion formula~\eqref{thm3_ope_recursive}, such that it really only depends on the classical action where one can freely integrate by parts as one would expect from a formal path integral treatment.

We also have derived recursion formulas for the quantum Slavnov-Taylor differential and the quantum antibracket, which are given by
\begin{theorem}
\label{thm5}
For an arbitrary massless, superficially renormalisable theory and up to an arbitrary, but fixed perturbation order, one can choose renormalisation conditions such that the derivative of the (components of the) quantum Slavnov-Taylor differential with respect to the coupling constant $g$ can be expressed as
\begin{splitequation}
\label{thm5_stq_recursive}
\hbar \partial_g \stQ_A{}^B &= \int \!\! \sum_{E\colon 1 \leq [\op_E] \leq 4} \mathcal{I}^E \left[ \sum_{C\colon [\op_C] \leq [\op_A]} \mathcal{C}^C_{E A}(y,0) \stQ_C{}^B - \!\! \sum_{C\colon [\op_B] \leq [\op_C] \leq [\op_A]+1} \stQ_A{}^C \mathcal{C}^B_{E C}(y,0) \right] \total^4 y \\
&+\quad \hbar \sum_{E\colon 1 \leq [\op_E] \leq 4} \mathcal{I}^E \widetilde{\bvq}^B_{EA} \eqend{,} \raisetag{1.7\baselineskip}
\end{splitequation}
and the derivative of the (components of the) quantum antibracket with respect to the coupling constant $g$ can be expressed as
\begin{splitequation}
\label{thm5_bvq_recursive}
\hbar \partial_g \widetilde{\bvq}^B_{A_1 A_2} = \int \!\! \sum_{E\colon 1 \leq [\op_E] \leq 4} \mathcal{I}^E \Bigg[ &\sum_{C\colon [\op_C] \leq [\op_{A_1}]} \widetilde{\bvq}^B_{C A_2} \mathcal{C}^C_{E A_1}(y,0) + \sum_{C\colon [\op_C] \leq [\op_{A_2}]} \widetilde{\bvq}^B_{A_1 C} \mathcal{C}^C_{E A_2}(y,0) \\
&\quad- \sum_{C\colon [\op_C] = [\op_A]+[\op_B]-3} \widetilde{\bvq}^C_{A_1 A_2} \mathcal{C}^B_{E C}(y,0) \Bigg] \total^4 y \eqend{.}
\end{splitequation}
In both formulas, the coefficients $\widetilde{\bvq}^F_{AB}$ are obtained from the expansion of
\begin{equation}
\int \left( \op_A(x), \op_B(y) \right)_\hbar \total^4 x = \sum_{C\colon [\op_C] = [\op_A]+[\op_B]-3} \widetilde{\bvq}^C_{AB} \op_C(y)
\end{equation}
in monomials. They are related to the coefficients $\bvq^{C,w}_{AB}$ defined in equation~\eqref{bvq_matrix_def} via
\begin{equation}
\widetilde{\bvq}^F_{AB} \equiv \sum_{C\colon [\op_C] \leq [\op_A]+[\op_B]-3} \sum_{w\colon \abs{w} = [\op_A]+[\op_B]-[\op_C]-3} (-1)^\abs{w} \bvq^{C,w}_{A B} \delta\left( \op_F, \partial^w \op_C \right) \eqend{,}
\end{equation}
where the expression $\delta\left( \op_B, \partial^w \op_C \right)$ counts the number of times the composite operator $\op_B$ appears in the expansion of $\partial^w \op_C$ in monomials, \eg, $\delta\left( \phi \partial \phi, \partial \phi^2 \right) = 2$. As in Theorem~\ref{thm3}, all composite operators are $g$-independent monomials, and the equations are to be understood in the sense of formal power series in the coupling constant $g$ as it arises in perturbation theory. Furthermore, the $\mathcal{I}^E$ are the same coefficients appearing in Theorem~\ref{thm3}.
\end{theorem}
These recursion formulas are of the same general form as the one for the OPE coefficients themselves~\eqref{thm3_ope_recursive}, and since for the determination of the interaction operator $\opint$ in higher orders in $g$ we need to use the condition~\eqref{thm4_stq_opint}, it is necessary to solve all recursion formulas together. Nevertheless, one can achieve a simplification by setting the antifields to zero. In the classical theory, and thus especially for the free theory where $g = 0$, the antibracket then vanishes, and the recursion formula~\eqref{thm5_bvq_recursive} shows that the quantum antibracket then vanishes to all orders, which also simplifies the recursion formula for the quantum Slavnov-Taylor differential~\eqref{thm5_stq_recursive}.

\bigskip

The rest of the article is structured as follows: in section~\ref{sec_framework} we give a short overview of the flow equation framework which we use to obtain the above results. In section~\ref{sec_previous}, we comment on the tree structures which were introduced in Ref.~\cite{froebhollandhollands2015} and which are necessary to state precise bounds on functionals with and without insertions of composite operators derived in that reference, and which we reproduce here for completeness, together with the pertinent Ward identities for functionals with insertions of composite operators. In section~\ref{sec_bounds} we derive additional bounds on yet other functionals with insertions of composite operators (partly connected, and with oversubtractions), which are needed for the proof of the main theorems~\ref{thm1}--\ref{thm5}. Section~\ref{sec_ope} gives our definition of the OPE coefficients, and establishes the asymptotic existence of the OPE, which is Theorem~\ref{thm1}, a simple corollary of Proposition~\ref{thm_ope_asymptotic} in the physical limit $\Lambda_0 \to \infty$. The recursion formula for the OPE coefficients, Theorem~\ref{thm3} (a simple corollary of Proposition~\ref{thm_coeffs_g} in the physical limit), is proven in section~\ref{sec_param}, where also the interaction operator $\opint$ is defined, and where we also show associativity, Theorem~\ref{thm6}, as a simple corollary of Proposition~\ref{thm_ope_associative} in the physical limit. Section~\ref{sec_gauge} considers the interplay between the OPE and gauge theories, and proves the Ward identity for the OPE coefficients, Theorem~\ref{thm2} (a simple corollary of Proposition~\ref{thm_ope_gaugeinv_strong}), and the recursion formulas for the quantum Slavnov-Taylor differential and the quantum antibracket, Theorem~\ref{thm5} (a simple corollary of Propositions~\ref{thm_stq_g} and~\ref{thm_bvq_g}), as well as further properties of the interaction operator, Theorem~\ref{thm4} (a simple corollary of Propositions~\ref{thm_l0_g} and~\ref{thm_opg_stq}).

\section{The flow equation framework}
\label{sec_framework}

Our proofs are obtained in the framework of the Wilson-Wegner-Polchinski-Wetterich renormalisation group flow equations~\cite{polchinski1984,wetterich1993,kopper1998,mueller2003,kopper2007}. The main objects of interest in this framework are the generating functionals of connected and amputated correlation functions with and without insertions of composite operators, in Euclidean quantum field theory. We refrain here from giving a full (and thus lengthy) exposition of this framework and just provide an overview containing the main points which are necessary for the following sections; the reader without previous experience is referred to the works~\cite{kopper1998,mueller2003,kopper2007,froebhollandhollands2015} for a more detailed explanation.

Regarding notation, we use a standard multiindex notation where $\vec{w} = (w_1,\ldots,w_n)$ and $w_i = (w_i^1,\ldots,w_i^4)$ with $w_i^\alpha \geq 0$,
\begin{equation}
\partial^\vec{w} f(\vec{q}) \equiv \partial^{w_1^1}_{q_1^1} \cdots \partial^{w_n^4}_{q_n^4} f(q_1, \ldots, q_n) \eqend{,}
\end{equation}
\begin{equation}
\abs{\vec{w}} \equiv \sum_{i=1}^n \abs{w_i} \equiv \sum_{i=1}^n \sum_{\alpha=1}^4 w_i^\alpha \eqend{,}
\end{equation}
and
\begin{equation}
\vec{w}! \equiv \prod_{i=1}^n w_i! \equiv \prod_{i=1}^n \prod_{\alpha=1}^4 w_i^\alpha! \eqend{.}
\end{equation}
To reduce notational clutter, we further stipulate that the vector $\vec{q}$ will always either have $n$ or $m+n$ entries, depending on context, while $\vec{q}_\rho \equiv \{ q_i \vert i \in \rho \}$. We use a condensed $L^2$ inner product notation, defining
\begin{equation}
\left\langle A, B \right\rangle \equiv \int A(x) B(x) \total^4 x
\end{equation}
and
\begin{equation}
A \ast B \equiv \int A(x-y) B(y) \total^4 y \eqend{.}
\end{equation}
We also use the positive part of the logarithm, defined as
\begin{equation}
\ln_+ x \equiv \sup( \ln x, 0 ) = \ln \sup(x,1) \eqend{,}
\end{equation}
which satisfies the useful identity
\begin{equation}
\ln_+ (a x + b y) \leq \ln_+ (a+b) + \ln_+ x + \ln_+ y
\end{equation}
for all $a,b \geq 0$ and $x,y \in \mathbb{R}$. For later use, we further define
\begin{equation}
\label{h_def}
\abs{\vec{q}} \equiv \sup_{Q \subseteq \{q_1,\ldots,q_n\}} \abs{\sum_{q \in Q} q} \eqend{,}
\end{equation}
which measures the maximum possible sum of a set of momenta,
\begin{equation}
\label{eta_i_def}
\eta_{q_i}(\vec{q}) \equiv \inf_{Q \subseteq \{q_1,\ldots,q_{n-1}\}\setminus\{q_i\}} \abs{q_i + \sum_{q \in Q} q}
\end{equation}
which measures the exceptionality of a set of momenta which includes $q_i$,
\begin{equation}
\label{eta_def}
\eta(\vec{q}) \equiv \inf_{q \in \{q_1,\ldots,q_{n-1}\}} \eta_q(q_1,\ldots,q_n) \eqend{,}
\end{equation}
which is the smallest such exceptionality,
\begin{equation}
\label{bareta_i_def}
\bar{\eta}_{q_i}(\vec{q}) \equiv \inf_{Q \subseteq \{q_1,\ldots,q_n\}\setminus\{q_i\}} \abs{q_i + \sum_{q \in Q} q}
\end{equation}
which is the same as $\eta_{q_i}$ except that the sum runs over all momenta including $q_n$, and
\begin{equation}
\label{bareta_def}
\bar{\eta}(\vec{q}) \equiv \inf_{q \in \{q_1,\ldots,q_n\}} \bar{\eta}_q(\vec{q}) \eqend{.}
\end{equation}
If $n = 1$, we define $\eta_q(q) \equiv \abs{q}$, and if $n = 0$ we define $\bar{\eta}() \equiv \mu$, $\eta() \equiv 0$ and $\abs{\cdot} \equiv 0$. With these definitions, the special cases are covered by the general estimates.

Throughout the whole article, we denote the ultraviolet (UV) cutoff by $\Lambda_0$, the infrared (IR) cutoff by $\Lambda$ and the renormalisation scale by $\mu$, which satisfy $\Lambda_0 \geq \Lambda \geq 0$ and $\Lambda_0 \geq \mu > 0$. Furthermore, $c$, $c'$, \etc, will denote arbitrary positive constants, which may change even within an equation, and $\mathcal{P}$ denotes a polynomial with positive coefficients, which also may change.

The generating functionals depend on the cutoffs and we denote them by $L^{\Lambda, \Lambda_0}\left( \bigotimes_{k=1}^s \op_{A_k}(x_k) \right)$, where the $A_k$ are labels for the composite operators in the theory under study. While we do not make this dependence explicit in notation to shorten the formulas, they also depend on the fields appearing in the theory, which we denote collectively by $\phi_K$, where $K$ is an index that distinguishes the kind of field and, if they appear, tensor and Lie algebra indices. We also define a regularised covariance matrix/propagator
\begin{equation}
C_{KL}^{\Lambda, \Lambda_0} \equiv C_{KL}^{0, \infty} \ast \left( R^{\Lambda_0} - R^\Lambda \right) \eqend{,}
\end{equation}
where $R^\Lambda$ is a real, smooth, $\mathrm{E}(4)$ invariant regulator function $R^\Lambda$, in Fourier space analytic at $p=0$, which fulfils the properties
\begin{equations}[r_prop]
0 < R^\Lambda(p) &< 1 \quad\text{ for }\quad 0 < \Lambda < \infty \eqend{,} \\
R^0(p) &= 0 \eqend{,} \qquad R^\infty(p) = 1 \eqend{,} \label{r_prop_2} \\
R^\Lambda(p) &< R^{\Lambda_0}(p) \quad\text{ for }\quad \Lambda < \Lambda_0 \eqend{,} \\
\label{r_prop_bound} \abs{\partial^w \partial^k_\Lambda R^\Lambda(p)} &\leq c \sup(\abs{p},\Lambda)^{-k-\abs{w}} \mathe^{-\frac{\abs{p}^2}{2 \Lambda^2}} \eqend{.}
\end{equations}
A simple example of such a regulator is given by
\begin{equation}
\label{r_example}
R^\Lambda(p) = \mathe^{-\frac{\abs{p}^2}{\Lambda^2}} \eqend{,}
\end{equation}
which is the one used in most previous works, but the results are independent of the actual choice as long as it fulfils the conditions~\eqref{r_prop}. To illustrate the above definitions, we take Yang-Mills theory based on a semi-simple Lie algebra, where
\begin{equation}
\left\{ \phi_K \right\} = \left\{ A_{a\mu}, B_a, \bar{c}_a, c_a \right\}
\end{equation}
with the components of the vector potential $A_{a\mu}$, the ghost and antighost $\bar{c}_a$ and $c_a$ and the auxiliary Nakanishi-Lautrup field $B_a$~\cite{nakanishi1966,lautrup1967,kugoojima1979}, where $\mu$ is a tensor index and $a$ the Lie algebra index relative to an arbitrary chosen basis in the Lie algebra. In Fourier space, the regularised covariance matrix reads
\begin{equation}
\label{cov_def}
C^{\Lambda, \Lambda_0}_{KL}(p) = \delta_{ab} \begin{pmatrix} \delta_{\mu\nu} + (\xi^{-1}-1) p_\mu p_\nu / p^2 & 0 & 0 & 0 \\ 0 & 0 & - 1 & 0 \\ 0 & 1 & 0 & 0 \\ 0 & 0 & 0 & p^2 \end{pmatrix} \frac{R^{\Lambda_0}(p) - R^\Lambda(p)}{p^2} \eqend{,}
\end{equation}
where $\xi$ is the usual gauge parameter of the linear covariant $R_\xi$ gauges. Note that this differs from the usual covariance/propagator by a linear field redefinition that makes the covariance matrix~\eqref{cov_def} positive definite, which is a technical necessity in the flow equation framework. As usual, we also assign an engineering dimension to the basic fields, which for technical reasons we assume to be $\geq 1$. For Yang-Mills theory, this can be realised with the assignment
\begin{equation}
[A_{a\mu}] = [\bar{c}_a] = [c_a] = 1 \eqend{,} \qquad [B_a] = 2 \eqend{,}
\end{equation}
and we consider power-counting renormalisable theories, meaning that the engineering dimension of all terms in the interaction Lagrangian is equal to or lower than the space-time dimension, $4$. Since the terms in the action which are quadratic in the fields define the covariance~\eqref{cov_def} and the interaction involves at least three fields, we obtain that $1 \leq [\phi_K] \leq 3$. It then follows from property~\eqref{r_prop_bound} of the regulator that
\begin{equation}
\label{prop_abl}
\abs{\partial^w \partial_\Lambda C^{\Lambda, \Lambda_0}_{KL}(p)} \leq c \sup(\abs{p}, \Lambda)^{-5+[\phi_K]+[\phi_L]-\abs{w}} \, \mathe^{-\frac{\abs{p}^2}{2 \Lambda^2}} \eqend{,}
\end{equation}
which is a bound that will be important later on. The engineering dimension of composite operators is given by the sum of the engineering dimension of its constituent fields, and we denote by $\Delta$ the minimal difference between the engineering dimensions of two different composite operators. Since the dimension of a derivative is given by $[\partial] = 1$, we always have $\Delta \leq 1$, and we assume that $\Delta > 0$ for technical reasons. As an example, for a theory with Dirac fermions $\psi$ we have $[\psi] = 3/2$ in four dimensions, and then $\Delta = 1/2$.

The generating functionals without insertions of composite operators then fulfil the flow equation
\begin{equation}
\label{l_0op_flow}
\partial_\Lambda L^{\Lambda, \Lambda_0} = \frac{\hbar}{2} \left\langle \frac{\delta}{\delta \phi_K}, \left( \partial_\Lambda C^{\Lambda, \Lambda_0}_{KL} \right) \ast \frac{\delta}{\delta \phi_L} \right\rangle L^{\Lambda, \Lambda_0} - \frac{1}{2} \left\langle \frac{\delta}{\delta \phi_K} L^{\Lambda, \Lambda_0}, \left( \partial_\Lambda C^{\Lambda, \Lambda_0}_{KL} \right) \ast \frac{\delta}{\delta \phi_L} L^{\Lambda, \Lambda_0} \right\rangle \eqend{,}
\end{equation}
while the generating functionals with insertions fulfil the flow equation
\begin{splitequation}
\label{l_sop_flow}
\partial_\Lambda L^{\Lambda, \Lambda_0}\left( \bigotimes_{k=1}^s \op_{A_k} \right) &= \frac{\hbar}{2} \left\langle \frac{\delta}{\delta \phi_K}, \left( \partial_\Lambda C^{\Lambda, \Lambda_0}_{KL} \right) \ast \frac{\delta}{\delta \phi_L} \right\rangle L^{\Lambda, \Lambda_0}\left( \bigotimes_{k=1}^s \op_{A_k} \right) \\
&\qquad- \left\langle \frac{\delta}{\delta \phi_K} L^{\Lambda, \Lambda_0}, \left( \partial_\Lambda C^{\Lambda, \Lambda_0}_{KL} \right) \ast \frac{\delta}{\delta \phi_L} L^{\Lambda, \Lambda_0}\left( \bigotimes_{k=1}^s \op_{A_k} \right) \right\rangle \\
&\qquad- \hspace{-1em}\sum_{\subline{\alpha \cup \beta = \{1, \ldots, s\} \\ \alpha \neq \emptyset \neq \beta}} \left\langle \frac{\delta}{\delta \phi_K} L^{\Lambda, \Lambda_0}\left( \bigotimes_{k\in\alpha} \op_{A_k} \right) , \left( \partial_\Lambda C^{\Lambda, \Lambda_0}_{KL} \right) \ast \frac{\delta}{\delta \phi_L} L^{\Lambda, \Lambda_0}\left( \bigotimes_{k\in\beta} \op_{A_k} \right) \right\rangle \eqend{,}
\end{splitequation}
which involves an extra source term depending on functionals with a smaller number of operator insertions. To uniquely determine the functionals, one needs to specify boundary conditions, which are obtained at $\Lambda = \Lambda_0$ where the generating functionals coincide with the bare interaction Lagrangian, including all counterterms which are necessary to cancel the UV divergences appearing in $L^{\Lambda, \Lambda_0}$ in the limit $\Lambda_0 \to \infty$.

One now expands the generating functionals in a formal power series in $\hbar$ (the loop expansion), and in the fields $\phi$. For gauge theories, it is also useful to introduce antifields $\phi^\ddag_L$, which are sources for the BRST transformation of the fields, and where again the index $L$ distinguishes the kind of antifield and, if they appear, tensor and Lie algebra indices, and also expand the generating functionals in the antifields. It is further useful to pass to Fourier space, and remove an overall Dirac $\delta$ distribution which enforces momentum conservation from the expansion coefficients. The expansion coefficients of the functionals without insertions of composite operators, \ie, the connected, amputated correlation functions of basic fields (and antifields), are denoted by
\begin{equation}
\mathcal{L}^{\Lambda, \Lambda_0, l}_{\vec{K} \vec{L}^\ddag}(\vec{q}) \eqend{,}
\end{equation}
where $l \in \mathbb{N}_0$ denotes the order in $\hbar$, and $\vec{K}$ and $\vec{L}^\ddag$ denote the fields and antifields (also called ``external legs''), which depend on the momenta $\vec{q}$. We will always assume that there are $\abs{\vec{K}} = m$ basic fields and $\abs{\vec{L}^\ddag} = n$ antifields. Because of overall momentum conservation, we have $\sum_{i=1}^{m+n} q_i = 0$ such that $q_{m+n}$ is not an independent variable, but will not make this explicit to shorten the notation. The expansion coefficients of the functionals with insertions of composite operators are the connected correlation functions of basic fields, antifields and the composite operators, where however only the basic fields and antifields are amputated, and we denote them by
\begin{equation}
\mathcal{L}^{\Lambda, \Lambda_0, l}_{\vec{K} \vec{L}^\ddag}\left( \bigotimes_{k=1}^s \op_{A_k}(x_k); \vec{q} \right) \eqend{.}
\end{equation}
Since we do not perform a Fourier transform of the position of the composite operator insertions, overall momentum is not conserved for those functionals and we do not remove any Dirac $\delta$ distribution enforcing momentum conservation from these functionals. Nevertheless, Euclidean translation invariance leads to the shift property
\begin{equation}
\label{func_sop_shift}
\mathcal{L}^{\Lambda, \Lambda_0, l}_{\vec{K} \vec{L}^\ddag}\left( \bigotimes_{k=1}^s \op_{A_k}(x_k); \vec{q} \right) = \mathe^{- \mathi y \sum_{i=1}^{m+n} q_i} \mathcal{L}^{\Lambda, \Lambda_0, l}_{\vec{K} \vec{L}^\ddag}\left( \bigotimes_{k=1}^s \op_{A_k}(x_k - y); \vec{q} \right) \eqend{.}
\end{equation}

For the expansion coefficients, we obtain from~\eqref{l_0op_flow} the hierarchy of flow equations
\begin{splitequation}
\label{l_0op_flow_hierarchy}
&\partial_\Lambda \mathcal{L}^{\Lambda, \Lambda_0, l}_{\vec{K} \vec{L}^\ddag}(\vec{q}) = \frac{c}{2} \int \left( \partial_\Lambda C^{\Lambda, \Lambda_0}_{MN}(-p) \right) \mathcal{L}^{\Lambda, \Lambda_0, l-1}_{MN \vec{K} \vec{L}^\ddag}(p, -p, \vec{q}) \frac{\total^4 p}{(2\pi)^4} \\
&\quad- \sum_{\subline{\sigma \cup \tau = \{1, \ldots, m\} \\ \rho \cup \varsigma = \{1, \ldots, n\}}} \sum_{l'=0}^l \frac{c_{\sigma\tau\rho\varsigma}}{2} \mathcal{L}^{\Lambda, \Lambda_0, l'}_{\vec{K}_\sigma \vec{L}_\rho^\ddag M}(\vec{q}_\sigma,\vec{q}_\rho,-k) \left( \partial_\Lambda C^{\Lambda, \Lambda_0}_{MN}(k) \right) \mathcal{L}^{\Lambda, \Lambda_0, l-l'}_{N \vec{K}_\tau \vec{L}_\varsigma^\ddag}(k,\vec{q}_\tau,\vec{q}_\varsigma)
\end{splitequation}
with
\begin{equation}
\label{k_def}
k \equiv \sum_{i \in \sigma \cup \rho} q_i = - \sum_{i \in \tau \cup \varsigma} q_i \eqend{,}
\end{equation}
where $c$ and $c_{\sigma\tau\rho\varsigma}$ are some constants stemming from the anticommutating nature of fermionic fields and antifields, if they appear among the $\vec{K}$ or $\vec{L}^\ddag$. One can now inductively prove bounds for the expansion coefficients, since the functionals on the right-hand side are either of lower order in $\hbar$, or, if they are of the same order as the left-hand side, have a smaller number of external fields and antifields. The only exception can appear for the functionals in the second line of~\eqref{l_0op_flow_hierarchy}, when $l' = 0$ and the first functional has only one or two external fields or antifields, or when $l' = l$ and the second functional has only one or two external fields or antifields. However, since we are working with amputated functionals, these vanish identically. Then induction thus ascends in $m+n+2l$, and for fixed $m+n+2l$, ascends in $l$. A similar hierarchy of flow equations is obtained from~\eqref{l_sop_flow} for the functionals with composite operator insertions, which we however refrain from spelling out in detail. Since the source term in~\eqref{l_sop_flow} involves functionals with a lower number of insertions, the induction also ascends in the number of insertions $s$, or more specifically, in $m+n+2l+2s$.

While a priori all boundary conditions are fixed at $\Lambda = \Lambda_0$, where the functionals are equal to the bare interaction Lagrangian, one can also set boundary conditions at some other scale, \eg, some fixed renormalisation scale $\Lambda = \mu$. This can be done since there is a one-to-one correspondence between conditions at $\Lambda = \Lambda_0$ and conditions given for some other value of $\Lambda$, namely
\begin{equation}
\mathcal{L}^{\Lambda_0, \Lambda_0, l}_{\vec{K} \vec{L}^\ddag}(\vec{q}) = \mathcal{L}^{\Lambda, \Lambda_0, l}_{\vec{K} \vec{L}^\ddag}(\vec{q}) + \int_{\Lambda}^{\Lambda_0} \partial_\lambda \mathcal{L}^{\lambda, \Lambda_0, l}_{\vec{K} \vec{L}^\ddag}(\vec{q}) \total \lambda \eqend{,}
\end{equation}
where the $\lambda$ derivative is given by the right-hand side of the flow equation~\eqref{l_0op_flow_hierarchy}, which is already completely determined by previous induction steps. Moreover, one can even fix boundary conditions at some fixed momenta, and then use Taylor's theorem with integral remainder to reach arbitrary momenta. In this case, one has to also bound functionals with momentum derivatives $\vec{w}$, starting for some large enough $\abs{\vec{w}}$ and then descending in $\abs{\vec{w}}$. In this way, the integral remainder is bounded in a previous induction step, and the inductive scheme closes if one can fix the functionals for some large $\abs{\vec{w}}$. Concretely, restricting to power-counting renormalisable theories, the interaction Lagrangian only contains terms of engineering dimension $\leq 4$, such that for irrelevant functionals which have $\abs{\vec{K}} + \abs{\vec{L}^\ddag} + \abs{\vec{w}} > 4$ we have vanishing boundary conditions at $\Lambda = \Lambda_0$ for all momenta. Marginal functionals for which $\abs{\vec{K}} + \abs{\vec{L}^\ddag} + \abs{\vec{w}} = 4$ have their boundary condition fixed at $\Lambda = \mu$ and vanishing momenta for simplicity, while relevant functionals with $\abs{\vec{K}} + \abs{\vec{L}^\ddag} + \abs{\vec{w}} < 4$ must be given vanishing boundary conditions at $\Lambda = 0$ and vanishing momenta to avoid unphysical IR divergences. For the marginal functionals, it would be also possible to give boundary conditions at $\Lambda = 0$, but then one has to restrict to non-exceptional momenta to avoid IR divergences -- which are physical in this case, and cannot be avoided. In general, the correspondence between different sets of boundary conditions will be quite complicated, but it is unique at least in perturbation theory since the difference will always have been bounded in previous induction steps. Note also that by fixing the boundary conditions in the way described, it is unnecessary to determine the precise form of the counterterms, which are automatically generated by the flow and can be obtained by evaluating the functionals at the (unphysical) point $\Lambda = \Lambda_0$. The physical limit is obtained as $\Lambda \to 0$ and $\Lambda_0 \to \infty$, and giving arbitrary, finite boundary conditions for the marginal functionals, this limit is finite for non-exceptional momenta as shown in the next section. Similarly, for the functionals with insertions of a composite operator $\op_A$ one fixes vanishing boundary conditions at $\Lambda = \Lambda_0$ for all momenta for the irrelevant functionals with $\abs{\vec{K}} + \abs{\vec{L}^\ddag} + \abs{\vec{w}} > [\op_A]$, arbitrary finite boundary conditions at $\Lambda = \mu$ and vanishing momenta for the marginal functionals with $\abs{\vec{K}} + \abs{\vec{L}^\ddag} + \abs{\vec{w}} = [\op_A]$ and vanishing boundary conditions at $\Lambda = \mu$ and vanishing momenta for the relevant functionals with $\abs{\vec{K}} + \abs{\vec{L}^\ddag} + \abs{\vec{w}} < [\op_A]$.

\section{Previous bounds}
\label{sec_previous}

The bounds that were inductively proven in Ref.~\cite{froebhollandhollands2015} for functionals without and with insertions of composite operators are expressed using fully reduced weighted trees. These trees essentially correspond to the tree-level Feynman graph for the corresponding correlation function, and thus make precise the notion that in perturbation theory, correlation functions are only modified logarithmically by loop corrections. However, in order to simplify the proofs, the concrete form of the trees is independent of any real $n$-point interaction that may be present in the theory, and they do not depend in any way on spin, colour or Lorentz indices. In this way, the bounds are very general and apply to any massless theory (as well as to massive theories, but there they are not optimal).

\subsection{Trees}

Concretely, our trees are the usual connected graphs consisting of vertices and lines but without loops, which we assume that that reader is familiar with. As in Ref.~\cite{froebhollandhollands2015}, we define
\begin{definition}
A weighted tree $T$ of order $(m+n,r)$ has $m+n$ external and $r$ internal vertices, and a tree $T^*$ of order $(m+n,r)$ has $m+n$ external, $r$ internal and one special vertex. External vertices have valency $1$ (\ie, exactly one line is incident to them), and may only be connected by a line to internal or special vertices. Internal vertices have valency between $1$ and $4$, and special vertices may have any valency. We require that $m+n \geq 1$ and $r \geq 1$ for trees $T$ (\ie, at least one external and one internal vertex), but do not impose further conditions on trees $T^*$ (\ie, a tree may consist only of the special vertex). The external vertices are numbered from $1$ to $m+n$, and a momentum $q$ is assigned to each of them. The momenta assigned to the lines are determined by imposing momentum conservation at each vertex (which is allowed because later on we will have $q_{m+n} = - \sum_{i=1}^{m+n-1} q_i$), except for the special vertex. Afterwards, internal vertices are assigned the momentum with highest absolute value among the momenta assigned to all lines incident to that vertex. Furthermore, we associate to each external vertex $v_e$ an index $K_e$ (or alternatively $L^\ddag_e$) and a dimension $[v_e] = [\phi_{K_e}] \in [1,3]$ (or $[v_e] = [\phi^\ddag_{L_e}] \in [1,3]$), and an overall derivative multiindex $\vec{w}$ to the tree.
\end{definition}
To reduce notational clutter, we will use $T$ also for a generic tree, when it is clear from the context which tree is meant (\ie, with or without a special vertex), or if a formula applies to all trees. To each tree, we assign a weight factor which appears in the bounds.
\begin{definition}
The weight factor $\mathsf{G}^{T,\vec{w}}_{\vec{K} \vec{L}^\ddag; [v_p]}(\vec{q}; \mu, \Lambda)$ associated to a tree $T$ is given by multiplying the weight factors assigned to each vertex and line of $T$ given in table~\ref{table_weights}, the so-called particular weight factor given by
\begin{equation}
\label{particular_weight}
\mathsf{G}^p(\vec{q}; \Lambda) = \sup(\abs{\vec{q}}, \mu, \Lambda)^{[v_p]}
\end{equation}
for a dimension $[v_p] \in \mathbb{R}$, and the derivative weight factor $\mathsf{G}^\vec{w}(\vec{q}; \Lambda)$, given by
\begin{equation}
\label{gw_def}
\mathsf{G}^\vec{w}(\vec{q}; \Lambda) \equiv \prod_{i=1}^{m+n} \begin{cases} \sup(\eta_{q_i}(\vec{q}), \Lambda)^{-\abs{w_i}} & \text{for trees $T$} \\ \sup(\bar{\eta}_{q_i}(\vec{q}), \Lambda)^{-\abs{w_i}} & \text{for trees $T^*$} \eqend{.} \end{cases}
\end{equation}
Note that for the trees $T$ with momentum conservation we require $w_{m+n} = 0$ in order to be consistent with the definition of $\eta_{q_i}$~\eqref{eta_i_def}. Since in this case the last momentum is determined by overall momentum conservation $q_{m+n} = - \sum_{i=1}^{m+n-1} q_i$, derivatives with respect to $q_n$ can be converted into derivatives with respect to the other $q_i$, and no problem arises.
\end{definition}
\begin{table}[ht]
\begin{center}
\begin{tabular}{cll}
\toprule
\multicolumn{2}{c}{Component} & \multicolumn{1}{c}{Associated weight} \\
\midrule
\raisebox{.3em}{\includegraphics{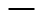}} & line $l$ & $\mathsf{G}^l(q; \mu, \Lambda) = \sup(\abs{q},\Lambda)^{-2}$ \\
\includegraphics{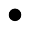} & external vertex $v_e$ & $\mathsf{G}^{v_e}(q; \mu, \Lambda) = \sup(\abs{q},\Lambda)^{3-[v_e]}$ \\
\includegraphics{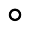} & internal vertex $v_i$ of valence $k$ & $\mathsf{G}^{v_i}(q; \mu, \Lambda) = \sup(\abs{q},\Lambda)^{4-k}$ \\
\raisebox{-.25\height}{\includegraphics{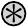}} & special vertex $v_s$ of valence $k$ & $\mathsf{G}^{v_s}(q; \mu, \Lambda) = \sup(\mu,\Lambda)^{-k}$ \\
\bottomrule
\end{tabular}
\end{center}
\caption{Weights $\mathsf{G}$ assigned to components of a tree of order $(n,r)$. $q$ always refers to the momentum associated to the component.}
\label{table_weights}
\end{table}
The tree itself is also assigned a dimension $[T]$, given by the sum of the exponents of all weight factors
\begin{equation}
[T] \equiv \sum_{v_e} ( 3 - [v_e] ) + \sum_{v_i} ( 4 - k_i ) + [v_p] - k_s - 2 N_l - \abs{\vec{w}} \eqend{,}
\end{equation}
where the sums run over all external vertices $v_e$ and all internal vertices $v_i$ with $k_i$ the valency of $v_i$, and $N_l$ is the number of lines and $k_s$ the valency of the special vertex (or $k_s = 0$ if no special vertex exists). This dimension measures the scaling of the tree weight, \ie, we have
\begin{equation}
\label{tree_scaling}
\lim_{\Lambda \to \infty} \mathsf{G}^{T,\vec{w}}_{\vec{K} \vec{L}^\ddag; [v_p]}(\vec{q}; \mu, \Lambda) \Lambda^{-[T]} = 1 \eqend{.}
\end{equation}
It is possible to obtain a simpler expression for the tree dimension, given by
\begin{lemma}
The tree dimension $[T]$ can be expressed as
\begin{splitequation}
\label{t_dim_def}
[T] &= 4 + [v_p] - \sum_{v_e} [v_e] - \abs{\vec{w}} = 4 + [v_p] - [\vec{K}] - [\vec{L}^\ddag] - \abs{\vec{w}} \\
[T^*] &= [v_p] - \sum_{v_e} [v_e] - \abs{\vec{w}} = [v_p] - [\vec{K}] - [\vec{L}^\ddag] - \abs{\vec{w}} \eqend{.}
\end{splitequation}
\end{lemma}
The proof of this lemma and the remaining ones of this section can be found in Ref.~\cite{froebhollandhollands2015}. We call a tree relevant if $[T] > 0$, marginal if $[T] = 0$ and irrelevant if $[T] < 0$. Since the smallest difference in engineering dimensions is given by $\Delta$, obviously $[T]$ is a multiple of $\Delta$, and we have the bounds $[T] \geq \Delta$ for relevant and $[T] \leq - \Delta$ for irrelevant trees.

One then defines various operations on trees, which may change its associated weight. To not prolong the exposition, here we only present a short overview and refer the reader to Ref.~\cite{froebhollandhollands2015} for more details. There are three different reduction operations:
\begin{itemize}
\item Remove an internal vertex of valence $2$, and fuse the incident lines into one line
\item Remove an internal vertex of valence $1$ and the incident line if it is connected to an internal vertex
\item Remove an internal vertex of valence $1$ and the incident line if it is connected to a particular or special vertex.
\end{itemize}
For all these operations, one easily calculates that $\mathsf{G}^T \leq \mathsf{G}^{T'}$ where $T'$ is the tree obtained from $T$ by the reduction, and thus bounds can only deteriorate by passing to the reduced tree. A fully reduced tree is one where all possible reduction operations have been done, and we denote the set of fully reduced trees $T$ of order $(m+n,r)$ with arbitrary $r$ by $\mathcal{T}_{m+n}$, while the corresponding set of fully reduced special trees $T^*$ is denoted by $\mathcal{T}^*_{m+n}$. It is then easy to see that these sets are finite for any fixed number $m+n$ of external vertices, since the fully reduced trees only contain internal vertices of valence $3$ and $4$ (except for $\mathcal{T}_1$, which consists of one tree with a one-valent vertex, and $\mathcal{T}_2$, which consists of one tree with a two-valent vertex).

It is also possible to fuse two trees: If both trees have a special vertex, fusing is done by merging the two special vertices into one. If only one or neither of both trees has a special vertex, fusion is only possible if $T_1$ has an external vertex $v_1$ with momentum $-k$ and $T_2$ has an external vertex $v_2$ with momentum $k$. If $T_1$ has no special vertex, $v_1$ must be the last external vertex of $T_1$, and if $T_2$ has no special vertex, $v_2$ must be the first external vertex of $T_2$. The trees are then fused by removing both external vertices and combining the incident lines into one line. For the change in weights, we obtain
\begin{lemma}
Fusing two trees $T_1^*$ and $T_2^*$ with special vertices into a tree $T^*$, the change in weight factors can be estimated by
\begin{equation}
\label{gw_fused_1_est}
\mathsf{G}^{T_1^*,\vec{w}_1}_{\vec{K}_1 \vec{L}^\ddag_1; [v_{p1}]}(\vec{q}_1; \mu, \Lambda) \mathsf{G}^{T_2^*,\vec{w}_2}_{\vec{K}_2 \vec{L}^\ddag_2; [v_{p2}]}(\vec{q}_2; \mu, \Lambda) \leq \mathsf{G}^{T^*,\vec{w}_1+\vec{w}_2}_{\vec{K}_1 \vec{L}^\ddag_1 \vec{K}_2 \vec{L}^\ddag_2; [v_{p1}]+[v_{p2}]}(\vec{q}_1, \vec{q}_2; \mu, \Lambda) \eqend{.}
\end{equation}
Fusing two trees $T_1$ and $T_2$ where at most one has a special vertex, under the assumption that the derivative weight factor acting on the momentum $k$ (of the line of $T_2$ which is combined in the fusion) was obtained by deriving the functional that is bounded by $T_2$ w.r.t. some specific $q_j$ (since $k = \sum_{i=1}^{n_1} q_i$), the change in weight factors is given by
\begin{equation}
\label{gw_fused_2_est}
\mathsf{G}^{T_1,\vec{w}_1}_{\vec{K}_1 \vec{L}^\ddag_1 M; [v_{p1}]}(\vec{q}_1, - k; \mu, \Lambda) \mathsf{G}^{T_2,\vec{w}_2}_{N \vec{K}_2 \vec{L}^\ddag_2; [v_{p2}]}(k, \vec{q}_2; \mu, \Lambda) \leq \frac{\mathsf{G}^{T,\vec{w}_1+\vec{w}_2}_{\vec{K}_1 \vec{L}^\ddag_1 \vec{K}_2 \vec{L}^\ddag_2; [v_{p1}]+[v_{p2}]}(\vec{q}_1, \vec{q}_2; \mu, \Lambda)}{\sup(\abs{k},\Lambda)^{[v_M]+[v_N]-4}} \eqend{.}
\end{equation}
\end{lemma}
Furthermore, for an external vertex with vanishing momentum we also amputate this vertex and the corresponding incident line. The change in weights for the amputation is given by
\begin{lemma}
Amputating an external vertex $v$ from a tree $T$ without a special vertex, the change in weights can be estimated by
\begin{equation}
\label{amputate}
\mathsf{G}^{T,\vec{w}}_{M \vec{K} \vec{L}^\ddag; [v_p]}(0, \vec{q}; \mu, \Lambda) \leq \frac{\Lambda^{1-[v]}}{\sup(\inf(\mu, \eta(\vec{q})),\Lambda)} \mathsf{G}^{T',\vec{w}}_{\vec{K} \vec{L}^\ddag; [v_p]}(\vec{q}; \mu, \Lambda) \eqend{,}
\end{equation}
while for trees $T^*$ with a special vertex we have the same estimate with $\eta$ replaced by $\bar{\eta}$.
\end{lemma}
If necessary, one has to perform additional reduction operations afterwards to obtain a fully reduced tree.

Lastly we present some inequalities for weight factors which are necessary to integrate the flow equations in $\Lambda$:
\begin{lemma}
For $\lambda \geq \Lambda$ and any tree $T$ with $[T] \leq 0$, we have
\begin{equation}
\label{t_irr_ineq2}
\mathsf{G}^{T,\vec{w}}_{\vec{K} \vec{L}^\ddag}(\vec{q}; \mu, \lambda) \leq \left( \frac{\sup(\inf(\mu, \eta(\vec{q})), \Lambda)}{\sup(\inf(\mu, \eta(\vec{q})), \lambda)} \right)^\epsilon \mathsf{G}^{T,\vec{w}}_{\vec{K} \vec{L}^\ddag}(\vec{q}; \mu, \Lambda)
\end{equation}
for any $0 \leq \epsilon \leq -[T]$. For trees $T^*$ with a special vertex where momentum is not conserved, the same estimate is valid, with $\eta$ replaced by $\bar{\eta}$.
\end{lemma}
\begin{lemma}
For $0 \leq t \leq 1$ and any tree $T$ with $[T] \geq 0$, we have
\begin{equation}
\label{t_rel_ineq1}
\mathsf{G}^{T,\vec{w}}_{\vec{K} \vec{L}^\ddag}(t \vec{q}; \Lambda, \Lambda) \leq \mathsf{G}^{T,\vec{w}}_{\vec{K} \vec{L}^\ddag}(\vec{q}; \Lambda, \Lambda) \eqend{.}
\end{equation}
\end{lemma}
\begin{lemma}
For $\Lambda \leq \lambda \leq \mu$ and any tree $T^*$ with a special vertex with $[T] \geq 0$, we have
\begin{equation}
\label{t_rel_ineq3}
\mathsf{G}^{T^*,\vec{w}}_{\vec{K} \vec{L}^\ddag}(\vec{q}; \mu, \lambda) \leq \mathsf{G}^{T^*,\vec{w}}_{\vec{K} \vec{L}^\ddag}(\vec{q}; \mu, \Lambda) \eqend{.}
\end{equation}
\end{lemma}

\subsection{Bounds on functionals without and with one insertion of a composite operator}

For the functionals without insertions of composite operators, from Ref.~\cite{froebhollandhollands2015} we have
\begin{proposition}
\label{thm_l0}
For all multiindices $\vec{w}$, at each order $l$ in perturbation theory and for an arbitrary number $m$ of external fields $\vec{K}$ and $n$ antifields $\vec{L}^\ddag$, we have the bound
\begin{equation}
\label{bound_l0}
\abs{ \partial^\vec{w} \mathcal{L}^{\Lambda, \Lambda_0, l}_{\vec{K} \vec{L}^\ddag}(\vec{q}) } \leq \sum_{T \in \mathcal{T}_{m+n}} \mathsf{G}^{T,\vec{w}}_{\vec{K} \vec{L}^\ddag}(\vec{q}; \mu, \Lambda) \,\mathcal{P}\left( \ln_+ \frac{\sup\left( \abs{\vec{q}}, \mu \right)}{\sup(\inf(\mu, \eta(\vec{q})), \Lambda)}, \ln_+ \frac{\Lambda}{\mu} \right) \eqend{,}
\end{equation}
where $\mathcal{P}$ is a polynomial with non-negative coefficients (depending on $m,n,l,\abs{\vec{w}}$ and the renormalisation conditions). The sum runs over all fully reduced trees $T$ of order $(m+n,r)$ with arbitrary $r$ (the number of internal vertices), where the dimension of the external vertices is given by the dimension of the corresponding operator (\ie, $[v_1] = [\phi_{K_1}]$, \etc).
\end{proposition}
This shows uniform boundedness of the functionals as $\Lambda_0 \to \infty$, as long as the external momenta are non-exceptional such that $\eta(\vec{q}) > 0$. Since it is well known that correlation functions in massless theories are IR-divergent in Fourier space for exceptional momenta, this is the expected result. For convergence one further needs bounds on the derivative with respect to $\Lambda_0$ (since in principle the functionals could oscillate within the bounds). These bounds were also proven in Ref.~\cite{froebhollandhollands2015}, but we do not state them here since they are not needed later on. We also note that while these bounds apply for arbitrary (finite) boundary conditions for the marginal functionals, to obtain proper Ward identities for gauge theories it may be necessary to choose special conditions, or to ``repair'' inappropriately chosen ones subsequently by an additional finite renormalisation, as explained in Ref.~\cite{froebhollandhollands2015}.

For functionals with one insertion of a composite operator $\op_A$, we first use the shift property~\eqref{func_sop_shift} with $y = x$ to obtain a functional with one insertion at $x = 0$. We then have
\begin{proposition}
\label{thm_l1}
For all multiindices $\vec{w}$, at each order $l$ in perturbation theory and for an arbitrary number $m$ of external fields $\vec{K}$ and $n$ antifields $\vec{L}^\ddag$ and any composite operator $\op_A$, we have the bound
\begin{splitequation}
\label{bound_l1}
&\abs{ \partial^\vec{w} \mathcal{L}^{\Lambda, \Lambda_0, l}_{\vec{K} \vec{L}^\ddag}\left( \op_A(0); \vec{q} \right) } \leq \sup\left( 1, \frac{\abs{\vec{q}}}{\sup(\mu, \Lambda)} \right)^{g^{(1)}([\op_A],m+n+2l,\abs{\vec{w}})} \\
&\qquad\times \sum_{T^* \in \mathcal{T}^*_{m+n}} \mathsf{G}^{T^*,\vec{w}}_{\vec{K} \vec{L}^\ddag; [\op_A]}(\vec{q}; \mu, \Lambda) \, \mathcal{P}\left( \ln_+ \frac{\sup\left( \abs{\vec{q}}, \mu \right)}{\sup(\inf(\mu, \bar{\eta}(\vec{q})), \Lambda)}, \ln_+ \frac{\Lambda}{\mu} \right) \eqend{,}
\end{splitequation}
where the sum runs over all fully reduced trees $T^*$ with one special vertex where momentum is not conserved.
\end{proposition}
This bound shows uniform boundedness as long as $\bar{\eta}(\vec{q}) > 0$, and to show convergence again an additional bound on the $\Lambda_0$ derivative is needed, which is stated in Ref.~\cite{froebhollandhollands2015}. In contrast to the case without insertions, this bound involves a loop-order dependent so-called large momentum factor with a function $g^{(s)}$ defined for $[\op] \geq 0$, $r \geq 0$ and $s \geq 1$ by
\begin{equation}
\label{gs_def}
g^{(s)}([\op],r,\abs{w}) \equiv ([\op]+s) (r+3s-3) + \sup([\op]+s-\abs{w},0) \eqend{,}
\end{equation}
which for all $u+v \leq w$ and all $w'$ fulfils the properties
\begin{equations}
g^{(s)}([\op],r,\abs{v}) &\leq g^{(s)}([\op],r+1,\abs{w}) \eqend{,} \label{gs_prop_1} \\
g^{(s)}([\op],r,\abs{w}) &\leq g^{(s)}([\op],r,\abs{v}) \eqend{,} \label{gs_prop_1a} \\
g^{(s)}([\op],r,\abs{w}+1) + 1 &\leq g^{(s)}([\op],r,\abs{w}) \qquad\text{ for } \abs{w} \leq [\op]+s-1 \eqend{,} \label{gs_prop_2} \\
g^{(s)}([\op],r,\abs{w}) &\leq g^{(s)}([\op],r+r',\abs{w}) - r' ([\op]+s) \eqend{,} \label{gs_prop_2a} \\
g^{(s)}([\op],r,\abs{u}) + g^{(s')}([\op'],r',\abs{v}) &\leq g^{(s+s')}([\op]+[\op'], r+r'-2, \abs{w'}) - ([\op]+[\op']+s+s') \eqend{.} \label{gs_prop_3}
\end{equations}

To prove these bounds, one needs a couple of Lemmas, which we here reproduce from Ref.~\cite{froebhollandhollands2015} since they are also needed later for the proofs in Section~\ref{sec_bounds}.

\begin{lemma}[Fractional derivatives]
\label{lemma_frac}
For rapidly decreasing $f(p)$ (such that we can perform integration by parts without boundary terms) with $\abs{\partial^w f(p)} \leq M^{-\abs{w}} \abs{f(p)}$, for $k \in \mathbb{N}_0$, $0 < \epsilon < 1$ and an arbitrary direction $\alpha \in \{1,2,3,4\}$, we have
\begin{equation}
\abs{\int \mathe^{-\mathi x p} f(p) \total^4 p} \leq ( \abs{x^\alpha} M )^{-k} \int \abs{f(p)} \total^4 p
\end{equation}
and
\begin{equation}
\abs{\int \mathe^{-\mathi x p} f(p) \total^4 p} \leq \frac{4}{1-\epsilon} ( \abs{x^\alpha} M )^{-k+\epsilon} \int \left( \frac{\abs{p^\alpha}}{M} \right)^{-1+\epsilon} \left( 1 + \frac{\abs{p^\alpha}}{M} \right) \abs{f(p)} \total^4 p \eqend{.}
\end{equation}
\end{lemma}

\begin{lemma}[$\Lambda$ integration]
\label{lemma_lambdaint}
For $a_0, A, K, L \geq 0$, $k,l \in \mathbb{N}_0$ and $m < -1$, we have
\begin{equation}
\int_{a_0}^{a_1} \sup(A, x)^m \ln_+^k \frac{K}{x} \ln_+^l \frac{x}{L} \total x \leq \sup(A, a_0)^{m+1} \, \mathcal{P}\left( \ln_+ \frac{\sup(K, A)}{\sup(a_0, \inf(A,L))}, \ln_+ \frac{a_0}{L} \right) \eqend{.}
\end{equation}
\end{lemma}

\begin{lemma}[$\Lambda$ integration, part 2]
\label{lemma_lambdaint2}
For $m > -1$, $a_1 \geq a_0 \geq 0$, $b,K,L \geq 0$ and $k,l \in \mathbb{N}_0$ we have
\begin{equation}
\int_{a_0}^{a_1} \sup(x,b)^m \ln_+^k \frac{K}{x} \ln_+^l \frac{x}{L} \total x \leq \sup(c,a_1)^{m+1} \, \mathcal{P}\left( \ln_+ \frac{K}{\sup(c,a_1)}, \ln_+ \frac{a_1}{L} \right) \eqend{.}
\end{equation}
for any $c \geq b$.
\end{lemma}

\begin{lemma}[$\Lambda$ integration, part 3]
\label{lemma_lambdaint3}
For $b \geq a \geq 0$, $c,K \geq 0$ and $k \in \mathbb{N}_0$ we have
\begin{equation}
\int_a^b \frac{1}{\sup(c,x)} \ln_+^k \frac{K}{x} \total x \leq \mathcal{P}\left( \ln_+ \frac{\sup(K, b)}{\sup(c,a)} \right) \eqend{.}
\end{equation}
\end{lemma}

\begin{lemma}[$p$ integration]
\label{lemma_pint}
For any function $f(x) \geq 0$ such that
\begin{equation}
\int \mathe^{- \alpha \abs{x}^2} f(x) \total^4 x < \infty
\end{equation}
for all $\alpha > 0$, and for $\beta_i \geq 1$ we have
\begin{equation}
\int \mathe^{-\alpha \abs{x}^2} f(x) \prod_{i=1}^n \sup(\abs{x + a_i}, \beta_i)^{m_i} \total^4 x \leq c \prod_{i=1}^n \sup(\abs{a_i}, \beta_i)^{m_i}
\end{equation}
for some positive constant $c$.
\end{lemma}

\begin{lemma}[$p$ integration, part 2]
\label{lemma_pint2}
For $\delta_i > 0$, $\gamma_i \geq 1$ and with the other assumptions as in Lemma~\ref{lemma_pint}, we have
\begin{splitequation}
\label{pint2_hyp}
&\int \mathe^{-\alpha \abs{x}^2} f(x) \prod_{k=1}^s \sup\left( \abs{\vec{b}_k, x, -x}, \gamma_k \right)^{\delta_k} \prod_{j=1}^t \sup\left( \eta_{d_j}(\vec{d},x,-x), \gamma_{s+j} \right)^{-\delta_{s+j}} \prod_{i=1}^n \sup(\abs{x + a_i}, \beta_i)^{m_i} \total^4 x \\
&\qquad\leq c \prod_{k=1}^s \sup\left( \abs{\vec{b}_k}, \gamma_k \right)^{\delta_k} \prod_{j=1}^t \sup\left( \eta_{d_j}(\vec{d}), \gamma_{s+j} \right)^{-\delta_{s+j}} \prod_{i=1}^n \sup(\abs{a_i}, \beta_i)^{m_i} \eqend{,}
\end{splitequation}
where $\eta_j$ is defined in equation~\eqref{eta_i_def}. The same estimate is valid if we replace $\eta_j$ by $\bar{\eta}_j$, defined in equation~\eqref{bareta_i_def}.
\end{lemma}


\subsection{Ward identities}

Gauge invariance of the generating functionals in the physical limit $\Lambda \to 0$, $\Lambda_0 \to \infty$ is expressed using certain Ward identities which relate different functionals with each other. For a general choice of boundary conditions, the Ward identities are violated by anomalous terms. However, as asserted in Theorem~\ref{thm2}, if an appropriate equivariant cohomology is empty, it is possible to choose boundary conditions such that the anomalous terms vanish. The procedure is explained in detail in Ref.~\cite{froebhollandhollands2015}, and here we only state the identities which are important later on. For this, we need the classical Slavnov-Taylor differential of the free theory $\st_0$; as an example, its action on fields is given by equation~\eqref{stq_qed_free} for QED. Since in general its action on fields and antifields is linear, we can apply it to the generating functionals without obtaining divergences, even in the physical limit. The Ward identities that have been proven in Ref.~\cite{froebhollandhollands2015} under the assumptions of Theorem~\ref{thm2} are then given by
\begin{equation}
\label{ward_l0}
\st_0 L^{0, \infty} = 0
\end{equation}
for the functionals without insertions and
\begin{equation}
\label{ward_l1}
\st_0 L^{0, \infty}\left( \op_A(x) \right) = L^{0, \infty}\left( \left( \stq \op_A \right)(x) \right) = \sum_B \stQ_A{}^B L^{0, \infty}\left( \op_B(x) \right)
\end{equation}
for the functionals with one insertion of a composite operator, with the quantum Slavnov-Taylor differential $\stq$~\eqref{stq_matrix_def}. For the functionals with more than one insertion of a composite operator, instead of the connected functionals $L^{\Lambda, \Lambda_0}\left( \bigotimes_{k=1}^s \op_{A_k}(x_k) \right)$ it is more useful for later calculations to look at the disconnected functionals
\begin{equation}
\label{g_sop_in_ls}
G^{\Lambda, \Lambda_0}\left( \bigotimes_{k=1}^s \op_{A_k}(x_k) \right) \equiv \sum_{k=1}^s (-\hbar)^{s-k} \sum_{\subline{\sigma_1 \cup \cdots \cup \sigma_k = \{1,\ldots,s\} \\ \sigma_i \neq \emptyset}} \prod_{i=1}^k L^{\Lambda, \Lambda_0}\left( \bigotimes_{j \in \sigma_i} \op_{A_j}(x_j) \right) \eqend{.}
\end{equation}
For those functionals, we have the Ward identity
\begin{splitequation}
\label{ward_gs}
&\st_0 G^{0, \infty}\left( \bigotimes_{k=1}^s \op_{A_k}(x_k) \right) = \sum_{l=1}^s G^{0, \infty}\left( \bigotimes_{k \in \{1,\ldots,s\} \setminus \{l\}} \op_{A_k}(x_k) \otimes (\stq \op_{A_l})(x_l) \right) \\
&\qquad\qquad- \hbar \sum_{1 \leq l < l' \leq s} G^{0, \infty}\left( \bigotimes_{k \in \{1,\ldots,s\} \setminus \{l,l'\}} \op_{A_k}(x_k) \otimes \left( \op_{A_l}(x_l), \op_{A_{l'}}(x_{l'}) \right)_\hbar \right) \\
&\quad= \sum_{l=1}^s \sum_{E\colon [\op_E] \leq [\op_{A_l}]+1} \stQ_{A_l}{}^E G^{0, \infty}\left( \bigotimes_{k \in \{1,\ldots,s\} \setminus \{l\}} \op_{A_k}(x_k) \otimes \op_E(x_l) \right) \\
&\qquad- \hbar \sum_{\subline{1 \leq l < l' \leq s \\ E\colon [\op_E] \leq [\op_{A_l}]+[\op_{A_{l'}}]-3}} G^{0, \infty}\left( \bigotimes_{k \in \{1,\ldots,s\} \setminus \{l,l'\}} \op_{A_k}(x_k) \otimes \op_E(x_l) \right) \sum_w \bvq^{E,w}_{A_l A_{l'}} \partial^w_{x_l} \delta^4(x_l-x_{l'}) \eqend{,}
\end{splitequation}
which additionally involves the quantum antibracket $( \cdot, \cdot )_\hbar$~\eqref{bvq_matrix_def}. These Ward identities form the base for the proof of gauge invariance of the OPE coefficients, Theorem~\ref{thm2}, and the recursion formulas for the quantum Slavnov-Taylor differential and the quantum antibracket, Theorem~\ref{thm5}. Note that if we would have proven convergence of the OPE as done for scalar field theory~\cite{hollandskopper2012,hollandetal2014}, a simple application of $\st_0$ to both sides of the OPE~\eqref{thm1_ope} would directly give the gauge invariance of the coefficients~\eqref{thm2_ward}. In our case, since we only have asymptotic equality, this results only in asymptotic gauge invariance, and we need an additional step to prove Theorem~\ref{thm2}.

\section{Bounds on functionals}
\label{sec_bounds}

We have to derive bounds on certain combinations of the connected functionals with $s$ insertions of composite operators. All these functionals have divergences when some of the points of the operator insertions scale together, and we also need to consider more regular versions of these functionals where these divergences are subtracted up to a certain order $D$. Instead of taking combinations of the connected functionals, we can also define these new functionals through a flow equation and boundary conditions. To state these conditions succinctly, we introduce the derivative operator $\mathcal{D}^A_{\vec{p}}$ by
\begin{equation}
\label{der_op_def}
\mathcal{D}^{A}_{\vec{p}} F\left( \phi, \phi^\ddag \right) \equiv \left[ \frac{\mathi^{\abs{\vec{w}}+\abs{\vec{w}^\ddag}}}{(\vec{w}+\vec{w}^\ddag)!} \partial^\vec{w}_\vec{q} \partial^{\vec{w}^\ddag}_{\vec{q}^\ddag} \frac{\delta}{\delta \phi_{\vec{K}'}(\vec{q})} \frac{\delta}{\delta \phi_{\vec{L}^{\ddag\prime}}(\vec{q}^\ddag)} F\left( \phi, \phi^\ddag \right) \right]_{\phi = \phi^\ddag = 0, \vec{q}+\vec{q}^\ddag = \vec{p}} \eqend{.}
\end{equation}
The boundary conditions for functionals with one operator insertion can then be written in the short form
\begin{equations}[l_1op_bdy_d]
\mathcal{D}^B_{\vec{0}} L^{\mu, \Lambda_0}\left( \op_A(0) \right) &= \delta^B_A \delta_{l,0} \quad\text{ for } [\op_B] \leq [\op_A] \eqend{,} \\
\mathcal{D}^B_{\vec{q}} L^{\Lambda_0, \Lambda_0}\left( \op_A(0) \right) &= \rlap{0}\phantom{\delta^B_A \delta_{l,0}} \quad\text{ for } [\op_B] > [\op_A] \eqend{.}
\end{equations}

We then first define the so-called oversubtracted, almost disconnected functionals with $s$ insertions of composite operators $F^{\Lambda, \Lambda_0}_D\left( \bigotimes_{k=1}^s \op_{A_k} \right)$ by the flow equation
\begin{splitequation}
\label{f_sop_flow}
\partial_\Lambda F^{\Lambda, \Lambda_0}_D\left( \bigotimes_{k=1}^s \op_{A_k}(x_k) \right) &= \frac{\hbar}{2} \left\langle \frac{\delta}{\delta \phi_K}, \left( \partial_\Lambda C^{\Lambda, \Lambda_0}_{KL} \right) \ast \frac{\delta}{\delta \phi_L} \right\rangle F^{\Lambda, \Lambda_0}_D\left( \bigotimes_{k=1}^s \op_{A_k}(x_k) \right) \\
&\quad- \left\langle \frac{\delta}{\delta \phi_K} L^{\Lambda, \Lambda_0}, \left( \partial_\Lambda C^{\Lambda, \Lambda_0}_{KL} \right) \ast \frac{\delta}{\delta \phi_L} F^{\Lambda, \Lambda_0}_D\left( \bigotimes_{k=1}^s \op_{A_k}(x_k) \right) \right\rangle \\
&\quad- \sum_{1 \leq k < k' \leq s} \left\langle \frac{\delta}{\delta \phi_K} L^{\Lambda, \Lambda_0}\left( \op_{A_k}(x_k) \right), \left( \partial_\Lambda C^{\Lambda, \Lambda_0}_{KL} \right) \ast \frac{\delta}{\delta \phi_L} L^{\Lambda, \Lambda_0}\left( \op_{A_{k'}}(x_{k'}) \right) \right\rangle \\
&\hspace{10em}\times \prod_{k'' \in \{1, \ldots, s\} \setminus \{k,k'\}} L^{\Lambda, \Lambda_0}\left( \op_{A_{k''}}(x_{k''}) \right) \eqend{,}
\end{splitequation}
and the boundary conditions
\begin{equations}[f_sop_bdy_d]
\mathcal{D}^B_{\vec{0}} F^{\mu, \Lambda_0}_D\left( \bigotimes_{k=1}^s \op_{A_k}(x_k) \right) &= 0 \quad\text{ for } [\op_B] < D \eqend{,} \\
\mathcal{D}^B_{\vec{q}} F^{\Lambda_0, \Lambda_0}_D\left( \bigotimes_{k=1}^s \op_{A_k}(x_k) \right) &= 0 \quad\text{ for } [\op_B] \geq D
\end{equations}
with $x_s = 0$ understood. For $D = 0$, these are the disconnected correlation functions but with the completely disconnected part removed. The oversubtracted disconnected functionals, which we call $G^{\Lambda, \Lambda_0}_D\left( \bigotimes_{k=1}^s \op_{A_k} \right)$, are obtained by adding back the completely disconnected part
\begin{equation}
\label{g_sop_def}
G^{\Lambda, \Lambda_0}_D\left( \bigotimes_{k=1}^s \op_{A_k}(x_k) \right) \equiv \prod_{k=1}^s L^{\Lambda, \Lambda_0}\left( \op_{A_k}(x_k) \right) - \hbar F^{\Lambda, \Lambda_0}_D\left( \bigotimes_{k=1}^s \op_{A_k}(x_k) \right) \eqend{.}
\end{equation}
For $D = 0$, these reduce to the usual disconnected functionals with composite operator insertions~\eqref{g_sop_in_ls}, since both sides of that equation fulfil the same linear flow equation and the same boundary conditions and thus agree, as can be easily checked. We also need functionals where only a subset of divergences is oversubtracted, and first define the almost disconnected, partially oversubtracted functionals $H^{\Lambda_0, \Lambda_0}_D\left( \bigotimes_{k=1}^{s'} \op_{A_k}(x_k); \bigotimes_{l=s'+1}^s \op_{A_l}(x_l) \right)$ by the flow equation
\begin{splitequation}
\label{h_sop_flow}
&\partial_\Lambda H^{\Lambda, \Lambda_0}_D\left( \bigotimes_{k=1}^{s'} \op_{A_k}(x_k); \bigotimes_{l=s'+1}^s \op_{A_l}(x_l) \right) \\
&= \frac{\hbar}{2} \left\langle \frac{\delta}{\delta \phi_K}, \left( \partial_\Lambda C^{\Lambda, \Lambda_0}_{KL} \right) \ast \frac{\delta}{\delta \phi_L} \right\rangle H^{\Lambda, \Lambda_0}_D\left( \bigotimes_{k=1}^{s'} \op_{A_k}(x_k); \bigotimes_{l=s'+1}^s \op_{A_l}(x_l) \right) \\
&\qquad- \left\langle \frac{\delta}{\delta \phi_K} L^{\Lambda, \Lambda_0}, \left( \partial_\Lambda C^{\Lambda, \Lambda_0}_{KL} \right) \ast \frac{\delta}{\delta \phi_L} H^{\Lambda, \Lambda_0}_D\left( \bigotimes_{k=1}^{s'} \op_{A_k}(x_k); \bigotimes_{l=s'+1}^s \op_{A_l}(x_l) \right) \right\rangle \\
&\qquad- \left\langle \frac{\delta}{\delta \phi_K} G^{\Lambda, \Lambda_0}_D\left( \bigotimes_{k=1}^{s'} \op_{A_k}(x_k) \right), \left( \partial_\Lambda C^{\Lambda, \Lambda_0}_{KL} \right) \ast \frac{\delta}{\delta \phi_L} G^{\Lambda, \Lambda_0}\left( \bigotimes_{l=s'+1}^s \op_{A_l}(x_l) \right) \right\rangle
\end{splitequation}
and the boundary conditions
\begin{equation}
\label{h_sop_bdy}
\mathcal{D}^B_{\vec{q}} H^{\Lambda_0, \Lambda_0}_D\left( \bigotimes_{k=1}^{s'} \op_{A_k}(x_k); \bigotimes_{l=s'+1}^s \op_{A_l}(x_l) \right) = 0 \eqend{.}
\end{equation}
These functionals contain all contributions where at least one pair of composite operators $\op_{A_i}$, $\op_{A_j}$ with $1 \leq i \leq s' < j \leq s$ belongs to the same connected functional, and we will see later that those divergences are softened where the first subset of operators scales together. Lastly, we define the partially oversubtracted, disconnected functionals $G^{\Lambda, \Lambda_0}_D\left( \bigotimes_{k=1}^{s'} \op_{A_k}(x_k); \bigotimes_{l=s'+1}^s \op_{A_l}(x_l) \right)$ by
\begin{splitequation}
\label{g_ssop_def}
G^{\Lambda, \Lambda_0}_D\left( \bigotimes_{k=1}^{s'} \op_{A_k}(x_k); \bigotimes_{l=s'+1}^s \op_{A_l}(x_l) \right) &\equiv G^{\Lambda, \Lambda_0}_D\left( \bigotimes_{k=1}^{s'} \op_{A_k}(x_k) \right) G^{\Lambda, \Lambda_0}\left( \bigotimes_{l=s'+1}^s \op_{A_l}(x_l) \right) \\
&\quad- \hbar H^{\Lambda, \Lambda_0}_D\left( \bigotimes_{k=1}^{s'} \op_{A_k}(x_k); \bigotimes_{l=s'+1}^s \op_{A_l}(x_l) \right) \eqend{;}
\end{splitequation}
which also have softened divergences as the first subset of operators scales together. While in the end we are only interested in the oversubtracted, disconnected functionals $G^{\Lambda, \Lambda_0}_D\left( \bigotimes_{k=1}^s \op_{A_k}(x_k) \right)$ and the partially oversubtracted, disconnected functionals $G^{\Lambda, \Lambda_0}_D\left( \bigotimes_{k=1}^{s'} \op_{A_k}(x_k); \bigotimes_{l=s'+1}^s \op_{A_l}(x_l) \right)$, the plethora of other functionals is needed for technical reasons. In the next subsections, we derive suitable bounds for these functionals.

\subsection{Bounds for \texorpdfstring{$F$}{F} functionals}

In this section, we want to derive bounds for the oversubtracted, almost disconnected functionals $F^{\Lambda, \Lambda_0}_D\left( \bigotimes_{k=1}^s \op_{A_k}(x_k) \right)$ which are sharp with respect to the dependence on the positions $x_k$ of the composite operator insertions. Expanding the functionals in the flow equation~\eqref{f_sop_flow} in external fields and antifields and $\hbar$, we arrive at
\begin{splitequation}
\label{fs_flow_hierarchy}
&\partial_\Lambda \mathcal{F}^{\Lambda, \Lambda_0, l}_{\vec{K} \vec{L}^\ddag; D}\left( \bigotimes_{k=1}^s \op_{A_k}; \vec{q} \right) = \frac{c}{2} \int \left( \partial_\Lambda C^{\Lambda, \Lambda_0}_{MN}(-p) \right) \mathcal{F}^{\Lambda, \Lambda_0, l-1}_{MN \vec{K} \vec{L}^\ddag; D}\left( \bigotimes_{k=1}^s \op_{A_k}; p,-p,\vec{q} \right) \frac{\total^4 p}{(2\pi)^4} \\
&\quad- \sum_{\subline{\sigma \cup \tau = \{1, \ldots, m\} \\ \rho \cup \varsigma = \{1, \ldots, n\} }} \sum_{l'=0}^l c_{\sigma\tau\rho\varsigma} \mathcal{L}^{\Lambda, \Lambda_0, l'}_{\vec{K}_\sigma \vec{L}_\rho^\ddag M}(\vec{q}_\sigma,\vec{q}_\rho,-k) \left( \partial_\Lambda C^{\Lambda, \Lambda_0}_{MN}(k) \right) \mathcal{F}^{\Lambda, \Lambda_0, l-l'}_{N \vec{K}_\tau \vec{L}_\varsigma^\ddag; D}\left( \bigotimes_{k=1}^s \op_{A_k}; k,\vec{q}_\tau,\vec{q}_\varsigma \right) \\
&\quad- \sum_{1 \leq k < k' \leq s} \sum_{\subline{\sigma_1 \cup \cdots \cup \sigma_s = \{1, \ldots, m\} \\ \rho_1 \cup \cdots \cup \rho_s = \{1, \ldots, n\} }} \sum_{l_1 + \cdots + l_s = l} c_{\{\sigma_i\} \{\rho_j\}} \int \mathcal{L}^{\Lambda, \Lambda_0, l_k}_{\vec{K}_{\sigma_k} \vec{L}_{\rho_k}^\ddag M}\left( \op_{A_k}; \vec{q}_{\sigma_k},\vec{q}_{\rho_k},p \right) \left( \partial_\Lambda C^{\Lambda, \Lambda_0}_{MN}(-p) \right) \\
&\hspace{4em}\times \mathcal{L}^{\Lambda, \Lambda_0, l_{k'}}_{N \vec{K}_{\sigma_{k'}} \vec{L}_{\rho_{k'}}^\ddag}\left( \op_{A_{k'}};-p,\vec{q}_{\sigma_{k'}},\vec{q}_{\rho_{k'}} \right) \frac{\total^4 p}{(2\pi)^4} \prod_{k'' \in \{1,\ldots,s\}\setminus\{k,k'\}} \mathcal{L}^{\Lambda, \Lambda_0, l_{k''}}_{\vec{K}_{\sigma_{k''}} \vec{L}_{\rho_{k''}}^\ddag}\left( \op_{A_{k''}};\vec{q}_{\sigma_{k''}},\vec{q}_{\rho_{k''}} \right) \eqend{,}
\end{splitequation}
where the momentum $k$ is defined in~\eqref{k_def}. We then want to show
\begin{proposition}
\label{thm_fs}
At each order $l$ in perturbation theory and for an arbitrary number $m$ of external fields $\vec{K}$ and $n$ antifields $\vec{L}^\ddag$, the almost disconnected functionals with $s \geq 2$ insertions of arbitrary (non-integrated) composite operators $\op_{A_i}$, oversubtracted at order $D$ with $0 \leq D \leq [\op_\vec{A}]$, satisfy the bound
\begin{splitequation}
\label{bound_fs}
&\abs{ \partial^\vec{w} \mathcal{F}^{\Lambda, \Lambda_0, l}_{\vec{K} \vec{L}^\ddag; D}\left( \bigotimes_{k=1}^s \op_{A_k}(x_k); \vec{q} \right) } \leq \mu^{[\op_\vec{A}]-D+\epsilon} \sup\left( 1, \frac{\abs{\vec{q}}}{\sup(\mu,\Lambda)} \right)^{g^{(s)}\left( [\op_\vec{A}], m+n+2l, \abs{\vec{w}} \right)} \\
&\qquad\times \Xi^{\Lambda, \Lambda_1}_{\sup(\abs{\vec{w}},D), [\op_\vec{A}]-D+\epsilon}(\vec{x}) \sum_{T^* \in \mathcal{T}^*_{m+n}} \mathsf{G}^{T^*,\vec{w}}_{\vec{K} \vec{L}^\ddag; D-\epsilon}(\vec{q}; \mu, \Lambda) \, \mathcal{P}\left( \ln_+ \frac{\sup\left( \abs{\vec{q}}, \mu \right)}{\sup(\inf(\mu, \bar{\eta}(\vec{q})), \Lambda)}, \ln_+ \frac{\Lambda}{\mu} \right)
\end{splitequation}
for $x_s = 0$, with the functions
\begin{equations}[xi_lambda_def]
\Xi^{\Lambda, \Lambda_1}_{p, p'}(\vec{x}) &\equiv \begin{cases} \Xi^{(1)}_{p, p', \rho}(\vec{x}) & \Lambda \geq \Lambda_1 \\ \sup\left[ \Xi^{(1)}_{p, p', \rho}(\vec{x}), \Xi^{(2)}_p(\vec{x}) \right] & \Lambda < \Lambda_1 \eqend{,} \end{cases} \\
\Xi^{(1)}_{p, p', \rho}(\vec{x}) &\equiv \left( \frac{\sup_{i \in \{1,\ldots,s\}} \abs{x_i - x_s}}{\inf_{1 \leq k < k' \leq s} \abs{x_k - x_{k'}}} \right)^p \left( \mu \inf_{1 \leq k < k' \leq s} \abs{x_k-x_{k'}} \right)^{-p'-\rho} \left( \frac{\mu}{\Lambda_1} \right)^{p'+\rho} \eqend{,} \\
\Xi^{(2)}_p(\vec{x}) &\equiv \sup\left( 1, \mu \sup_{i \in \{1,\ldots,s\}} \abs{x_i-x_s} \right)^p
\end{equations}
for all $0 < \epsilon < \Delta$, an arbitrary $0 < \Lambda_1 \leq \mu$ and an arbitrary $\rho \geq 0$, where
\begin{equation}
[\op_\vec{A}] \equiv \sum_{k=1}^s [\op_{A_k}] \eqend{.}
\end{equation}
\end{proposition}
Note that the coefficients of the polynomial depend on $\epsilon$, and diverge as $\epsilon \to 0$.

For the $\Lambda$ derivative, we want to prove the bounds
\begin{splitequation}
\label{bound_fs_lambdaderiv}
&\abs{ \partial_\Lambda \partial^\vec{w} \mathcal{F}^{\Lambda, \Lambda_0, l}_{\vec{K} \vec{L}^\ddag; D}\left( \bigotimes_{k=1}^s \op_{A_k}(x_k); \vec{q} \right) } \leq \frac{1}{\sup(\inf(\mu, \bar{\eta}(\vec{q})),\Lambda)} \mu^{[\op_\vec{A}]-D+\epsilon} \, \Xi^{\Lambda, \Lambda_1}_{\sup(\abs{\vec{w}},D), [\op_\vec{A}]-D+\epsilon}(\vec{x}) \\
&\quad\times \sup\left( 1, \frac{\abs{\vec{q}}}{\sup(\mu,\Lambda)} \right)^{g^{(s)}\left( [\op_\vec{A}], m+n+2l, \abs{\vec{w}} \right)} \sum_{T^* \in \mathcal{T}^*_{m+n}} \mathsf{G}^{T^*,\vec{w}}_{\vec{K} \vec{L}^\ddag; D-\epsilon}(\vec{q}; \mu, \Lambda) \, \mathcal{P}\left( \ln_+ \frac{\sup\left( \abs{\vec{q}}, \mu \right)}{\Lambda}, \ln_+ \frac{\Lambda}{\mu} \right) \eqend{,}
\end{splitequation}
a bound very similar to~\eqref{bound_fs}. Taking $w$ derivatives with respect to the momenta $q$ of the flow hierarchy~\eqref{fs_flow_hierarchy}, let us denote the linear term in the first line of the right-hand side by $F_1$, the quadratic term in the second line by $F_2$ and the source term in the last line by $F_3$.

To estimate the linear term, we use the bound~\eqref{bound_fs} for the functional and the bound~\eqref{prop_abl} for the covariance to obtain
\begin{splitequation}
\abs{F_1} &\leq \mu^{[\op_\vec{A}]-D+\epsilon} \, \Xi^{\Lambda, \Lambda_1}_{\sup(\abs{\vec{w}},D), [\op_\vec{A}]-D+\epsilon}(\vec{x}) \int \sup(\abs{p}, \Lambda)^{-5+[\phi_M]+[\phi_N]} \, \mathe^{-\frac{\abs{p}^2}{2 \Lambda^2}} \\
&\quad\times \sup\left( 1, \frac{\abs{\vec{q},p,-p}}{\sup(\mu,\Lambda)} \right)^{g^{(s)}\left( [\op_\vec{A}], m+n+2l, \abs{\vec{w}} \right)} \\
&\quad\times \sum_{T^* \in \mathcal{T}^*_{m+n+2}} \mathsf{G}^{T^*,\vec{w}}_{MN \vec{K} \vec{L}^\ddag; D-\epsilon}(p,-p,\vec{q}; \mu, \Lambda) \, \mathcal{P}\left( \ln_+ \frac{\sup\left( \abs{\vec{q},p,-p}, \mu \right)}{\Lambda}, \ln_+ \frac{\Lambda}{\mu} \right) \frac{\total^4 p}{(2\pi)^4} \eqend{,}
\end{splitequation}
since $\bar{\eta}(\vec{q},p,-p) = 0$. We then use the estimates
\begin{equations}[fs_qp_factor]
\abs{\vec{q},p,-p} &\leq 2 \abs{p} + \abs{\vec{q}} \eqend{,} \\
\ln_+ \frac{\sup\left( 2 \abs{p} + \abs{\vec{q}}, \mu \right)}{\Lambda} &\leq \ln_+ \frac{\sup\left( \abs{\vec{q}}, \mu \right)}{\Lambda} + \ln_+ \frac{\abs{p}}{\Lambda} + \ln 2
\end{equations}
to simplify the polynomial in logarithms and the large-momentum factor, and use Lemma~\ref{lemma_pint2} to perform the $p$ integral. This results in
\begin{splitequation}
\abs{F_1} &\leq \mu^{[\op_\vec{A}]-D+\epsilon} \, \Xi^{\Lambda, \Lambda_1}_{\sup(\abs{\vec{w}},D), [\op_\vec{A}]-D+\epsilon}(\vec{x}) \Lambda^{-1+[\phi_M]+[\phi_N]} \sup\left( 1, \frac{\abs{\vec{q}}}{\sup(\mu,\Lambda)} \right)^{g^{(s)}\left( [\op_\vec{A}], m+n+2l, \abs{\vec{w}} \right)} \\
&\quad\times \sum_{T^* \in \mathcal{T}^*_{m+n+2}} \mathsf{G}^{T^*,\vec{w}}_{MN \vec{K} \vec{L}^\ddag; D-\epsilon}(0,0,\vec{q}; \mu, \Lambda) \, \mathcal{P}\left( \ln_+ \frac{\sup\left( \abs{\vec{q}}, \mu \right)}{\Lambda}, \ln_+ \frac{\Lambda}{\mu} \right) \eqend{.} \raisetag{1.7\baselineskip}
\end{splitequation}
For each tree $T^*$ in the sum, we now amputate the first two external vertices with zero momentum corresponding to $M$ and $N$. The amputation gives us an extra factor~\eqref{amputate}
\begin{equation}
\frac{\Lambda^{2-[\phi_M]-[\phi_N]}}{\sup(\inf(\mu, \bar{\eta}(\vec{q})),\Lambda)^2} \leq \frac{\Lambda^{1-[\phi_M]-[\phi_N]}}{\sup(\inf(\mu, \bar{\eta}(\vec{q})),\Lambda)} \eqend{,}
\end{equation}
and the new tree $T^{*\prime}$ has $m+n$ external vertices such that $T^{*\prime} \in \mathcal{T}_{m+n}$, and thus we obtain a bound of the form~\eqref{bound_fs_lambdaderiv} for $F_1$.

For the quadratic term, we obtain using the bounds~\eqref{bound_fs}, \eqref{bound_l0} and~\eqref{prop_abl}
\begin{splitequation}
\abs{F_2} &\leq \sum_{\subline{\sigma \cup \tau = \{1, \ldots, m\} \\ \rho \cup \varsigma = \{1, \ldots, n\}}} \sum_{l'=0}^l \sum_{\vec{u}+\vec{v} \leq \vec{w}} \mathe^{-\frac{\abs{k}^2}{2 \Lambda^2}} \mu^{[\op_\vec{A}]-D+\epsilon} \, \Xi^{\Lambda, \Lambda_1}_{\sup(\abs{\vec{v}},D), [\op_\vec{A}]-D+\epsilon}(\vec{x}) \\
&\quad\times \sum_{T \in \mathcal{T}_{\abs{\sigma} + \abs{\rho}+1}} \mathsf{G}^{T,\vec{u}}_{\vec{K}_\sigma \vec{L}_\rho^\ddag M}(\vec{q}_\sigma,\vec{q}_\rho,-k; \mu, \Lambda) \,\mathcal{P}\left( \ln_+ \frac{\sup\left( \abs{\vec{q}_\sigma,\vec{q}_\rho,-k}, \mu \right)}{\Lambda}, \ln_+ \frac{\Lambda}{\mu} \right) \\
&\quad\times \sup\left( 1, \frac{\abs{k,\vec{q}_\tau,\vec{q}_\varsigma}}{\sup(\mu,\Lambda)} \right)^{g^{(s)}\left( [\op_\vec{A}], \abs{\tau}+\abs{\varsigma}+1+2(l-l'), \abs{\vec{v}} \right)} \sup(\abs{k}, \Lambda)^{-5+[\phi_M]+[\phi_N]-\abs{\vec{w}}+\abs{\vec{u}}+\abs{\vec{v}}} \\
&\quad\times \sum_{T^* \in \mathcal{T}^*_{\abs{\tau}+\abs{\varsigma}+1}} \mathsf{G}^{T^*,\vec{v}}_{N \vec{K}_\tau \vec{L}_\varsigma^\ddag; D-\epsilon}(k,\vec{q}_\tau,\vec{q}_\varsigma; \mu, \Lambda) \, \mathcal{P}\left( \ln_+ \frac{\sup\left( \abs{k,\vec{q}_\tau,\vec{q}_\varsigma}, \mu \right)}{\Lambda}, \ln_+ \frac{\Lambda}{\mu} \right) \eqend{,} \raisetag{2.1\baselineskip}
\end{splitequation}
since the definition of $k$~\eqref{k_def} shows that $\bar{\eta}(k,\vec{q}_\tau,\vec{q}_\varsigma) = 0$ and we estimate $\eta(\vec{q}_\sigma,\vec{q}_\rho,-k) \geq 0$. Using that
\begin{equation}
\abs{\vec{q}_\sigma,\vec{q}_\rho,-k}, \abs{k,\vec{q}_\tau,\vec{q}_\varsigma} \leq \abs{\vec{q}}
\end{equation}
we can estimate the polynomials in logarithms and the large-momentum factor. Since we have
\begin{equation}
\abs{\tau}+\abs{\varsigma}+1+2(l-l') \leq m+n+2l-1
\end{equation}
(obvious for $l' > 0$, while for $l' = 0$ we may assume that $\abs{\tau}+\abs{\varsigma} \leq m+n-1$ since otherwise the functional without operator insertions, and thus $F_2$, vanishes), property~\eqref{gs_prop_1} of $g^{(s)}$ shows us that
\begin{equation}
g^{(s)}\left( [\op_\vec{A}], \abs{\tau}+\abs{\varsigma}+1+2(l-l'), \abs{\vec{v}} \right) \leq g^{(s)}\left( [\op_\vec{A}], m+n+2l, \abs{\vec{w}} \right)
\end{equation}
which allows us to estimate the large-momentum factor. We then fuse the trees according to the estimate~\eqref{gw_fused_2_est}, obtaining
\begin{splitequation}
\abs{F_2} &\leq \sum_{\subline{\sigma \cup \tau = \{1, \ldots, m\} \\ \rho \cup \varsigma = \{1, \ldots, n\}}} \sum_{l'=0}^l \sum_{\vec{u}+\vec{v} \leq \vec{w}} \mu^{[\op_\vec{A}]-D+\epsilon} \, \Xi^{\Lambda, \Lambda_1}_{\sup(\abs{\vec{v}},D), [\op_\vec{A}]-D+\epsilon}(\vec{x}) \, \mathe^{-\frac{\abs{k}^2}{2 \Lambda^2}} \sup(\abs{k}, \Lambda)^{-1-\abs{\vec{w}}+\abs{\vec{u}}+\abs{\vec{v}}} \\
&\quad\times \sup\left( 1, \frac{\abs{\vec{q}}}{\sup(\mu,\Lambda)} \right)^{g^{(s)}\left( [\op_\vec{A}], m+n+2l, \abs{\vec{w}} \right)} \sum_{T^* \in \mathcal{T}^*_{m+n}} \mathsf{G}^{T^*,\vec{v}+\vec{u}}_{\vec{K} \vec{L}^\ddag; D-\epsilon}(\vec{q}; \mu, \Lambda) \, \mathcal{P}\left( \ln_+ \frac{\sup\left( \abs{\vec{q}}, \mu \right)}{\Lambda}, \ln_+ \frac{\Lambda}{\mu} \right) \eqend{.}
\end{splitequation}
Since $\Xi^{\Lambda, \Lambda_1}_{p, p'} \leq \Xi^{\Lambda, \Lambda_1}_{p+q,p'}$ for all $q \geq 0$, we can estimate this factor changing $\sup(\abs{\vec{v}},D)$ to $\sup(\abs{\vec{w}},D)$. It then remains to change the $\vec{u}+\vec{v}$ derivatives acting on the tree to $\vec{w}$ derivatives. Since $\bar{\eta}_{q_i}(\vec{q}) \leq \abs{k}$ for any $i$, we have
\begin{equation}
\label{fs_estfuse_a}
\sup(\abs{k},\Lambda)^{-\abs{\vec{w}}+\abs{\vec{u}}+\abs{\vec{v}}} \leq \prod_{i=1}^{m+n} \sup(\bar{\eta}_{q_i}(\vec{q}),\Lambda)^{-\abs{(\vec{w}-\vec{u}-\vec{v})_i}}
\end{equation}
and thus (remembering the definition of the derivative weight factor~\eqref{gw_def})
\begin{equation}
\label{fs_estfuse_b}
\sup(\abs{k},\Lambda)^{-\abs{\vec{w}}+\abs{\vec{u}}+\abs{\vec{v}}} \mathsf{G}^{T^*,\vec{u}+\vec{v}}_{\vec{K} \vec{L}^\ddag; D-\epsilon}(\vec{q}; \mu, \Lambda) \leq \mathsf{G}^{T^*,\vec{w}}_{\vec{K} \vec{L}^\ddag; D-\epsilon}(\vec{q}; \mu, \Lambda) \eqend{.}
\end{equation}
The last estimate
\begin{equation}
\label{fs_estexp}
\sup(\abs{k}, \Lambda)^{-1} \, \mathe^{-\frac{\abs{k}^2}{2 \Lambda^2}} \leq \sup(\inf(\mu, \bar{\eta}(\vec{q})), \Lambda)^{-1}
\end{equation}
then gives us the required bound~\eqref{bound_fs_lambdaderiv}.

To estimate the source term, we use the shift property~\eqref{func_sop_shift} to bring all operator insertions to $x_k = 0$ and then use the bounds~\eqref{bound_l1} such that
\begin{splitequation}
\abs{F_3} &\leq \sum_{1 \leq k < k' \leq s} \sum_{\subline{\sigma_1 \cup \cdots \cup \sigma_s = \{1, \ldots, m\} \\ \rho_1 \cup \cdots \cup \rho_s = \{1, \ldots, n\} }} \sum_{l_1 + \cdots + l_s = l} c_{\{\sigma_i\} \{\rho_j\}} \sum_{\vec{v}_1 + \cdots + \vec{v}_s = \vec{w}} c_{\{\vec{v}_i\} \vec{w}} \prod_{i=1}^s \sum_{\vec{u}_i \leq \vec{v}_i} c_{\vec{u}_i \vec{v}_i} \\
&\quad\times \prod_{k'' \in \{1,\ldots,s\}\setminus\{k,k'\}} \abs{ \partial^{\vec{u}_{k''}} \mathe^{- \mathi x_{k''} \left( \sum_{i \in \sigma_{k''} \cup \rho_{k''}} q_i \right)} } \abs{ \partial^{\vec{v}_{k''} - \vec{u}_{k''}} \mathcal{L}^{\Lambda, \Lambda_0, l_{k''}}_{\vec{K}_{\sigma_{k''}} \vec{L}_{\rho_{k''}}^\ddag}\left( \op_{A_{k''}}(0); \vec{q}_{\sigma_{k''}},\vec{q}_{\rho_{k''}} \right) } \\
&\quad\times \abs{ \partial^{\vec{u}_k} \mathe^{- \mathi x_k \sum_{i \in \sigma_k \cup \rho_k} q_i} } \abs{ \partial^{\vec{u}_{k'}} \mathe^{- \mathi x_{k'} \sum_{i \in \sigma_{k'} \cup \rho_{k'}} q_i} } \left\lvert \int \left( \partial^{\vec{v}_k - \vec{u}_k} \mathcal{L}^{\Lambda, \Lambda_0, l_k}_{\vec{K}_{\sigma_k} \vec{L}_{\rho_k}^\ddag M}\left( \op_{A_k}(0); \vec{q}_{\sigma_k},\vec{q}_{\rho_k},p \right) \right) \right. \\
&\hspace{4em}\times \left. \mathe^{- \mathi (x_k-x_{k'}) p} \left( \partial_\Lambda C^{\Lambda, \Lambda_0}_{MN}(-p) \right) \left( \partial^{\vec{v}_{k'} - \vec{u}_{k'}} \mathcal{L}^{\Lambda, \Lambda_0, l_{k'}}_{N \vec{K}_{\sigma_{k'}} \vec{L}_{\rho_{k'}}^\ddag}\left( \op_{A_{k'}}(0); -p,\vec{q}_{\sigma_{k'}},\vec{q}_{\rho_{k'}} \right) \right) \frac{\total^4 p}{(2\pi)^4} \right\rvert \eqend{.}
\end{splitequation}
We now have to distinguish between the cases $\Lambda \geq \Lambda_1$ and $\Lambda < \Lambda_1$. In the first case, we would like to use
\begin{equation}
\exp\left( - \mathi (x_k-x_{k'}) p \right) = \frac{\mathi^{\abs{a}}}{(x_k-x_{k'})^a} \partial_p^a \exp\left( - \mathi (x_k-x_{k'}) p \right)
\end{equation}
to generate additional $p$ derivatives and integrate them by parts. This then generates additional negative powers of $\Lambda$, which are needed to integrate the flow equation downwards. Concretely, the bound~\eqref{bound_l1} together with the definition of the derivative weight factor~\eqref{gw_def} show that we obtain an extra factor
\begin{equation}
\sup\left( 1, \frac{\abs{\vec{q},p}}{\sup(\mu, \Lambda)} \right)^{g^{(1)}([\op_A],m+n+2l,\abs{w}+1)-g^{(1)}([\op_A],m+n+2l,\abs{w})} \sup(\bar{\eta}_p(\vec{q},p), \Lambda)^{-1}
\end{equation}
for each additional derivative w.r.t. $p$ on a functional with one insertion. Property~\eqref{gs_prop_1a} of $g^{(s)}$ shows that $g^{(s)}$ can only decrease for more derivatives, and thus we can simply estimate this factor against $\Lambda^{-1}$. Similarly, for each additional derivative acting on the derivative of the covariance we obtain a factor of
\begin{equation}
\sup(\abs{p}, \Lambda)^{-1} \leq \Lambda^{-1}
\end{equation}
according to the bound~\eqref{prop_abl}. We can therefore choose a direction $\alpha \in \{1,2,3,4\}$ such that $\abs{(x_k-x_{k'})^\alpha} \geq \abs{x_k-x_{k'}}/2$ (which is always possible), use Lemma~\ref{lemma_frac} with $M = \Lambda$ to estimate the $p$ integral, the bound~\eqref{bound_l1} for all other functionals with one operator insertion and the bound~\eqref{prop_abl} for the covariance to obtain
\begin{splitequation}
\abs{F_3} &\leq \sum_{1 \leq k < k' \leq s} \sum_{\subline{\sigma_1 \cup \cdots \cup \sigma_s = \{1, \ldots, m\} \\ \rho_1 \cup \cdots \cup \rho_s = \{1, \ldots, n\} }} \sum_{l_1 + \cdots + l_s = l} \sum_{\vec{v}_1 + \cdots + \vec{v}_s = \vec{w}} \prod_{i=1}^s \sum_{\vec{u}_i \leq \vec{v}_i} \abs{x_i}^\abs{\vec{u}_i} \\
&\quad\times \prod_{k'' \in \{1,\ldots,s\}\setminus\{k,k'\}} \sup\left( 1, \frac{\abs*[\big]{\vec{q}_{\sigma_{k''}},\vec{q}_{\rho_{k''}}}}{\sup(\mu, \Lambda)} \right)^{g^{(1)}\left( [\op_{A_{k''}}],\abs{\sigma_{k''}}+\abs{\rho_{k''}}+2l_{k''},\abs{\vec{v}_{k''}} - \abs{\vec{u}_{k''}} \right)} \\
&\quad\qquad\times \sum_{T^* \in \mathcal{T}^*_{\abs*[]{\sigma_{k''}}+\abs*[]{\rho_{k''}}}} \mathsf{G}^{T^*,\vec{v}_{k''} - \vec{u}_{k''}}_{\vec{K}_{\sigma_{k''}} \vec{L}_{\rho_{k''}}^\ddag; [\op_{A_{k''}}]}(\vec{q}_{\sigma_{k''}},\vec{q}_{\rho_{k''}}; \mu, \Lambda) \, \mathcal{P}\left( \ln_+ \frac{\sup\left( \abs*[\big]{\vec{q}_{\sigma_{k''}},\vec{q}_{\rho_{k''}}}, \mu \right)}{\Lambda}, \ln_+ \frac{\Lambda}{\mu} \right) \\
&\quad\times ( \abs{(x_k-x_{k'})^\alpha} \Lambda )^{-\nu+\epsilon} \int \left( \frac{\abs{p^\alpha}}{\Lambda} \right)^{-1+\epsilon} \left( 1 + \frac{\abs{p^\alpha}}{\Lambda} \right) \sup(\abs{p}, \Lambda)^{-5+[\phi_M]+[\phi_N]} \, \mathe^{-\frac{\abs{p}^2}{2 \Lambda^2}} \\
&\quad\qquad\times \sum_{T^* \in \mathcal{T}^*_{\abs*[]{\sigma_k}+\abs*[]{\rho_k}+1}} \mathsf{G}^{T^*,\vec{v}_k - \vec{u}_k}_{\vec{K}_{\sigma_k} \vec{L}_{\rho_k}^\ddag M; [\op_{A_k}]}(\vec{q}_{\sigma_k},\vec{q}_{\rho_k},p; \mu, \Lambda) \, \mathcal{P}\left( \ln_+ \frac{\sup\left( \abs*[\big]{\vec{q}_{\sigma_k},\vec{q}_{\rho_k},p}, \mu \right)}{\Lambda}, \ln_+ \frac{\Lambda}{\mu} \right) \\
&\quad\qquad\times \sum_{T^* \in \mathcal{T}^*_{\abs*[]{\sigma_{k'}}+\abs*[]{\rho_{k'}}+1}} \mathsf{G}^{T^*,\vec{v}_{k'} - \vec{u}_{k'}}_{N \vec{K}_{\sigma_{k'}} \vec{L}_{\rho_{k'}}^\ddag; [\op_{A_{k'}}]}(-p,\vec{q}_{\sigma_{k'}},\vec{q}_{\rho_{k'}}; \mu, \Lambda) \, \mathcal{P}\left( \ln_+ \frac{\sup\left( \abs*[\big]{-p,\vec{q}_{\sigma_{k'}},\vec{q}_{\rho_{k'}}}, \mu \right)}{\Lambda}, \ln_+ \frac{\Lambda}{\mu} \right) \\
&\quad\qquad\times \sup\left( 1, \frac{\abs*[\big]{\vec{q}_{\sigma_k},\vec{q}_{\rho_k},p}}{\sup(\mu, \Lambda)} \right)^{g^{(1)}\left( [\op_{A_k}],\abs{\sigma_k}+\abs{\rho_k}+1+2l_k,\abs{\vec{v}_k} - \abs{\vec{u}_k} \right)} \\
&\quad\qquad\times \sup\left( 1, \frac{\abs*[\big]{-p,\vec{q}_{\sigma_{k'}},\vec{q}_{\rho_{k'}}}}{\sup(\mu, \Lambda)} \right)^{g^{(1)}\left( [\op_{A_{k'}}],\abs{\sigma_{k'}}+\abs{\rho_{k'}}+1+2l_{k'},\abs{\vec{v}_{k'}} - \abs{\vec{u}_{k'}} \right)} \total^4 p \eqend{,} \raisetag{1.3\baselineskip}
\end{splitequation}
for any $\nu \in \mathbb{N}$ and all $0 < \epsilon < 1$, but since only the combination $-\nu+\epsilon$ appears, we can assume the above to hold for any $\nu > \epsilon$. In the second case where $\Lambda < \Lambda_1$, we do not integrate by parts and obtain the same estimate except for the factor
\begin{equation}
\label{fs_ibp_factor}
( \abs{(x_k-x_{k'})^\alpha} \Lambda )^{-\nu+\epsilon} \left( \frac{\abs{p^\alpha}}{\Lambda} \right)^{-1+\epsilon} \left( 1 + \frac{\abs{p^\alpha}}{\Lambda} \right) \eqend{.}
\end{equation}
We then fuse the trees according to the estimate~\eqref{gw_fused_1_est}, and use that
\begin{equation}
\abs{\vec{q}_{\sigma_k},\vec{q}_{\rho_k},p}, \abs{-p,\vec{q}_{\sigma_{k'}},\vec{q}_{\rho_{k'}}}, \abs{\vec{q}_{\sigma_{k''}},\vec{q}_{\rho_{k''}}} \leq \abs{\vec{q},p,-p}
\end{equation}
and the estimates~\eqref{fs_qp_factor} to fuse the polynomials in logarithms and the large momentum factors. We then use property~\eqref{gs_prop_3} of $g^{(s)}$ and estimate
\begin{splitequation}
&g^{(1)}\left( [\op_{A_k}],\abs{\sigma_k}+\abs{\rho_k}+1+2l_k,\abs{\vec{v}_k} - \abs{\vec{u}_k} \right) + g^{(1)}\left( [\op_{A_{k'}}],\abs{\sigma_{k'}}+\abs{\rho_{k'}}+1+2l_{k'},\abs{\vec{v}_{k'}} - \abs{\vec{u}_{k'}} \right) \\
&\quad\leq g^{(2)}\left( [\op_{A_k}]+[\op_{A_{k'}}], \abs{\sigma_k}+\abs{\rho_k}+\abs{\sigma_{k'}}+\abs{\rho_{k'}}+2(l_k+l_{k'}), \abs{\vec{v}_k} + \abs{\vec{v}_{k'}} \right) - \left( [\op_{A_k}]+[\op_{A_{k'}}]+2 \right) \eqend{,}
\end{splitequation}
and succesively the same estimate until all $g^{(1)}$ are combined. This results in
\begin{splitequation}
\abs{F_3} &\leq \sum_{1 \leq k < k' \leq s} \sum_{\vec{u}_1 + \cdots + \vec{u}_s \leq \vec{w}} \prod_{i=1}^s \abs{x_i}^\abs{\vec{u}_i} \left( \abs{(x_k-x_{k'})^\alpha} \Lambda \right)^{-\nu+\epsilon} \int \left( \frac{\abs{p^\alpha}}{\Lambda} \right)^{-1+\epsilon} \left( 1 + \frac{\abs{p^\alpha}}{\Lambda} \right) \\
&\quad\times \sup(\abs{p}, \Lambda)^{-5+[\phi_M]+[\phi_N]} \, \mathe^{-\frac{\abs{p}^2}{2 \Lambda^2}} \sup\left( 1, \frac{\abs{\vec{q},p,-p}}{\sup(\mu, \Lambda)} \right)^{g^{(s)}\left( [\op_\vec{A}], m+n-(s-2)+2l, \abs{\vec{w}} \right) - [\op_\vec{A}] - s} \\
&\qquad\times \sum_{T^* \in \mathcal{T}^*_{m+n+2}} \mathsf{G}^{T^*,\vec{w}-\vec{u}}_{\vec{K} \vec{L}^\ddag M N; [\op_\vec{A}]}(\vec{q},p,-p; \mu, \Lambda) \, \mathcal{P}\left( \ln_+ \frac{\sup\left( \abs{\vec{q}}, \mu \right)}{\Lambda}, \ln_+ \frac{\abs{p}}{\Lambda}, \ln_+ \frac{\Lambda}{\mu} \right) \total^4 p \eqend{.}
\end{splitequation}
with $\vec{u} = \sum_{i=1}^s \vec{u}_i$ if $\Lambda \geq \Lambda_1$, and the same estimate except for the factor~\eqref{fs_ibp_factor} if $\Lambda < \Lambda_1$.

We can now perform the $p$ integral using Lemma~\ref{lemma_pint2}, and estimate
\begin{equation}
\left( \abs{(x_k-x_{k'})^\alpha} \Lambda \right)^{-\nu+\epsilon} \leq 2^{-\nu+\epsilon} \left( \abs{x_k-x_{k'}} \Lambda \right)^{-\nu+\epsilon} = 2^{-\nu+\epsilon} \left( \abs{x_k-x_{k'}} \mu \right)^{-\nu+\epsilon} \left( \frac{\Lambda}{\mu} \right)^{-\nu+\epsilon} \eqend{.}
\end{equation}
The particular weight of the tree is changed from $[\op_\vec{A}]$ to $D-\epsilon$, which results in an extra factor~\eqref{particular_weight}
\begin{equation}
\sup(\abs{\vec{q}}, \mu, \Lambda)^{[\op_\vec{A}]-D+\epsilon} = \sup\left( 1, \frac{\abs{\vec{q}}}{\sup(\mu, \Lambda)} \right)^{[\op_\vec{A}]-D+\epsilon} \mu^{[\op_\vec{A}]-D+\epsilon} \sup\left(1, \frac{\Lambda}{\mu} \right)^{[\op_\vec{A}]-D+\epsilon} \eqend{,}
\end{equation}
we amputate the external legs corresponding to $M$ and $N$ obtaining an extra factor of~\eqref{amputate}
\begin{equation}
\label{amputate_p_factor}
\frac{\Lambda^{2-[\phi_M]-[\phi_N]}}{\sup(\inf(\mu, \bar{\eta}(\vec{q})),\Lambda)^2} \leq \frac{\Lambda^{1-[\phi_M]-[\phi_N]}}{\sup(\inf(\mu, \bar{\eta}(\vec{q})),\Lambda)} \eqend{,}
\end{equation}
and we change the $\vec{w}-\vec{u}$ derivatives acting on the tree to $\vec{w}$ derivatives, which gives a factor of~\eqref{gw_def}
\begin{equation}
\Lambda^\abs{\vec{u}} = \mu^\abs{\vec{u}} \left( \frac{\Lambda}{\mu} \right)^\abs{\vec{u}} \leq \mu^\abs{\vec{u}} \left[ \Theta(\Lambda_1-\Lambda) + \Theta(\Lambda-\Lambda_1) \left( \frac{\Lambda}{\mu} \right)^\abs{\vec{u}} \right]
\end{equation}
since the $\vec{u}$ derivatives are with respect to the momentum $p$ which now vanishes such that $\bar{\eta}_p(\vec{q}, p, -p) = 0$, and since $\Lambda_1 \leq \mu$. This results in
\begin{splitequation}
\abs{F_3} &\leq \frac{1}{\sup(\inf(\mu, \bar{\eta}(\vec{q})),\Lambda)} \sum_{1 \leq k < k' \leq s} \sum_{\vec{u}_1 + \cdots + \vec{u}_s \leq \vec{w}} \prod_{i=1}^s \left( \mu \abs{x_i} \right)^\abs{\vec{u}_i} \mu^{[\op_\vec{A}]-D+\epsilon} \sum_{T^* \in \mathcal{T}^*_{m+n}} \mathsf{G}^{T^*,\vec{w}}_{\vec{K} \vec{L}^\ddag; D-\epsilon}(\vec{q}; \mu, \Lambda) \\
&\quad\times \left[ \Theta(\Lambda_1-\Lambda) + \Theta(\Lambda-\Lambda_1) \left( \mu \abs{x_k-x_{k'}} \right)^{-\nu+\epsilon} \sup\left( 1, \frac{\Lambda}{\mu} \right)^{[\op_\vec{A}]-D+\epsilon} \left( \frac{\Lambda}{\mu} \right)^{\abs{\vec{u}}-\nu+\epsilon} \right] \\
&\quad\times \sup\left( 1, \frac{\abs{\vec{q}}}{\sup(\mu,\Lambda)} \right)^{g^{(s)}\left( [\op_\vec{A}], m+n-(s-2)+2l, \abs{\vec{w}} \right)-s-D+\epsilon} \mathcal{P}\left( \ln_+ \frac{\sup\left( \abs{\vec{q}}, \mu \right)}{\Lambda}, \ln_+ \frac{\Lambda}{\mu} \right) \eqend{.} \raisetag{1.3\baselineskip}
\end{splitequation}
Since $s \geq 2$, we can use property~\eqref{gs_prop_2a} of $g^{(s)}$ to estimate
\begin{equation}
\sup\left( 1, \frac{\abs{\vec{q}}}{\sup(\mu,\Lambda)} \right)^{g^{(s)}\left( [\op_\vec{A}], m+n-(s-2)+2l, \abs{\vec{w}} \right)-s-D+\epsilon} \leq \sup\left( 1, \frac{\abs{\vec{q}}}{\sup(\mu,\Lambda)} \right)^{g^{(s)}\left( [\op_\vec{A}], m+n+2l, \abs{\vec{w}} \right)} \eqend{.}
\end{equation}
Furthermore, we now choose $\nu = [\op_\vec{A}]-D+2\epsilon+\abs{\vec{u}}+\rho$ with $\rho \geq 0$, which fulfils $\nu > \epsilon$ since $D \leq [\op_\vec{A}]$ such that
\begin{equation}
\sup\left( 1, \frac{\Lambda}{\mu} \right)^{[\op_\vec{A}]-D+\epsilon} \left( \frac{\Lambda}{\mu} \right)^{\abs{\vec{u}}-\nu+\epsilon} = \sup\left( 1, \frac{\mu}{\Lambda} \right)^{[\op_\vec{A}]-D+\epsilon} \left( \frac{\Lambda}{\mu} \right)^{-\rho} \leq \left( \frac{\mu}{\Lambda_1} \right)^{[\op_\vec{A}]-D+\epsilon+\rho} \eqend{,}
\end{equation}
since $\Lambda \geq \Lambda_1$. Remembering that $x_s = 0$, the last estimates
\begin{equation}
\sum_{\vec{u}_1 + \cdots + \vec{u}_s \leq \vec{w}} \prod_{i=1}^s \left( \mu \abs{x_i} \right)^\abs{\vec{u}_i} \leq \sum_{\vec{u}_1 + \cdots + \vec{u}_s \leq \vec{w}} \prod_{i=1}^s \left( \mu \sup_{j \in \{1,\ldots,s\}} \abs{x_j} \right)^\abs{\vec{u}_i} \leq c \left( \mu \sup_{i \in \{1,\ldots,s\}} \abs{x_i} \right)^\abs{\vec{u}} \eqend{,}
\end{equation}
\begin{equation}
\sum_{1 \leq k < k' \leq s} \left( \mu \sup_{i \in \{1,\ldots,s\}} \abs{x_i} \right)^\abs{\vec{u}} \left( \mu \abs{x_k-x_{k'}} \right)^{-\abs{\vec{u}}} \leq \left( \frac{\sup_{i \in \{1,\ldots,s\}} \abs{x_i}}{\inf_{1 \leq k < k' \leq s} \abs{x_k-x_{k'}}} \right)^\abs{\vec{w}} \eqend{,}
\end{equation}
\begin{equation}
\left( \mu \sup_{i \in \{1,\ldots,s\}} \abs{x_i} \right)^\abs{\vec{u}} \leq \sup\left( 1, \mu \sup_{i \in \{1,\ldots,s\}} \abs{x_i} \right)^\abs{\vec{w}} \eqend{,}
\end{equation}
\begin{equation}
\left( \mu \abs{x_k-x_{k'}} \right)^{-\nu+\epsilon} \leq \left( \mu \inf_{1 \leq k < k' \leq s} \abs{x_k-x_{k'}} \right)^{-\nu+\epsilon}
\end{equation}
since $\nu > \epsilon$ and
\begin{equation}
\abs{\vec{w}} \leq \sup(\abs{\vec{w}},D)
\end{equation}
then result in a bound of the required form~\eqref{bound_fs_lambdaderiv} for $F_3$.

For irrelevant functionals, where $[\vec{K}] + [\vec{L}^\ddag] + \abs{\vec{w}} \geq D$ and thus $[T^*] = D - \epsilon - [\vec{K}] - [\vec{L}^\ddag] - \abs{\vec{w}} \leq - \epsilon < 0$ for the dimension of the trees appearing in the bound, we have vanishing boundary conditions at $\Lambda = \Lambda_0$, and integrate the bound~\eqref{bound_fs_lambdaderiv} downwards in $\Lambda$. This gives
\begin{splitequation}
&\abs{ \partial^\vec{w} \mathcal{F}^{\Lambda, \Lambda_0, l}_{\vec{K} \vec{L}^\ddag; D}\left( \bigotimes_{k=1}^s \op_{A_k}(x_k); \vec{q} \right)} \leq \int_\Lambda^{\Lambda_0} \abs{\partial_\lambda \partial^\vec{w} \mathcal{F}^{\lambda, \Lambda_0, l}_{\vec{K} \vec{L}^\ddag; D}\left( \bigotimes_{k=1}^s \op_{A_k}(x_k); \vec{q} \right)} \total \lambda \\
&\quad\leq \mu^{[\op_\vec{A}]-D+\epsilon} \int_\Lambda^{\Lambda_0} \Xi^{\lambda, \Lambda_1}_{\sup(\abs{\vec{w}},D), [\op_\vec{A}]-D+\epsilon}(\vec{x}) \sup\left( 1, \frac{\abs{\vec{q}}}{\sup(\mu,\lambda)} \right)^{g^{(s)}\left( [\op_\vec{A}], m+n+2l, \abs{\vec{w}} \right)} \\
&\qquad\times \frac{1}{\sup(\inf(\mu, \bar{\eta}(\vec{q})),\lambda)} \sum_{T^* \in \mathcal{T}^*_{m+n}} \mathsf{G}^{T^*,\vec{w}}_{\vec{K} \vec{L}^\ddag; D-\epsilon}(\vec{q}; \mu, \lambda) \, \mathcal{P}\left( \ln_+ \frac{\sup\left( \abs{\vec{q}}, \mu \right)}{\lambda}, \ln_+ \frac{\lambda}{\mu} \right) \total \lambda \eqend{.}
\end{splitequation}
By definition, we always have $\Xi^{\lambda, \Lambda_1} \leq \Xi^{\Lambda, \Lambda_1}$. Since $[T^*] \leq - \epsilon$, we can then use the estimate~\eqref{t_irr_ineq2} to estimate the tree weight factor at the lower bound $\lambda = \Lambda$, and the large momentum factor is trivially estimated there. An application of Lemma~\ref{lemma_lambdaint} to the remaining integral then gives the bound~\eqref{bound_fs}.

For relevant and marginal functionals, which have $[\vec{K}] + [\vec{L}^\ddag] + \abs{\vec{w}} < D$, boundary conditions are given at $\Lambda = \mu$ and vanishing momenta. Since the smallest difference in operator dimensions is given by $\Delta > 0$, we even have $[\vec{K}] + [\vec{L}^\ddag] + \abs{\vec{w}} \leq D - \Delta$ in this case, and thus $[T^*] \geq \Delta - \epsilon > 0$ if we restrict to $\epsilon < \Delta$. We then first bound the functionals at zero momentum for all $\Lambda \geq \mu$, then extend this bound to arbitrary momenta using Taylor's theorem, and then integrate the bound downwards to $\Lambda < \mu$. Naturally, the boundary conditions at $\Lambda = \mu$ have to be compatible with the bounds~\eqref{bound_fs}. For vanishing momenta, we thus have
\begin{splitequation}
&\abs{ \partial^\vec{w} \mathcal{F}^{\Lambda, \Lambda_0, l}_{\vec{K} \vec{L}^\ddag; D}\left( \bigotimes_{k=1}^s \op_{A_k}(x_k); \vec{0} \right)} \leq \abs{ \partial^\vec{w} \mathcal{F}^{\mu, \Lambda_0, l}_{\vec{K} \vec{L}^\ddag; D}\left( \bigotimes_{k=1}^s \op_{A_k}(x_k); \vec{0} \right)} + \int_\mu^\Lambda \abs{\partial_\lambda \partial^\vec{w} \mathcal{F}^{\lambda, \Lambda_0, l}_{\vec{K} \vec{L}^\ddag; D}\left( \bigotimes_{k=1}^s \op_{A_k}(x_k); \vec{0} \right)} \total \lambda \\
&\quad\leq \mu^{[\op_\vec{A}]-D+\epsilon} \, \Xi^{\Lambda, \Lambda_1}_{\sup(\abs{\vec{w}},D), [\op_\vec{A}]-D+\epsilon}(\vec{x}) \\
&\quad\qquad\times \sum_{T^* \in \mathcal{T}^*_{m+n}} \left[ c \, \mathsf{G}^{T^*,\vec{w}}_{\vec{K} \vec{L}^\ddag; D-\epsilon}(\vec{0}; \mu, \mu) + \int_\mu^\Lambda \frac{1}{\lambda} \mathsf{G}^{T^*,\vec{w}}_{\vec{K} \vec{L}^\ddag; D-\epsilon}(\vec{0}; \mu, \lambda) \mathcal{P}\left( \ln_+ \frac{\lambda}{\mu} \right) \total \lambda \right] \raisetag{1.7\baselineskip}
\end{splitequation}
since $\Xi^{\Lambda, \Lambda_1}$ is independent of $\Lambda$ for all $\Lambda \geq \Lambda_1$ and we have $\Lambda_1 \leq \mu$. At zero momentum and for $\Lambda \geq \mu$, we simply have
\begin{equation}
\mathsf{G}^{T^*,\vec{w}}_{\vec{K} \vec{L}^\ddag; D-\epsilon}(\vec{0}; \mu, \lambda) = \lambda^{[T^*]} \eqend{.}
\end{equation}
Since $[T^*] > 0$, we can estimate the first tree weight factor by
\begin{equation}
\mu^{[T^*]} \leq \Lambda^{[T^*]} = \mathsf{G}^{T^*,\vec{w}}_{\vec{K} \vec{L}^\ddag; D-\epsilon}(\vec{0}; \mu, \Lambda) \eqend{,}
\end{equation}
and an application of Lemma~\ref{lemma_lambdaint2} to the $\lambda$ integral results in
\begin{splitequation}
\label{bound_fs_zeromom}
\abs{ \partial^\vec{w} \mathcal{F}^{\Lambda, \Lambda_0, l}_{\vec{K} \vec{L}^\ddag; D}\left( \bigotimes_{k=1}^s \op_{A_k}(x_k); \vec{0} \right)} &\leq \mu^{[\op_\vec{A}]-D+\epsilon} \, \Xi^{\Lambda, \Lambda_1}_{\sup(\abs{\vec{w}},D), [\op_\vec{A}]-D+\epsilon}(\vec{x}) \\
&\quad\times \sum_{T^* \in \mathcal{T}^*_{m+n}} \mathsf{G}^{T^*,\vec{w}}_{\vec{K} \vec{L}^\ddag; D-\epsilon}(\vec{0}; \mu, \Lambda) \, \mathcal{P}\left( \ln_+ \frac{\Lambda}{\mu} \right) \eqend{.}
\end{splitequation}
If no external legs are present, this is already the correct bound~\eqref{bound_fs} and we are done. Otherwise, we extend this bound to general momenta and use Taylor's formula with integral remainder, which reads
\begin{splitequation}
\label{fs_taylor}
\partial^\vec{w} \mathcal{F}^{\Lambda, \Lambda_0, l}_{\vec{K} \vec{L}^\ddag; D}\left( \bigotimes_{k=1}^s \op_{A_k}(x_k); \vec{q} \right) &= \partial^\vec{w} \mathcal{F}^{\Lambda, \Lambda_0, l}_{\vec{K} \vec{L}^\ddag; D}\left( \bigotimes_{k=1}^s \op_{A_k}(x_k); \vec{0} \right) \\
&\quad+ \sum_{i=1}^{m+n-1} \sum_{\alpha=1}^4 \int_0^1 \frac{\partial k_i^\alpha(t)}{\partial t} \partial_{k_i^\alpha} \partial^\vec{w} \mathcal{F}^{\Lambda, \Lambda_0, l}_{\vec{K} \vec{L}^\ddag; D}\left( \bigotimes_{k=1}^s \op_{A_k}(x_k); \vec{k}(t) \right) \total t
\end{splitequation}
for some path $\vec{k}(t)$ with $\vec{k}(0) = 0$ and $\vec{k}(1) = \vec{q}$, where the result is independent of the path taken. Furthermore, the second term involves a functional which has one momentum derivative more. We thus need to bound the functionals in increasing order of relevancy; for the least relevant functional the second term is then irrelevant and has thus already be bounded, and then we ascend in relevancy. We then take the simple path $k_i(t) = t q_i$, take the absolute value of equation~\eqref{fs_taylor} and insert the bounds~\eqref{bound_fs_zeromom} and~\eqref{bound_fs}. Since the functional is relevant or marginal, we have $[\vec{K}] + [\vec{L}^\ddag] + \abs{\vec{w}} \leq D - \Delta$, and since there is at least one external leg we have $[\vec{K}] + [\vec{L}^\ddag] \geq 1$, such that $\abs{\vec{w}} \leq D - 1$ follows and thus
\begin{equation}
\sup(\abs{\vec{w}}+1,D) = D = \sup(\abs{\vec{w}},D) \eqend{,}
\end{equation}
such that
\begin{splitequation}
&\abs{ \partial^\vec{w} \mathcal{F}^{\Lambda, \Lambda_0, l}_{\vec{K} \vec{L}^\ddag; D}\left( \bigotimes_{k=1}^s \op_{A_k}(x_k); \vec{q} \right)} \leq \mu^{[\op_\vec{A}]-D+\epsilon} \, \Xi^{\Lambda, \Lambda_1}_{\sup(\abs{\vec{w}},D), [\op_\vec{A}]-D+\epsilon}(\vec{x}) \\
&\quad\times \Bigg[ \sum_{T^* \in \mathcal{T}^*_{m+n}} \mathsf{G}^{T^*,\vec{w}}_{\vec{K} \vec{L}^\ddag; D-\epsilon}(\vec{0}; \mu, \Lambda) \, \mathcal{P}\left( \ln_+ \frac{\Lambda}{\mu} \right) + \sum_{i=1}^{m+n-1} \int_0^1 \abs{q_i} \sup\left( 1, \frac{t \abs{\vec{q}}}{\Lambda} \right)^{g^{(s)}\left( [\op_\vec{A}], m+n+2l, \abs{\vec{w}}+1 \right)} \\
&\qquad\quad\times \sum_{T^* \in \mathcal{T}^*_{m+n}} \mathsf{G}^{T^*,\vec{w}+\tilde{\vec{w}}}_{\vec{K} \vec{L}^\ddag; D-\epsilon}(t \vec{q}; \mu, \Lambda) \, \mathcal{P}\left( \ln_+ \frac{\sup\left( t \abs{\vec{q}}, \mu \right)}{\Lambda}, \ln_+ \frac{\Lambda}{\mu} \right) \total t \Bigg] \eqend{,}
\end{splitequation}
where $\tilde{\vec{w}}$ is a multiindex which has only one non-zero entry corresponding to the additional $k_i^\alpha$ derivative.

We then extract the derivative weight factor corresponding to the extra derivative of the second term, which reads
\begin{equation}
\sup(\bar{\eta}_{q_i}(t \vec{q}),\Lambda)^{-1} \eqend{.}
\end{equation}
The remaining tree weight factors fulfill $[T^*] > 0$ since we deal with relevant functionals, and since $\Lambda \geq \mu$ the tree weight factors actually do not depend on $\mu$ and we can estimate them at $t = 1$ using the estimate~\eqref{t_rel_ineq1}. Since $\abs{\vec{w}}+1 \leq D \leq [\op_\vec{A}] \leq [\op_\vec{A}]+1$ and $s \geq 2$, we can use property~\eqref{gs_prop_2} of $g^{(s)}$ to estimate the large momentum factor
\begin{splitequation}
\frac{\abs{q_i}}{\sup(\bar{\eta}_{q_i}(t \vec{q}),\Lambda)} \sup\left( 1, \frac{t \abs{\vec{q}}}{\Lambda} \right)^{g^{(s)}\left( [\op_\vec{A}], m+n+2l, \abs{\vec{w}}+1 \right)} &\leq \frac{\abs{q_i}}{\Lambda} \sup\left( 1, \frac{\abs{\vec{q}}}{\Lambda} \right)^{g^{(s)}\left( [\op_\vec{A}], m+n+2l, \abs{\vec{w}}+1 \right)} \\
&\leq \sup\left( 1, \frac{\abs{\vec{q}}}{\sup(\mu, \Lambda)} \right)^{g^{(s)}\left( [\op_\vec{A}], m+n+2l, \abs{\vec{w}} \right)} \eqend{.}
\end{splitequation}
For the polynomial in logarithms, we use
\begin{equation}
\ln_+ \frac{\sup\left( t \abs{\vec{q}}, \mu \right)}{\Lambda} \leq \ln_+ \frac{\sup\left( \abs{\vec{q}}, \mu \right)}{\Lambda} = \ln_+ \frac{\sup\left( \abs{\vec{q}}, \mu \right)}{\sup(\inf(\mu, \bar{\eta}(\vec{q})), \Lambda)} \eqend{,}
\end{equation}
and the remaining $t$ integral is trivially done, such that we obtain the bounds~\eqref{bound_fs} also in this case. We then integrate the flow equation down to $\Lambda < \mu$, taking these bounds are boundary conditions. This gives
\begin{splitequation}
&\abs{ \partial^\vec{w} \mathcal{F}^{\Lambda, \Lambda_0, l}_{\vec{K} \vec{L}^\ddag; D}\left( \bigotimes_{k=1}^s \op_{A_k}(x_k); \vec{q} \right)} \leq \abs{\partial_\lambda \partial^\vec{w} \mathcal{F}^{\mu, \Lambda_0, l}_{\vec{K} \vec{L}^\ddag; D}\left( \bigotimes_{k=1}^s \op_{A_k}(x_k); \vec{q} \right)} + \int_\Lambda^{\mu} \abs{\partial_\lambda \partial^\vec{w} \mathcal{F}^{\lambda, \Lambda_0, l}_{\vec{K} \vec{L}^\ddag; D}\left( \bigotimes_{k=1}^s \op_{A_k}(x_k); \vec{q} \right)} \total \lambda \\
&\quad\leq \mu^{[\op_\vec{A}]-D+\epsilon} \, \Xi^{\mu, \Lambda_1}_{\sup(\abs{\vec{w}},D), [\op_\vec{A}]-D+\epsilon}(\vec{x}) \sup\left( 1, \frac{\abs{\vec{q}}}{\mu} \right)^{g^{(s)}\left( [\op_\vec{A}], m+n+2l, \abs{\vec{w}} \right)} \\
&\quad\qquad\times \sum_{T^* \in \mathcal{T}^*_{m+n}} \mathsf{G}^{T^*,\vec{w}}_{\vec{K} \vec{L}^\ddag; D-\epsilon}(\vec{q}; \mu, \mu) \, \mathcal{P}\left( \ln_+ \frac{\sup\left( \abs{\vec{q}}, \mu \right)}{\mu} \right) \\
&\quad+ \mu^{[\op_\vec{A}]-D+\epsilon} \int_\Lambda^\mu \Xi^{\lambda, \Lambda_1}_{\sup(\abs{\vec{w}},D), [\op_\vec{A}]-D+\epsilon}(\vec{x}) \sup\left( 1, \frac{\abs{\vec{q}}}{\sup(\mu,\lambda)} \right)^{g^{(s)}\left( [\op_\vec{A}], m+n+2l, \abs{\vec{w}} \right)} \\
&\quad\qquad\times \frac{1}{\sup(\inf(\mu, \bar{\eta}(\vec{q})),\lambda)} \sum_{T^* \in \mathcal{T}^*_{m+n}} \mathsf{G}^{T^*,\vec{w}}_{\vec{K} \vec{L}^\ddag; D-\epsilon}(\vec{q}; \mu, \lambda) \, \mathcal{P}\left( \ln_+ \frac{\sup\left( \abs{\vec{q}}, \mu \right)}{\lambda}, \ln_+ \frac{\lambda}{\mu} \right) \total \lambda \eqend{.} \raisetag{1.8\baselineskip}
\end{splitequation}
Since by definition $\Xi^{\lambda, \Lambda_1} \leq \Xi^{\Lambda, \Lambda_1}$ for all $\Lambda$, we can estimate this factor at the lower bound. Since $\Lambda < \mu$, the large-momentum factor is trivially estimated, and the trees are estimated at $\lambda = \Lambda$ using the estimate~\eqref{t_rel_ineq3}. For the first polynomial in logarithms we use
\begin{equation}
\ln_+ \frac{\sup\left( \abs{\vec{q}}, \mu \right)}{\mu} = \ln_+ \frac{\sup\left( \abs{\vec{q}}, \mu \right)}{\sup(\mu, \Lambda)} \leq \ln_+ \frac{\sup\left( \abs{\vec{q}}, \mu \right)}{\sup(\inf(\mu, \bar{\eta}(\vec{q})), \Lambda)} \eqend{,}
\end{equation}
and the remaining $\lambda$ integral can be done applying Lemma~\ref{lemma_lambdaint3}, such that we arrive at the bounds~\eqref{bound_fs}.

We next need bounds for Taylor-expanded functionals, given by
\begin{proposition}
\label{thm_fs_taylor}
At each order $l$ in perturbation theory and for an arbitrary number $m$ of external fields $\vec{K}$ and $n$ antifields $\vec{L}^\ddag$, the partially connected functionals with $s \geq 2$ insertions of arbitrary (non-integrated) composite operators $\op_{A_i}$, oversubtracted at order $D$ with $D > [\op_\vec{A}]$, satisfy the bound
\begin{splitequation}
\label{bound_fs_taylor}
&\abs{ \partial^\vec{w} \left( 1 - \sum_{n < D - [\op_\vec{A}]} \mathcal{T}^n_{\vec{x} \to \vec{0}} \right) \mathcal{F}^{\Lambda, \Lambda_0, l}_{\vec{K} \vec{L}^\ddag; D}\left( \bigotimes_{k=1}^s \op_{A_k}(x_k); \vec{q} \right) } \leq \mu^{[\op_\vec{A}]+r-D+\epsilon} \sup_{i\in\{1,\ldots,s\}} \abs{x_i}^r \\
&\quad\times \Xi^{\Lambda, \Lambda_1}_{\sup(\abs{\vec{w}},D), [\op_\vec{A}]+r-D+\epsilon}(\vec{x}) \ \sup\left( 1, \frac{\abs{\vec{q}}}{\sup(\mu,\Lambda)} \right)^{g^{(s)}\left( [\op_\vec{A}]+r, m+n+2l, \abs{\vec{w}} \right)} \\
&\quad\times \sum_{T^* \in \mathcal{T}^*_{m+n}} \mathsf{G}^{T^*,\vec{w}}_{\vec{K} \vec{L}^\ddag; D-\epsilon}(\vec{q}; \mu, \Lambda) \, \mathcal{P}\left( \ln_+ \frac{\sup\left( \abs{\vec{q}}, \mu \right)}{\sup(\inf(\mu, \bar{\eta}(\vec{q})), \Lambda)}, \ln_+ \frac{\Lambda}{\mu} \right)
\end{splitequation}
for all $0 < \epsilon < \Delta$, where $r$ is the smallest integer $\geq D-[\op_\vec{A}]$ and where we take $\rho = 0$ for the function $\Xi^{\Lambda, \Lambda_1}$ defined by equation~\eqref{xi_lambda_def}. 
\end{proposition}
A Taylor expansion with remainder gives
\begin{splitequation}
&\left( 1 - \sum_{n < D - [\op_\vec{A}]} \mathcal{T}^n_{\vec{x} \to \vec{0}} \right) F^{\Lambda, \Lambda_0}_D\left( \bigotimes_{k=1}^s \op_{A_k}(x_k) \right) = \int_0^1 \frac{r}{(1-t)} \mathcal{T}^r_{\vec{x} \to t \vec{x}} F^{\Lambda, \Lambda_0}_D\left( \bigotimes_{k=1}^s \op_{A_k}(x_k) \right) \total t \\
&\qquad\qquad= r \sum_{\abs{\vec{a}} = r} \frac{\vec{x}^\vec{a}}{\vec{a}!} \int_0^1 (1-t)^{r-1} \left[ \partial^\vec{a}_\vec{x} F^{\Lambda, \Lambda_0}_D\left( \bigotimes_{k=1}^s \op_{A_k}(x_k) \right) \right]_{\vec{x} \to t \vec{x}} \total t \eqend{,} \raisetag{1.7\baselineskip}
\end{splitequation}
where the sum only runs over multiindices $\vec{a}$ not involving the last coordinate $x_s = 0$ (since otherwise $\vec{x}^\vec{a} = 0$). We can then use Proposition~\ref{thm_lowenstein_2} to pull the derivatives inside the functional and obtain
\begin{equation}
\left( 1 - \sum_{n < D - [\op_\vec{A}]} \mathcal{T}^n_{\vec{x} \to \vec{0}} \right) F^{\Lambda, \Lambda_0}_D\left( \bigotimes_{k=1}^s \op_{A_k}(x_k) \right) = r \sum_{\abs{\vec{a}} = r} \frac{\vec{x}^\vec{a}}{\vec{a}!} \int_0^1 (1-t)^{r-1} F^{\Lambda, \Lambda_0}_D\left( \bigotimes_{k=1}^s \left( \partial^{a_k} \op_{A_k} \right)(t x_k) \right) \total t \eqend{.}
\end{equation}
Expanding in fields, antifields and $\hbar$ and using the bounds~\eqref{bound_fs}, it follows that
\begin{splitequation}
&\abs{ \partial^\vec{w} \left( 1 - \sum_{n < D - [\op_\vec{A}]} \mathcal{T}^n_{\vec{x} \to \vec{0}} \right) \mathcal{F}^{\Lambda, \Lambda_0, l}_{\vec{K} \vec{L}^\ddag; D}\left( \bigotimes_{k=1}^s \op_{A_k}(x_k); \vec{q} \right) } \leq \sup_{i\in\{1,\ldots,s\}} \abs{x_i}^r \mu^{[\op_\vec{A}]+r-D+\epsilon} \\
&\qquad\times \sup\left( 1, \frac{\abs{\vec{q}}}{\sup(\mu,\Lambda)} \right)^{g^{(s)}\left( [\op_\vec{A}]+r, m+n+2l, \abs{\vec{w}} \right)} \int_0^1 (1-t)^{r-1} \Xi^{\Lambda, \Lambda_1}_{\sup(\abs{\vec{w}},D), [\op_\vec{A}]+r-D+\epsilon}(t \vec{x}) \total t \\
&\qquad\times \sum_{T^* \in \mathcal{T}^*_{m+n}} \mathsf{G}^{T^*,\vec{w}}_{\vec{K} \vec{L}^\ddag; D-\epsilon}(\vec{q}; \mu, \Lambda) \, \mathcal{P}\left( \ln_+ \frac{\sup\left( \abs{\vec{q}}, \mu \right)}{\sup(\inf(\mu, \bar{\eta}(\vec{q})), \Lambda)}, \ln_+ \frac{\Lambda}{\mu} \right) \eqend{.} \raisetag{1.8\baselineskip}
\end{splitequation}
Since $0 \leq t \leq 1$, from the definition of $\Xi^{\Lambda, \Lambda_1}_{p,p'}$~\eqref{xi_lambda_def} we obtain the estimate
\begin{equation}
\Xi^{\Lambda, \Lambda_1}_{p,p'}\left( t \vec{x} \right) \leq t^{-p'} \Xi^{\Lambda, \Lambda_1}_{p,p'}\left( \vec{x} \right)
\end{equation}
upon taking $\rho = 0$, and since $D - [\op_\vec{A}] \leq r \leq D - [\op_\vec{A}]+1-\Delta$ (because $r$ is the smallest integer $\geq D - [\op_\vec{A}]$) and $0 < \epsilon < \Delta$ we have
\begin{equation}
\epsilon \leq p' = [\op_\vec{A}]+r-D+\epsilon \leq 1-\Delta+\epsilon < 1 \eqend{,}
\end{equation}
such that the $t$ integral converges absolutely, and we obtain the claimed bound~\eqref{bound_fs_taylor}.

\subsection{Bounds for \texorpdfstring{$H$}{H} functionals}

In this section, we want to derive bounds for the partially oversubtracted, almost disconnected functionals $H^{\Lambda, \Lambda_0}_D\left( \bigotimes_{k=1}^{s'} \op_{A_k}(x_k); \bigotimes_{l=s'+1}^s \op_{A_l}(x_l) \right)$ which are sufficiently sharp with respect to the dependence on the positions $x_k$ of the composite operator insertions. Expanding the functionals in the flow equation~\eqref{h_sop_flow} in external fields and antifields and $\hbar$, we arrive at
\begin{splitequation}
\label{hs_flow_hierarchy}
&\partial_\Lambda \mathcal{H}^{\Lambda, \Lambda_0, l}_{\vec{K} \vec{L}^\ddag; D}\left( \bigotimes_{k=1}^{s'} \op_{A_k}; \bigotimes_{k'=s'+1}^s \op_{A_{k'}}; \vec{q} \right) \\
&= \frac{c}{2} \int \left( \partial_\Lambda C^{\Lambda, \Lambda_0}_{MN}(-p) \right) \mathcal{H}^{\Lambda, \Lambda_0, l-1}_{MN \vec{K} \vec{L}^\ddag; D}\left( \bigotimes_{k=1}^{s'} \op_{A_k}; \bigotimes_{k'=s'+1}^s \op_{A_{k'}}; p,-p,\vec{q} \right) \frac{\total^4 p}{(2\pi)^4} \\
&\quad- \sum_{\subline{\sigma \cup \tau = \{1, \ldots, m\} \\ \rho \cup \varsigma = \{1, \ldots, n\} }} \sum_{l'=0}^l c_{\sigma\tau\rho\varsigma} \mathcal{L}^{\Lambda, \Lambda_0, l'}_{\vec{K}_\sigma \vec{L}_\rho^\ddag M}(\vec{q}_\sigma,\vec{q}_\rho,-k) \left( \partial_\Lambda C^{\Lambda, \Lambda_0}_{MN}(k) \right) \mathcal{H}^{\Lambda, \Lambda_0, l-l'}_{N \vec{K}_\tau \vec{L}_\varsigma^\ddag; D}\left( \bigotimes_{k=1}^{s'} \op_{A_k}; \bigotimes_{k'=s'+1}^s \op_{A_{k'}}; k,\vec{q}_\tau,\vec{q}_\varsigma \right) \\
&\quad- \sum_{\subline{\sigma \cup \tau = \{1, \ldots, m\} \\ \rho \cup \varsigma = \{1, \ldots, n\} }} \sum_{l' = 0}^l c_{\sigma\tau\rho\varsigma} \int \mathcal{G}^{\Lambda, \Lambda_0, l'}_{\vec{K}_\sigma \vec{L}_\rho^\ddag M; D}\left( \bigotimes_{k=1}^{s'} \op_{A_k}; \vec{q}_\sigma,\vec{q}_\rho,p \right) \left( \partial_\Lambda C^{\Lambda, \Lambda_0}_{MN}(-p) \right) \\
&\hspace{12em}\times \mathcal{G}^{\Lambda, \Lambda_0, l-l'}_{N \vec{K}_\tau \vec{L}_\varsigma^\ddag}\left( \bigotimes_{k'=s'+1}^s \op_{A_{k'}}; -p,\vec{q}_\tau,\vec{q}_\varsigma \right) \frac{\total^4 p}{(2\pi)^4} \eqend{.} \raisetag{1.7\baselineskip}
\end{splitequation}
Note that if $s' = 1$ or $s = s'+1$, the definition of the (oversubtracted) disconnected functionals $G^{\Lambda, \Lambda_0}_D$~\eqref{g_sop_def} appearing in the source term on the last line shows that they reduce to the functionals with one insertion of a composite operator $L^{\Lambda, \Lambda_0}$, since the functionals $F^{\Lambda, \Lambda_0}_D$ vanish in that case. Especially, for $s' = 1$ the first functional has a single insertion of a composite operator and is not oversubtracted at all, such that $D = 0$ in this case.

We then want to show
\begin{proposition}
\label{thm_hs}
At each order $l$ in perturbation theory and for an arbitrary number $m$ of external fields $\vec{K}$ and $n$ antifields $\vec{L}^\ddag$, the almost disconnected functionals with $s \geq 2$ insertions of arbitrary (non-integrated) composite operators $\op_{A_i}$, partially oversubtracted at order $D$ with $0 \leq D \leq [\op_\vec{A}'] \equiv \sum_{k=1}^{s'} [\op_{A_k}]$, satisfy the bound
\begin{splitequation}
\label{bound_hs}
&\abs{ \mathcal{H}^{\Lambda, \Lambda_0, l}_{\vec{K} \vec{L}^\ddag; D}\left( \bigotimes_{k=1}^{s'} \op_{A_k}(x_k); \bigotimes_{k'=s'+1}^s \op_{A_{k'}}(x_{k'}); \vec{q} \right) } \leq \mu^{[\op_\vec{A}]+\epsilon} \sup\left( 1, \frac{\abs{\vec{q}}}{\sup(\mu,\Lambda)} \right)^{g^{(s)}\left( [\op_\vec{A}], m+n+2l, 0 \right)} \\
&\qquad\times \varXi^{\Lambda, \Lambda_1}_{s',[\op_\vec{A}]+\epsilon}(\vec{x}) \sup\left[ 1, \Xi^{\Lambda, \Lambda_1}_{r+\rho, [\op_\vec{A}']-D+\epsilon}(x_1,\ldots,x_{s'}) \right] \sup\left[ 1, \Xi^{\Lambda, \Lambda_1}_{r+\rho, [\op_\vec{A}]-[\op_\vec{A}']+\epsilon}(x_{s'+1},\ldots,x_s) \right] \\
&\qquad\times \sum_{T^* \in \mathcal{T}^*_{m+n}} \mathsf{G}^{T^*,0}_{\vec{K} \vec{L}^\ddag; -\epsilon}(\vec{q}; \mu, \Lambda) \, \mathcal{P}\left( \ln_+ \frac{\sup\left( \abs{\vec{q}}, \mu \right)}{\sup(\inf(\mu, \bar{\eta}(\vec{q})), \Lambda)}, \ln_+ \frac{\Lambda}{\mu} \right) \eqend{,} \raisetag{1.7\baselineskip}
\end{splitequation}
with the function
\begin{equations}[varxi_lambda_def]
\varXi^{\Lambda, \Lambda_1}_{s',p}(\vec{x}) &= \begin{cases} \varXi^{(1)}_{s',p,\rho}(\vec{x}) & \Lambda \geq \Lambda_1 \\ \sup\left[ 1, \varXi^{(1)}_{s',p,\rho}(\vec{x}) \right] & \Lambda < \Lambda_1 \end{cases} \eqend{,} \\
\varXi^{(1)}_{s',p,\rho}(\vec{x}) &= \left( \frac{\mu}{\Lambda_1} \right)^{p+\rho} \sup_{1 \leq k \leq s' < k' \leq s} \left( \mu \abs{x_k-x_{k'}} \right)^{-p-\rho} \eqend{,}
\end{equations}
the functions $\Xi^{\Lambda, \Lambda_1}$ defined by equation~\eqref{xi_lambda_def} and for all $0 < \epsilon < \Delta$, an arbitrary $0 < \Lambda_1 \leq \mu$ and an arbitrary $\rho \geq 0$, where $r$ is the smallest integer strictly larger than $[\op_\vec{A}]$.
\end{proposition}
Note that as in Proposition~\ref{thm_fs} the coefficients of the polynomial depend on $\epsilon$, and diverge as $\epsilon \to 0$.

For the $\Lambda$ derivative, we want to prove the bounds
\begin{splitequation}
\label{bound_hs_lambdaderiv}
&\abs{ \partial_\Lambda \mathcal{H}^{\Lambda, \Lambda_0, l}_{\vec{K} \vec{L}^\ddag; D}\left( \bigotimes_{k=1}^{s'} \op_{A_k}(x_k); \bigotimes_{k'=s'+1}^s \op_{A_{k'}}(x_{k'}); \vec{q} \right) } \leq \frac{1}{\sup(\inf(\mu, \bar{\eta}(\vec{q})),\Lambda)} \mu^{[\op_\vec{A}]+\epsilon} \\
&\qquad\times \varXi^{\Lambda, \Lambda_1}_{s',[\op_\vec{A}]+\epsilon}(\vec{x}) \sup\left[ 1, \varXi^{\Lambda, \Lambda_1}_{r+\rho, [\op_\vec{A}']-D+\epsilon}(x_1,\ldots,x_{s'}) \right] \sup\left[ 1, \varXi^{\Lambda, \Lambda_1}_{r+\rho, [\op_\vec{A}]-[\op_\vec{A}']+\epsilon}(x_{s'+1},\ldots,x_s) \right] \\
&\qquad\times \sup\left( 1, \frac{\abs{\vec{q}}}{\sup(\mu,\Lambda)} \right)^{g^{(s)}\left( [\op_\vec{A}], m+n+2l, 0 \right)} \sum_{T^* \in \mathcal{T}^*_{m+n}} \mathsf{G}^{T^*,0}_{\vec{K} \vec{L}^\ddag; -\epsilon}(\vec{q}; \mu, \Lambda) \, \mathcal{P}\left( \ln_+ \frac{\sup\left( \abs{\vec{q}}, \mu \right)}{\Lambda}, \ln_+ \frac{\Lambda}{\mu} \right) \eqend{,}
\end{splitequation}
a bound very similar to~\eqref{bound_hs}. Let us denote the linear term in the first line of the right-hand side of the flow hierarchy~\eqref{hs_flow_hierarchy} by $F_1$, the quadratic term in the second line by $F_2$ and the source term in the last line by $F_3$.

To estimate the linear term, we use the bound~\eqref{bound_hs} for the functional and the bound~\eqref{prop_abl} for the covariance to obtain
\begin{splitequation}
\abs{F_1} &\leq \varXi^{\Lambda, \Lambda_1}_{s',[\op_\vec{A}]+\epsilon}(\vec{x}) \sup\left[ 1, \Xi^{\Lambda, \Lambda_1}_{r+\rho, [\op_\vec{A}']-D+\epsilon}(x_1,\ldots,x_{s'}) \right] \sup\left[ 1, \Xi^{\Lambda, \Lambda_1}_{r+\rho, [\op_\vec{A}]-[\op_\vec{A}']+\epsilon}(x_{s'+1},\ldots,x_s) \right] \\
&\quad\times \mu^{[\op_\vec{A}]+\epsilon} \int \sup(\abs{p}, \Lambda)^{-5+[\phi_M]+[\phi_N]} \, \mathe^{-\frac{\abs{p}^2}{2 \Lambda^2}} \sup\left( 1, \frac{\abs{\vec{q},p,-p}}{\sup(\mu,\Lambda)} \right)^{g^{(s)}\left( [\op_\vec{A}], m+n+2l, 0 \right)} \\
&\quad\times \sum_{T^* \in \mathcal{T}^*_{m+n+2}} \mathsf{G}^{T^*,0}_{MN \vec{K} \vec{L}^\ddag; -\epsilon}(p,-p,\vec{q}; \mu, \Lambda) \, \mathcal{P}\left( \ln_+ \frac{\sup\left( \abs{\vec{q},p,-p}, \mu \right)}{\Lambda}, \ln_+ \frac{\Lambda}{\mu} \right) \frac{\total^4 p}{(2\pi)^4} \eqend{,} \raisetag{1.7\baselineskip}
\end{splitequation}
since $\bar{\eta}(\vec{q},p,-p) = 0$. We then use the estimates~\eqref{fs_qp_factor} to simplify the polynomial in logarithms and the large-momentum factor, and use Lemma~\ref{lemma_pint2} to perform the $p$ integral. This results in
\begin{splitequation}
\abs{F_1} &\leq \varXi^{\Lambda, \Lambda_1}_{s',[\op_\vec{A}]+\epsilon}(\vec{x}) \sup\left[ 1, \Xi^{\Lambda, \Lambda_1}_{r+\rho, [\op_\vec{A}']-D+\epsilon}(x_1,\ldots,x_{s'}) \right] \sup\left[ 1, \Xi^{\Lambda, \Lambda_1}_{r+\rho, [\op_\vec{A}]-[\op_\vec{A}']+\epsilon}(x_{s'+1},\ldots,x_s) \right] \\
&\quad\times \mu^{[\op_\vec{A}]+\epsilon} \Lambda^{-1+[\phi_M]+[\phi_N]} \sup\left( 1, \frac{\abs{\vec{q}}}{\sup(\mu,\Lambda)} \right)^{g^{(s)}\left( [\op_\vec{A}], m+n+2l, 0 \right)} \\
&\quad\times \sum_{T^* \in \mathcal{T}^*_{m+n+2}} \mathsf{G}^{T^*,0}_{MN \vec{K} \vec{L}^\ddag; -\epsilon}(0,0,\vec{q}; \mu, \Lambda) \, \mathcal{P}\left( \ln_+ \frac{\sup\left( \abs{\vec{q}}, \mu \right)}{\Lambda}, \ln_+ \frac{\Lambda}{\mu} \right) \eqend{.} \raisetag{1.7\baselineskip}
\end{splitequation}
We then amputate the first two external vertices with zero momentum corresponding to $M$ and $N$, obtaining an extra factor~\eqref{amputate}
\begin{equation}
\frac{\Lambda^{2-[\phi_M]-[\phi_N]}}{\sup(\inf(\mu, \bar{\eta}(\vec{q})),\Lambda)^2} \leq \frac{\Lambda^{1-[\phi_M]-[\phi_N]}}{\sup(\inf(\mu, \bar{\eta}(\vec{q})),\Lambda)} \eqend{.}
\end{equation}
This then already gives a bound of the required form~\eqref{bound_hs_lambdaderiv} for $F_1$.

For the quadratic term, we obtain using the bounds~\eqref{bound_hs}, \eqref{bound_l0} and~\eqref{prop_abl}
\begin{splitequation}
\abs{F_2} &\leq \sum_{\subline{\sigma \cup \tau = \{1, \ldots, m\} \\ \rho \cup \varsigma = \{1, \ldots, n\}}} \sum_{l'=0}^l \mathe^{-\frac{\abs{k}^2}{2 \Lambda^2}} \mu^{[\op_\vec{A}]+\epsilon} \sup\left( 1, \frac{\abs{k,\vec{q}_\tau,\vec{q}_\varsigma}}{\sup(\mu,\Lambda)} \right)^{g^{(s)}\left( [\op_\vec{A}], \abs{\tau}+\abs{\varsigma}+1+2(l-l'), 0 \right)} \\
&\quad\times \sup(\abs{k}, \Lambda)^{-5+[\phi_M]+[\phi_N]} \, \varXi^{\Lambda, \Lambda_1}_{s',[\op_\vec{A}]+\epsilon}(\vec{x}) \\
&\quad\times \sup\left[ 1, \Xi^{\Lambda, \Lambda_1}_{r+\rho, [\op_\vec{A}']-D+\epsilon}(x_1,\ldots,x_{s'}) \right] \sup\left[ 1, \Xi^{\Lambda, \Lambda_1}_{r+\rho, [\op_\vec{A}]-[\op_\vec{A}']+\epsilon}(x_{s'+1},\ldots,x_s) \right] \\
&\quad\times \sum_{T \in \mathcal{T}_{\abs{\sigma} + \abs{\rho}+1}} \mathsf{G}^{T,0}_{\vec{K}_\sigma \vec{L}_\rho^\ddag M}(\vec{q}_\sigma,\vec{q}_\rho,-k; \mu, \Lambda) \,\mathcal{P}\left( \ln_+ \frac{\sup\left( \abs{\vec{q}_\sigma,\vec{q}_\rho,-k}, \mu \right)}{\Lambda}, \ln_+ \frac{\Lambda}{\mu} \right) \\
&\quad\times \sum_{T^* \in \mathcal{T}^*_{\abs{\tau}+\abs{\varsigma}+1}} \mathsf{G}^{T^*,0}_{N \vec{K}_\tau \vec{L}_\varsigma^\ddag; -\epsilon}(k,\vec{q}_\tau,\vec{q}_\varsigma; \mu, \Lambda) \, \mathcal{P}\left( \ln_+ \frac{\sup\left( \abs{k,\vec{q}_\tau,\vec{q}_\varsigma}, \mu \right)}{\Lambda}, \ln_+ \frac{\Lambda}{\mu} \right) \eqend{,} \raisetag{2.1\baselineskip}
\end{splitequation}
since the definition of $k$~\eqref{k_def} shows that $\bar{\eta}(k,\vec{q}_\tau,\vec{q}_\varsigma) = 0$ and we estimate $\eta(\vec{q}_\sigma,\vec{q}_\rho,-k) \geq 0$. Using that
\begin{equation}
\abs{\vec{q}_\sigma,\vec{q}_\rho,-k}, \abs{k,\vec{q}_\tau,\vec{q}_\varsigma} \leq \abs{\vec{q}}
\end{equation}
we can estimate the polynomials in logarithms and the large-momentum factor. Since we have
\begin{equation}
\abs{\tau}+\abs{\varsigma}+1+2(l-l') \leq m+n+2l-1
\end{equation}
(obvious for $l' > 0$, while for $l' = 0$ we may assume that $\abs{\tau}+\abs{\varsigma} \leq m+n-1$ since otherwise the functional without operator insertions, and thus $F_2$, vanishes), property~\eqref{gs_prop_1} of $g^{(s)}$ shows us that
\begin{equation}
g^{(s)}\left( [\op_\vec{A}], \abs{\tau}+\abs{\varsigma}+1+2(l-l'), 0 \right) \leq g^{(s)}\left( [\op_\vec{A}], m+n+2l, 0 \right)
\end{equation}
which allows us to estimate the large-momentum factor. Fusing the trees according to the estimate~\eqref{gw_fused_2_est} and using the estimate~\eqref{fs_estexp} then gives the required bound~\eqref{bound_fs_lambdaderiv} for $F_2$.

To estimate the source term, we use the shift property~\eqref{func_sop_shift} to bring the last operator insertion of the first (oversubtracted) functional to $x_{s'} = 0$, which results in
\begin{splitequation}
\label{bound_hs_f3_a}
\abs{F_3} &\leq \sum_{\subline{\sigma \cup \tau = \{1, \ldots, m\} \\ \rho \cup \varsigma = \{1, \ldots, n\} }} \sum_{l' = 0}^l c_{\sigma\tau\rho\varsigma} \Bigg\lvert \int \mathe^{- \mathi x_{s'} (p+\sum_{i \in \sigma\cup\rho} q_i)} \mathcal{G}^{\Lambda, \Lambda_0, l'}_{\vec{K}_\sigma \vec{L}_\rho^\ddag M; D}\left( \bigotimes_{k=1}^{s'} \op_{A_k}; \vec{q}_\sigma,\vec{q}_\rho,p \right) \\
&\hspace{4em}\times \left( \partial_\Lambda C^{\Lambda, \Lambda_0}_{MN}(-p) \right) \mathcal{G}^{\Lambda, \Lambda_0, l-l'}_{N \vec{K}_\tau \vec{L}_\varsigma^\ddag}\left( \bigotimes_{k'=s'+1}^s \op_{A_{k'}}; -p,\vec{q}_\tau,\vec{q}_\varsigma \right) \frac{\total^4 p}{(2\pi)^4} \Bigg\rvert \eqend{,}
\end{splitequation}
where now $x_{s'} = x_s = 0$. As in the proof of Theorem~\ref{thm_fs}, we now distinguish between the cases $\Lambda \geq \Lambda_1$ and $\Lambda < \Lambda_1$. In the first case, we would like to use
\begin{equation}
\label{bound_hs_shiftderivs}
\exp\left( - \mathi x_{s'} p \right) = \frac{\mathi^{\abs{a}}}{x_{s'}^a} \partial_p^a \exp\left( - \mathi x_{s'} p \right)
\end{equation}
to generate additional $p$ derivatives and integrate them by parts to obtain additional negative powers of $\Lambda$, and concretely we would like to use Lemma~\ref{lemma_frac}. For this, we need to find bounds on the $p$ derivatives of the terms appearing in the integral. First, the bound~\eqref{prop_abl} shows that for each derivative acting on the covariance we obtain an additional factor of
\begin{equation}
\sup(\abs{p}, \Lambda)^{-1} \leq \Lambda^{-1} \eqend{.}
\end{equation}
We would like to obtain similar factors when $p$ derivatives act on the other factors, for which we use the definition~\eqref{g_sop_def} to separate the completely disconnected part, which is a product of functionals with one insertion of a composite operator. For these terms, it will happen that $\phi_M(p)$ is part of a functional
\begin{equation}
\label{bound_hs_l_1op}
\mathcal{L}^{\Lambda, \Lambda_0, l''}_{\vec{K}_\kappa \vec{L}_\lambda^\ddag M}\left( \op_{A_k}(x_k-x_{s'}); \vec{q}_\kappa,\vec{q}_\lambda,p \right)
\end{equation}
for some collection of external legs $\kappa \subseteq \sigma$ and $\lambda \subseteq \rho$, the corresponding momenta $\vec{q}_\kappa$ and $\vec{q}_\lambda$, some $l'' \leq l'$ and the insertion at $x_k-x_{s'} \neq 0$ (since we shifted all insertions by $x_{s'}$). Similarly, for the completely disconnected part of the second (not oversubtracted) functional it will happen that $\phi_N(-p)$ is part of a functional
\begin{equation}
\label{bound_hs_l_1op_2}
\mathcal{L}^{\Lambda, \Lambda_0, l''}_{N \vec{K}_\kappa \vec{L}_\lambda^\ddag}\left( \op_{A_{k'}}(x_{k'}); -p,\vec{q}_\kappa,\vec{q}_\lambda \right)
\end{equation}
for some collection of external legs $\kappa \subseteq \tau$ and $\lambda \subseteq \varsigma$, the corresponding momenta $\vec{q}_\kappa$ and $\vec{q}_\lambda$, some $l'' \leq l$ and the insertion at $x_{k'} \neq 0$. In those cases, we use the shift property~\eqref{func_sop_shift} to bring the corresponding operator insertion to zero before generating $p$ derivatives. Thus, in general we would have to use
\begin{equation}
\label{bound_hs_shiftderivs2}
\exp\left[ - \mathi (x_k - x_{k'}) p \right] = \frac{\mathi^{\abs{a}}}{(x_k - x_{k'})^a} \partial_p^a \exp\left[ - \mathi (x_k - x_{k'}) p \right]
\end{equation}
for some $1 \leq k \leq s'$ and some $s'+1 \leq k' \leq s$ instead of equation~\eqref{bound_hs_shiftderivs}, but of course after taking the absolute value we can estimate all these contributions against the supremum over all $k$ and $k'$. Then the bound~\eqref{bound_l1} together with the definition of the derivative weight factor~\eqref{gw_def} show that we obtain an extra factor
\begin{equation}
\sup\left( 1, \frac{\abs{\vec{q}_\kappa,\vec{q}_\lambda,p}}{\sup(\mu, \Lambda)} \right)^{g^{(1)}([\op_{\vec{A}_k}],\abs{\kappa}+\abs{\lambda}+2l'',1)-g^{(1)}([\op_{\vec{A}_k}],\abs{\kappa}+\abs{\lambda}+2l'',0)} \sup(\bar{\eta}_p(\vec{q}_\kappa,\vec{q}_\lambda,p), \Lambda)^{-1}
\end{equation}
for each $p$ derivative acting on a functional~\eqref{bound_hs_l_1op} [or~\eqref{bound_hs_l_1op_2}]. Property~\eqref{gs_prop_1a} of $g^{(s)}$ shows that $g^{(s)}$ can only decrease for more derivatives, and thus we can simply estimate this factor against $\Lambda^{-1}$. Similarly, if at most $D$ derivatives with respect to $p$ act on the almost disconnected functional $\mathcal{F}^{\Lambda, \Lambda_0, l'}_{\vec{K}_\sigma \vec{L}_\rho^\ddag M; D}$, the bound~\eqref{bound_fs} together with the definition of the derivative weight factor~\eqref{gw_def} show that we obtain an extra factor
\begin{equation}
\sup\left( 1, \frac{\abs{\vec{q}_\sigma,\vec{q}_\rho,p}}{\sup(\mu, \Lambda)} \right)^{g^{(s')}([\op_\vec{A}'],\abs{\sigma}+\abs{\rho}+1+2l',1)-g^{(s')}([\op_\vec{A}'],\abs{\sigma}+\abs{\rho}+1+2l',0)} \sup(\bar{\eta}_p(\vec{q}_\sigma,\vec{q}_\rho,p), \Lambda)^{-1}
\end{equation}
for each additional $p$ derivative, which again by property~\eqref{gs_prop_1a} of $g^{(s)}$ can be estimated against $\Lambda^{-1}$. If more derivatives act, we would get additional problematic contributions from the term $\Xi^{\Lambda, \Lambda_1}_{\sup(\abs{\vec{w}},D), [\op_\vec{A}']-D+\epsilon}(\vec{x})$ appearing in the bound~\eqref{bound_fs}, which however can be circumvented by replacing this factor with
\begin{equation}
\label{bound_hs_frepl}
\Xi^{\Lambda, \Lambda_1}_{\sup(K, [\op_\vec{A}']-D+\epsilon}(\vec{x}) \geq \Xi^{\Lambda, \Lambda_1}_{\sup(\abs{\vec{w}},D), [\op_\vec{A}']-D+\epsilon}(\vec{x})
\end{equation}
for some fixed $K \geq \sup(\abs{\vec{w}},D)$. Thus, with this replacement understood, every $p$ derivative can be estimated by a factor $\Lambda^{-1}$, and we can therefore choose a direction $\alpha \in \{1,2,3,4\}$ such that $\abs{(x_k-x_{k'})^\alpha} \geq \abs{x_k-x_{k'}}/2$ (which is always possible) and use Lemma~\ref{lemma_frac} with $M = \Lambda$ to estimate the $p$ integral. We then first estimate the sum
\begin{splitequation}
&\mathcal{G}^{\Lambda, \Lambda_0, l'}_{\vec{K}_\sigma \vec{L}_\rho^\ddag M; D}\left( \bigotimes_{k=1}^{s'} \op_{A_k}(x_k); \vec{q}_\sigma,\vec{q}_\rho,p \right) = \mathcal{F}^{\Lambda, \Lambda_0, l'}_{\vec{K}_\sigma \vec{L}_\rho^\ddag M; D}\left( \bigotimes_{k=1}^{s'} \op_{A_k}(x_k); \vec{q}_\sigma,\vec{q}_\rho,p \right) \\
&\quad+ \sum_{l_1 + \cdots + l_{s'} = l'} \sum_{\subline{\{\kappa_i\}\colon \vec{K}_{\kappa_1} \cdots \vec{K}_{\kappa_{s'}} = \vec{K}_\sigma M \\ \{\lambda_i\}\colon \vec{L}_{\lambda_1}^\ddag \cdots \vec{L}_{\lambda_{s'}}^\ddag = \vec{L}_\rho^\ddag}} c_{\{l_i\}\{\vec{K}_{\kappa_i}\}\{\vec{L}_{\lambda_i}^\ddag\}} \prod_{k=1}^{s'} \mathcal{L}^{\Lambda, \Lambda_0, l_k}_{\vec{K}_{\kappa_k} \vec{L}_{\lambda_k}^\ddag}\left( \op_{A_k}(x_k); \vec{q}_{\kappa_k},\vec{q}_{\lambda_k} \right) \eqend{,}
\end{splitequation}
with $x_{s'} = 0$ understood for the first term. For the almost disconnected functional, we can simply use the bound~\eqref{bound_fs}, but with the replacement~\eqref{bound_hs_frepl} where we can take $K = \sup(\abs{a},D)$ since at most $\abs{a}$ derivatives with respect to $p$ act on this functional. For the second part, we first use the shift property~\eqref{func_sop_shift} to bring all operator insertions to $x_k = 0$, and then use the bound~\eqref{bound_l1} for each functional, resulting in
\begin{splitequation}
&\abs{\mathcal{G}^{\Lambda, \Lambda_0, l'}_{\vec{K}_\sigma \vec{L}_\rho^\ddag M; D}\left( \bigotimes_{k=1}^{s'} \op_{A_k}(x_k); \vec{q}_\sigma,\vec{q}_\rho,p \right)} \leq \mu^{[\op_\vec{A}']-D+\epsilon} \, \Xi^{\Lambda, \Lambda_1}_{\sup(\abs{a},D), [\op_\vec{A}']-D+\epsilon}(x_1, \ldots, x_{s'-1}, 0) \\
&\quad\times \sup\left( 1, \frac{\abs{\vec{q}_\sigma,\vec{q}_\rho,p}}{\sup(\mu,\Lambda)} \right)^{g^{(s')}\left( [\op_\vec{A}'], \abs{\sigma}+\abs{\rho}+1+2l', 0 \right)} \\
&\quad\times \sum_{T^* \in \mathcal{T}^*_{\abs{\sigma}+\abs{\rho}+1}} \mathsf{G}^{T^*,0}_{\vec{K}_\sigma \vec{L}_\rho^\ddag M; D-\epsilon}(\vec{q}_\sigma,\vec{q}_\rho,p; \mu, \Lambda) \, \mathcal{P}\left( \ln_+ \frac{\sup\left( \abs{\vec{q}_\sigma,\vec{q}_\rho,p}, \mu \right)}{\sup(\inf(\mu, \bar{\eta}(\vec{q}_\sigma,\vec{q}_\rho,p)), \Lambda)}, \ln_+ \frac{\Lambda}{\mu} \right) \\
&\quad+ \sum_{l_1 + \cdots + l_{s'} = l'} \sum_{\subline{\{\kappa_i\}\colon \vec{K}_{\kappa_1} \cdots \vec{K}_{\kappa_{s'}} = \vec{K}_\sigma M \\ \{\lambda_i\}\colon \vec{L}_{\lambda_1}^\ddag \cdots \vec{L}_{\lambda_{s'}}^\ddag = \vec{L}_\rho^\ddag}} \prod_{k=1}^{s'} \sup\left( 1, \frac{\abs{\vec{q}_{\kappa_k},\vec{q}_{\lambda_k}}}{\sup(\mu, \Lambda)} \right)^{g^{(1)}([\op_{A_k}],\abs{\kappa_k}+\abs{\lambda_k}+2l_k,0)} \\
&\qquad\times \sum_{T^* \in \mathcal{T}^*_{\abs{\kappa_k}+\abs{\lambda_k}}} \mathsf{G}^{T^*,0}_{\vec{K}_{\kappa_k} \vec{L}_{\lambda_k}^\ddag; [\op_{A_k}]}(\vec{q}_{\kappa_k},\vec{q}_{\lambda_k}; \mu, \Lambda) \, \mathcal{P}\left( \ln_+ \frac{\sup\left( \abs{\vec{q}_{\kappa_k},\vec{q}_{\lambda_k}}, \mu \right)}{\sup(\inf(\mu, \bar{\eta}(\vec{q}_{\kappa_k},\vec{q}_{\lambda_k})), \Lambda)}, \ln_+ \frac{\Lambda}{\mu} \right) \eqend{.}
\end{splitequation}
To bring the second term into the same form as the first, we use that
\begin{equations}
\abs{\vec{q}_{\kappa_k},\vec{q}_{\lambda_k}} &\leq \abs{\vec{q}_\sigma,\vec{q}_\rho,p} \eqend{,} \\
\bar{\eta}(\vec{q}_{\kappa_k},\vec{q}_{\lambda_k}), \bar{\eta}(\vec{q}_\sigma,\vec{q}_\rho,p) &\geq 0 \eqend{,}
\end{equations}
which allows us to obtain the same polynomials in logarithms and large momentum factors. Using then property~\eqref{gs_prop_3} of $g^{(s)}$ recursively we conclude that
\begin{equation}
\sum_{k=1}^{s'} g^{(1)}([\op_{A_k}],\abs{\kappa_k}+\abs{\lambda_k}+2l_k+2l_k,0) \leq g^{(s')}([\op_\vec{A}'], \abs{\sigma}+\abs{\rho}+1+2l'-2(s'-1), 0) - ([\op_\vec{A}']+s') \eqend{,}
\end{equation}
and using property~\eqref{gs_prop_2a} of $g^{(s)}$ we obtain
\begin{equation}
\sum_{k=1}^{s'} g^{(1)}([\op_{A_k}],\abs{\kappa_k}+\abs{\lambda_k}+2l_k+2l_k,0) \leq g^{(s')}([\op_\vec{A}'], \abs{\sigma}+\abs{\rho}+1+2l', 0) - (2s'-1) ([\op_\vec{A}']+s') \eqend{.}
\end{equation}
We then fuse the trees according to the estimate~\eqref{gw_fused_1_est}, and change the particular weight of the resulting tree from $[\op_\vec{A}]$ to $D-\epsilon$, which gives an extra factor of
\begin{equation}
\sup(\abs{\vec{q}_\sigma,\vec{q}_\rho,p},\mu,\Lambda)^{[\op_\vec{A}']-D+\epsilon} = \sup\left( 1, \frac{\abs{\vec{q}_\sigma,\vec{q}_\rho,p}}{\sup(\mu, \Lambda)} \right)^{[\op_\vec{A}']-D+\epsilon} \sup(\mu, \Lambda)^{[\op_\vec{A}']-D+\epsilon} \eqend{.}
\end{equation}
For the large momentum factor, we now estimate
\begin{equation}
g^{(s')}([\op_\vec{A}'], \abs{\sigma}+\abs{\rho}+1+2l', 0) - (2s'-1) ([\op_\vec{A}']+s')+[\op_\vec{A}']-D+\epsilon \leq g^{(s')}([\op_\vec{A}'], \abs{\sigma}+\abs{\rho}+1+2l', 0)
\end{equation}
since $s' \geq 1$, $D \geq 0$ and $\epsilon < 1$. Since also $\mu^{[\op_\vec{A}']-D+\epsilon} \leq \sup(\mu, \Lambda)^{[\op_\vec{A}']-D+\epsilon}$ and the function $\Xi^{\Lambda, \Lambda_1}$ is translation invariant such that we can reintroduce $x_{s'} \neq 0$ without changing the bound, we finally obtain
\begin{splitequation}
&\abs{\mathcal{G}^{\Lambda, \Lambda_0, l'}_{\vec{K}_\sigma \vec{L}_\rho^\ddag M; D}\left( \bigotimes_{k=1}^{s'} \op_{A_k}(x_k); \vec{q}_\sigma,\vec{q}_\rho,p \right)} \leq \sup(\mu, \Lambda)^{[\op_\vec{A}']-D+\epsilon} \sup\left[ 1, \Xi^{\Lambda, \Lambda_1}_{\sup(\abs{a},D), [\op_\vec{A}']-D+\epsilon}(x_1,\ldots,x_{s'}) \right] \\
&\quad\times \sup\left( 1, \frac{\abs{\vec{q}_\sigma,\vec{q}_\rho,p}}{\sup(\mu,\Lambda)} \right)^{g^{(s')}\left( [\op_\vec{A}'], \abs{\sigma}+\abs{\rho}+1+2l', 0 \right)} \\
&\quad\times \sum_{T^* \in \mathcal{T}^*_{\abs{\sigma}+\abs{\rho}+1}} \mathsf{G}^{T^*,0}_{\vec{K}_\sigma \vec{L}_\rho^\ddag M; D-\epsilon}(\vec{q}_\sigma,\vec{q}_\rho,p; \mu, \Lambda) \, \mathcal{P}\left( \ln_+ \frac{\sup\left( \abs{\vec{q}_\sigma,\vec{q}_\rho,p}, \mu \right)}{\Lambda}, \ln_+ \frac{\Lambda}{\mu} \right) \eqend{.} \raisetag{1.7\baselineskip}
\end{splitequation}
The same estimate is valid for the second functional in the source term, which is not oversubtracted and has $D = 0$. Thus, applying Lemma~\ref{lemma_frac} to the $p$ integral~\eqref{bound_hs_f3_a} using these bounds and the bound~\eqref{prop_abl} on the covariance we obtain the result
\begin{splitequation}
\abs{F_3} &\leq \sum_{\subline{\sigma \cup \tau = \{1, \ldots, m\} \\ \rho \cup \varsigma = \{1, \ldots, n\} }} \sum_{l' = 0}^l \left( \inf_{1 \leq k \leq s' < k' \leq s} \abs{(x_k-x_{k'})^\alpha} \Lambda \right)^{-\abs{a}+\delta} \sup\left[ 1, \Xi^{\Lambda, \Lambda_1}_{\sup(\abs{a},D), [\op_\vec{A}']-D+\epsilon}(x_1,\ldots,x_{s'}) \right] \\
&\quad\times \sup(\mu, \Lambda)^{[\op_\vec{A}]-D+2\epsilon} \sup\left[ 1, \Xi^{\Lambda, \Lambda_1}_{\abs{a}, [\op_\vec{A}]-[\op_\vec{A}']+\epsilon}(x_{s'+1},\ldots,x_s) \right] \int \left( \frac{\abs{p^\alpha}}{\Lambda} \right)^{-1+\delta} \left( 1 + \frac{\abs{p^\alpha}}{\Lambda} \right) \\
&\quad\times \sup\left( 1, \frac{\abs{\vec{q}_\sigma,\vec{q}_\rho,p}}{\sup(\mu,\Lambda)} \right)^{g^{(s')}\left( [\op_\vec{A}'], \abs{\sigma}+\abs{\rho}+1+2l', 0 \right)} \sup(\abs{p}, \Lambda)^{-5+[\phi_M]+[\phi_N]} \, \mathe^{-\frac{\abs{p}^2}{2 \Lambda^2}} \\
&\quad\times \sum_{T^* \in \mathcal{T}^*_{\abs{\sigma}+\abs{\rho}+1}} \mathsf{G}^{T^*,0}_{\vec{K}_\sigma \vec{L}_\rho^\ddag M; D-\epsilon}(\vec{q}_\sigma,\vec{q}_\rho,p; \mu, \Lambda) \, \mathcal{P}\left( \ln_+ \frac{\sup\left( \abs{\vec{q}_\sigma,\vec{q}_\rho,p}, \mu \right)}{\Lambda}, \ln_+ \frac{\Lambda}{\mu} \right) \\
&\quad\times \sum_{T^* \in \mathcal{T}^*_{\abs{\tau}+\abs{\varsigma}+1}} \mathsf{G}^{T^*,0}_{N \vec{K}_\tau \vec{L}_\varsigma^\ddag; -\epsilon}(-p,\vec{q}_\tau,\vec{q}_\varsigma; \mu, \Lambda) \, \mathcal{P}\left( \ln_+ \frac{\sup\left( \abs{-p,\vec{q}_\tau,\vec{q}_\varsigma}, \mu \right)}{\Lambda}, \ln_+ \frac{\Lambda}{\mu} \right) \\
&\quad\times \sup\left( 1, \frac{\abs{-p,\vec{q}_\tau,\vec{q}_\varsigma}}{\sup(\mu,\Lambda)} \right)^{g^{(s-s')}\left( [\op_\vec{A}]-[\op_\vec{A}'], \abs{\tau}+\abs{\varsigma}+1+2(l-l'), 0 \right)} \frac{\total^4 p}{(2\pi)^4} \raisetag{1.3\baselineskip}
\end{splitequation}
for all $0 < \delta < 1$. In the second case where $\Lambda < \Lambda_1$, we do not integrate by parts such that $\abs{a} = 0$ and obtain the same estimate except for the factor
\begin{equation}
\label{hs_ibp_factor}
\left( \inf_{1 \leq k \leq s' < k' \leq s} \abs{(x_k-x_{k'})^\alpha} \Lambda \right)^{-\abs{a}+\delta} \left( \frac{\abs{p^\alpha}}{\Lambda} \right)^{-1+\delta} \left( 1 + \frac{\abs{p^\alpha}}{\Lambda} \right) \eqend{.}
\end{equation}
We then fuse the trees according to the estimate~\eqref{gw_fused_1_est}, and use that
\begin{equation}
\abs{\vec{q}_\sigma,\vec{q}_\rho,p}, \abs{-p,\vec{q}_\tau,\vec{q}_\varsigma} \leq \abs{\vec{q},p,-p}
\end{equation}
and the estimates~\eqref{fs_qp_factor} to fuse the polynomials in logarithms and the large momentum factors, for which we furthermore use property~\eqref{gs_prop_3} of $g^{(s)}$ to obtain
\begin{splitequation}
&g^{(s-s')}\left( [\op_\vec{A}]-[\op_\vec{A}'], \abs{\tau}+\abs{\varsigma}+1+2(l-l'), 0 \right) + g^{(s')}\left( [\op_\vec{A}'], \abs{\sigma}+\abs{\rho}+1+2l', 0 \right) \\
&\leq g^{(s)}\left( [\op_\vec{A}], m+n+2l, 0 \right) - ([\op_\vec{A}]+s) \eqend{.}
\end{splitequation}
We can now perform the $p$ integral using Lemma~\ref{lemma_pint2}, and estimate
\begin{equation}
( \abs{(x_k-x_{k'})^\alpha} \Lambda )^{-\abs{a}+\delta} \leq 2^{-\abs{a}+\delta} ( \abs{x_k-x_{k'}} \Lambda )^{-\abs{a}+\delta} = 2^{-\abs{a}+\delta} ( \mu \abs{x_k-x_{k'}} )^{-\abs{a}+\delta} \left( \frac{\Lambda}{\mu} \right)^{-\abs{a}+\delta} \eqend{.}
\end{equation}
The particular weight of the tree is changed from $D-2\epsilon$ to $-\epsilon$, which results in an extra factor~\eqref{particular_weight}
\begin{equation}
\sup(\abs{\vec{q}}, \mu, \Lambda)^{D-\epsilon} = \sup\left( 1, \frac{\abs{\vec{q}}}{\sup(\mu, \Lambda)} \right)^{D-\epsilon} \sup(\mu, \Lambda)^{D-\epsilon} \eqend{,}
\end{equation}
and we amputate the external legs corresponding to $M$ and $N$ obtaining an extra factor of~\eqref{amputate_p_factor}. This results in
\begin{splitequation}
\abs{F_3} &\leq \frac{1}{\sup(\inf(\mu, \bar{\eta}(\vec{q})),\Lambda)} \left[ \Theta(\Lambda_1 - \Lambda) + \Theta(\Lambda - \Lambda_1) \left( \frac{\Lambda}{\mu} \right)^{-\abs{a}+\delta} \sup_{1 \leq k \leq s' < k' \leq s} \left( \mu \abs{x_k-x_{k'}} \right)^{-\abs{a}+\delta} \right] \\
&\quad\times \sup\left[ 1, \Xi^{\Lambda, \Lambda_1}_{\sup(\abs{a},D), [\op_\vec{A}']-D+\epsilon}(x_1,\ldots,x_{s'}) \right] \sup\left[ 1, \Xi^{\Lambda, \Lambda_1}_{\abs{a}, [\op_\vec{A}]-[\op_\vec{A}']+\epsilon}(x_{s'+1},\ldots,x_s) \right] \\
&\quad\times \sup(\mu, \Lambda)^{[\op_\vec{A}]+\epsilon} \sup\left( 1, \frac{\abs{\vec{q}}}{\sup(\mu,\Lambda)} \right)^{g^{(s)}\left( [\op_\vec{A}], m+n+2l, 0 \right) - ([\op_\vec{A}]+s)+D-\epsilon} \\
&\quad\times \sum_{T^* \in \mathcal{T}^*_{m+n+2}} \mathsf{G}^{T^*,0}_{\vec{K} \vec{L}^\ddag; -\epsilon}(\vec{q}; \mu, \Lambda) \, \mathcal{P}\left( \ln_+ \frac{\sup\left( \abs{\vec{q}}, \mu \right)}{\Lambda}, \ln_+ \frac{\Lambda}{\mu} \right) \raisetag{1.7\baselineskip}
\end{splitequation}
with $\abs{a} = 0$ if $\Lambda < \Lambda_1$. Let now $r$ be the smallest integer such that $r > [\op_\vec{A}]$, and take $\abs{a} = r + \rho$ with $\rho \geq 0$ and $\delta = r-[\op_\vec{A}]-\epsilon$. Since the smallest difference between two dimensions of composite operators is $\Delta$, we have $[\op_\vec{A}]+\Delta \leq r \leq [\op_\vec{A}]+1$ and from this $\Delta-\epsilon \leq \delta \leq 1-\epsilon$, such that with this choice we have $0 < \delta < 1$ since $0 < \epsilon < \Delta$. The last estimates
\begin{equation}
g^{(s)}\left( [\op_\vec{A}], m+n+2l, 0 \right) - ([\op_\vec{A}]+s)+D-\epsilon \leq g^{(s)}\left( [\op_\vec{A}], m+n+2l, 0 \right)
\end{equation}
since $D \leq [\op_A]$,
\begin{equation}
\sup(\mu, \Lambda)^{[\op_\vec{A}]+\epsilon} \leq \mu^{[\op_\vec{A}]+\epsilon} \left[ \Theta(\Lambda_1 - \Lambda) + \Theta(\Lambda - \Lambda_1) \sup\left( 1, \frac{\Lambda}{\mu} \right)^{[\op_\vec{A}]+\epsilon} \right]
\end{equation}
since $\Lambda_1 \leq \mu$,
\begin{equation}
\sup\left( 1, \frac{\Lambda}{\mu} \right)^{[\op_\vec{A}]+\epsilon} \left( \frac{\Lambda}{\mu} \right)^{-\rho-[\op_\vec{A}]-\epsilon} = \sup\left( 1, \frac{\mu}{\Lambda} \right)^{[\op_\vec{A}]+\epsilon} \left( \frac{\Lambda}{\mu} \right)^{-\rho} \leq \left( \frac{\mu}{\Lambda_1} \right)^{[\op_\vec{A}]+\epsilon+\rho}
\end{equation}
since $\Lambda \geq \Lambda_1$ for this term, and $\sup(r+\rho,D) = r+\rho$ since $r > [\op_\vec{A}] \geq D$ and $\rho \geq 0$ then give the bound~\eqref{bound_hs_lambdaderiv} also for the source term.

Since all functionals have vanishing boundary conditions at $\Lambda = \Lambda_0$, we now simply integrate the bound~\eqref{bound_hs_lambdaderiv} downwards in $\Lambda$. This gives
\begin{splitequation}
&\abs{ \mathcal{H}^{\Lambda, \Lambda_0, l}_{\vec{K} \vec{L}^\ddag; D}\left( \bigotimes_{k=1}^{s'} \op_{A_k}; \bigotimes_{k'=s'+1}^s \op_{A_{k'}}; \vec{q} \right)} \leq \int_\Lambda^{\Lambda_0} \abs{\partial_\lambda \mathcal{H}^{\lambda, \Lambda_0, l}_{\vec{K} \vec{L}^\ddag; D}\left( \bigotimes_{k=1}^{s'} \op_{A_k}; \bigotimes_{k'=s'+1}^s \op_{A_{k'}}; \vec{q} \right)} \total \lambda \\
&\quad\leq \mu^{[\op_\vec{A}]+\epsilon} \int_\Lambda^{\Lambda_0} \varXi^{\lambda, \Lambda_1}_{s',[\op_\vec{A}]+\epsilon}(\vec{x}) \sup\left[ 1, \Xi^{\lambda, \Lambda_1}_{r+\rho, [\op_\vec{A}']-D+\epsilon}(x_1,\ldots,x_{s'}) \right] \\
&\qquad\times \sup\left[ 1, \Xi^{\lambda, \Lambda_1}_{r+\rho, [\op_\vec{A}]-[\op_\vec{A}']+\epsilon}(x_{s'+1},\ldots,x_s) \right] \sup\left( 1, \frac{\abs{\vec{q}}}{\sup(\mu,\lambda)} \right)^{g^{(s)}\left( [\op_\vec{A}], m+n+2l, 0 \right)} \\
&\qquad\times \frac{1}{\sup(\inf(\mu, \bar{\eta}(\vec{q})),\lambda)} \sum_{T^* \in \mathcal{T}^*_{m+n}} \mathsf{G}^{T^*,0}_{\vec{K} \vec{L}^\ddag; -\epsilon}(\vec{q}; \mu, \lambda) \, \mathcal{P}\left( \ln_+ \frac{\sup\left( \abs{\vec{q}}, \mu \right)}{\lambda}, \ln_+ \frac{\lambda}{\mu} \right) \total \lambda \eqend{.}
\end{splitequation}
Since by definition $\varXi^{\lambda, \Lambda_1} \leq \varXi^{\Lambda, \Lambda_1}$ and $\Xi^{\lambda, \Lambda_1} \leq \Xi^{\Lambda, \Lambda_1}$, we can estimate these factors at the lower bound. Since $[T^*] \leq - \epsilon$, we can then use the estimate~\eqref{t_irr_ineq2} to estimate the tree weight factor at the lower bound $\lambda = \Lambda$, and the large momentum factor is trivially estimated there. An application of Lemma~\ref{lemma_lambdaint} to the remaining integral then gives the bound~\eqref{bound_hs}.

\subsection{Bounds for \texorpdfstring{$G$}{G} functionals}

The bounds for the oversubtracted disconnected functionals $G^{\Lambda, \Lambda_0}_D\left( \bigotimes_{k=1}^s \op_{A_k}(x_k) \right)$ and the partially oversubtracted disconnected functionals $G^{\Lambda, \Lambda_0}_D\left( \bigotimes_{k=1}^{s'} \op_{A_k}(x_k); \bigotimes_{l=s'+1}^s \op_{A_l}(x_l) \right)$ are simple corollaries of Theorems~\ref{thm_fs} and~\ref{thm_hs}. Concretely, we obtain
\begin{corollary}
\label{thm_gs}
At each order $l$ in perturbation theory and for an arbitrary number $m$ of external fields $\vec{K}$ and $n$ antifields $\vec{L}^\ddag$, and for $\Lambda \leq \Lambda_1$, the disconnected functionals with $s \geq 2$ insertions of arbitrary (non-integrated) composite operators $\op_{A_i}$, oversubtracted at order $D$ with $0 \leq D \leq [\op_\vec{A}]$, and for $\abs{\vec{w}} \leq [\op_\vec{A}]$ momentum derivatives satisfy the bound
\begin{splitequation}
\label{bound_gs}
&\abs{ \partial^\vec{w} \mathcal{G}^{\Lambda, \Lambda_0, l}_{\vec{K} \vec{L}^\ddag; D}\left( \bigotimes_{k=1}^s \op_{A_k}(x_k); \vec{q} \right) } \leq \mu^{[\op_\vec{A}]-D+\epsilon} \sup\left( 1, \frac{\abs{\vec{q}}}{\sup(\mu,\Lambda)} \right)^{g^{(s)}\left( [\op_\vec{A}], m+n+2l, \abs{\vec{w}} \right)} \\
&\qquad\times \Xi^{\Lambda, \Lambda_1}_{\sup(\abs{\vec{w}},D), [\op_\vec{A}]-D+\epsilon}(\vec{x}) \sum_{T^* \in \mathcal{T}^*_{m+n}} \mathsf{G}^{T^*,\vec{w}}_{\vec{K} \vec{L}^\ddag; D-\epsilon}(\vec{q}; \mu, \Lambda) \, \mathcal{P}\left( \ln_+ \frac{\sup\left( \abs{\vec{q}}, \mu \right)}{\sup(\inf(\mu, \bar{\eta}(\vec{q})), \Lambda)}, \ln_+ \frac{\Lambda}{\mu} \right)
\end{splitequation}
for all $0 < \epsilon < \Delta$, an arbitrary $0 < \Lambda_1 \leq \mu$ and with $\rho = 0$ for the function $\Xi^{\Lambda, \Lambda_1}$ defined by~\eqref{xi_lambda_def}.
\end{corollary}
\begin{corollary}
\label{thm_gs_taylor}
At each order $l$ in perturbation theory and for an arbitrary number $m$ of external fields $\vec{K}$ and $n$ antifields $\vec{L}^\ddag$, and for $\Lambda \leq \mu$, the disconnected functionals with $s \geq 2$ insertions of arbitrary (non-integrated) composite operators $\op_{A_i}$, oversubtracted at order $D$ with $D > [\op_\vec{A}]$, and for at most $\abs{\vec{w}} \leq D$ momentum derivatives, satisfy the bound
\begin{splitequation}
\label{bound_gs_taylor}
&\abs{ \partial^\vec{w} \left( 1 - \sum_{n < D - [\op_\vec{A}]} \mathcal{T}^n_{\vec{x} \to \vec{0}} \right) \mathcal{G}^{\Lambda, \Lambda_0, l}_{\vec{K} \vec{L}^\ddag; D}\left( \bigotimes_{k=1}^s \op_{A_k}(x_k); \vec{q} \right) } \leq \mu^{[\op_\vec{A}]+r-D+\epsilon} \sup_{i\in\{1,\ldots,s\}} \abs{x_i}^r \\
&\quad\times \Xi^{\Lambda, \Lambda_1}_{\sup(\abs{\vec{w}},D), [\op_\vec{A}]+r-D+\epsilon}(\vec{x}) \ \sup\left( 1, \frac{\abs{\vec{q}}}{\sup(\mu,\Lambda)} \right)^{g^{(s)}\left( [\op_\vec{A}]+r, m+n+2l, \abs{\vec{w}} \right)} \\
&\quad\times \sum_{T^* \in \mathcal{T}^*_{m+n}} \mathsf{G}^{T^*,\vec{w}}_{\vec{K} \vec{L}^\ddag; D-\epsilon}(\vec{q}; \mu, \Lambda) \, \mathcal{P}\left( \ln_+ \frac{\sup\left( \abs{\vec{q}}, \mu \right)}{\sup(\inf(\mu, \bar{\eta}(\vec{q})), \Lambda)}, \ln_+ \frac{\Lambda}{\mu} \right)
\end{splitequation}
for all $0 < \epsilon < \Delta$, an arbitrary $0 < \Lambda_1 \leq \mu$ and with $\rho = 0$ for the function $\Xi^{\Lambda, \Lambda_1}$ defined by~\eqref{xi_lambda_def}, where $r$ is the smallest integer larger than or equal to $D-[\op_\vec{A}]$.
\end{corollary}

Corollary~\ref{thm_gs} follows from Theorem~\ref{thm_fs} via the definition~\eqref{g_sop_def} if we can prove the bound for the completely disconnected part, the product of functionals with one insertion. Expanding in fields, antifields and $\hbar$, and taking some momentum derivatives we obtain from the bounds~\eqref{bound_l1} and the shift property~\eqref{func_sop_shift} that
\begin{splitequation}
&\abs{ \prod_{k=1}^s \sum_{\subline{l_1 + \cdots + l_s = l \\ \vec{w}_1 + \cdots + \vec{w}_s = \vec{w}}} \sum_{\subline{\vec{K}_1 + \cdots + \vec{K}_k = \vec{K} \\ \vec{L}^\ddag_1 + \cdots + \vec{L}^\ddag_k = \vec{L}^\ddag}} c_{\{l_i\}\{\vec{K}_i\}\{\vec{L}_i^\ddag\}\{\vec{w}_i\}} \partial^{\vec{w}_k} \mathcal{L}^{\Lambda, \Lambda_0, l_k}_{\vec{K}_k \vec{L}^\ddag_k}\left( \op_{A_k}(x_k); \vec{q}_k \right) } \\
&\quad\leq \prod_{k=1}^s \sum_{\subline{l_1 + \cdots + l_s = l \\ \vec{w}_1 + \cdots + \vec{w}_s = \vec{w}}} \sum_{\subline{\vec{K}_1 + \cdots + \vec{K}_s = \vec{K} \\ \vec{L}^\ddag_1 + \cdots + \vec{L}^\ddag_s = \vec{L}^\ddag}} \sum_{\vec{v}_k \leq \vec{w}_k} \abs{x_k}^{\abs{\vec{w}_k}-\abs{\vec{v}_k}} \sup\left( 1, \frac{\abs{\vec{q}_k}}{\sup(\mu, \Lambda)} \right)^{g^{(1)}([\op_{A_k}],\abs*[]{\vec{K}_k}+\abs*[]{\vec{L}^\ddag_k}+2l_k,\abs{\vec{v}_k})} \\
&\qquad\quad\times \sum_{T^* \in \mathcal{T}^*_{\abs*[]{\vec{K}_k}+\abs*[]{\vec{L}^\ddag_k}}} \mathsf{G}^{T^*,\vec{v}_k}_{\vec{K}_k \vec{L}^\ddag_k; [\op_{A_k}]}(\vec{q}_k; \mu, \Lambda) \, \mathcal{P}\left( \ln_+ \frac{\sup\left( \abs{\vec{q}_k}, \mu \right)}{\sup(\inf(\mu, \bar{\eta}(\vec{q}_k)), \Lambda)}, \ln_+ \frac{\Lambda}{\mu} \right) \raisetag{2.1\baselineskip}
\end{splitequation}
Since $\abs{\vec{q}_k} \leq \abs{\vec{q}}$ and $\bar{\eta}(\vec{q}_k) \geq \bar{\eta}(\vec{q})$, the polynomials in logarithms can be simply estimated. The same applies to the large-momentum factors, where we additionally need property~\eqref{gs_prop_3} of $g^{(s)}$ recursively to conclude that
\begin{equation}
\sum_{k=1}^s g^{(1)}([\op_{A_k}],\abs*[]{\vec{K}_k}+\abs*[]{\vec{L}^\ddag_k}+2l_k,\abs{\vec{v}_k}) \leq g^{(s)}([\op_\vec{A}], m+n+2l-2(s-1), \abs{\vec{w}}) - ([\op_\vec{A}]+s) \eqend{,}
\end{equation}
and using property~\eqref{gs_prop_2a} of $g^{(s)}$ we obtain
\begin{equation}
\sum_{k=1}^s g^{(1)}([\op_{A_k}],\abs*[]{\vec{K}_k}+\abs*[]{\vec{L}^\ddag_k}+2l_k,\abs{\vec{v}_k}) \leq g^{(s)}([\op_\vec{A}], m+n+2l, \abs{\vec{w}}) - (2s-1) ([\op_\vec{A}]+s) \eqend{.}
\end{equation}
We then convert the $\vec{v}_k$ derivatives acting on each tree to $\vec{w}_k$ derivatives, which gives an additional factor of~\eqref{gw_def} (since $\Lambda \leq \mu$)
\begin{splitequation}
\prod_{i=1}^{\abs*[]{\vec{K}_k}+\abs*[]{\vec{L}^\ddag_k}} \sup\left( \bar{\eta}_{q_{k,i}}(\vec{q}_k), \Lambda \right)^\abs{(\vec{w}_k-\vec{v}_k)_i} &\leq \sup\left( \abs{\vec{q}}, \Lambda \right)^{\abs{\vec{w}_k}-\abs{\vec{v}_k}} \leq \sup\left( \abs{\vec{q}}, \mu, \Lambda \right)^{\abs{\vec{w}_k}-\abs{\vec{v}_k}} \\
&= \sup\left( 1, \frac{\abs{\vec{q}}}{\sup(\mu, \Lambda)} \right)^{\abs{\vec{w}_k}-\abs{\vec{v}_k}} \mu^{\abs{\vec{w}_k}-\abs{\vec{v}_k}} \eqend{,}
\end{splitequation}
and fuse the trees using the estimate~\eqref{gw_fused_1_est}. This gives
\begin{splitequation}
&\abs{ \prod_{k=1}^s \sum_{\subline{l_1 + \cdots + l_s = l \\ \vec{w}_1 + \cdots + \vec{w}_s = \vec{w}}} \sum_{\subline{\vec{K}_1 + \cdots + \vec{K}_k = \vec{K} \\ \vec{L}^\ddag_1 + \cdots + \vec{L}^\ddag_k = \vec{L}^\ddag}} c_{\{l_i\}\{\vec{K}_i\}\{\vec{L}_i^\ddag\}\{\vec{w}_i\}} \partial^{\vec{w}_k} \mathcal{L}^{\Lambda, \Lambda_0, l_k}_{\vec{K}_k \vec{L}^\ddag_k}\left( \op_{A_k}(x_k); \vec{q}_k \right) } \\
&\quad\leq \prod_{k=1}^s \sum_{\vec{w}_1 + \cdots + \vec{w}_s = \vec{w}} \sum_{\vec{v}_k \leq \vec{w}_k} \left( \mu \abs{x_k} \right)^{\abs{\vec{w}_k}-\abs{\vec{v}_k}} \sup\left( 1, \frac{\abs{\vec{q}}}{\sup(\mu, \Lambda)} \right)^{g^{(s)}([\op_\vec{A}], m+n+2l, \abs{\vec{w}}) - (2s-1) ([\op_\vec{A}]+s) + \abs{\vec{w}}-\abs{\vec{v}}} \\
&\qquad\quad\times \sum_{T^* \in \mathcal{T}^*_{m+n}} \mathsf{G}^{T^*,\vec{w}}_{\vec{K} \vec{L}^\ddag; [\op_\vec{A}]}(\vec{q}; \mu, \Lambda) \, \mathcal{P}\left( \ln_+ \frac{\sup\left( \abs{\vec{q}}, \mu \right)}{\sup(\inf(\mu, \bar{\eta}(\vec{q})), \Lambda)}, \ln_+ \frac{\Lambda}{\mu} \right) \eqend{.}
\end{splitequation}
We then change the dimension of the tree from $[\op_\vec{A}]$ to $D-\epsilon$, which gives an extra factor of
\begin{equation}
\sup\left( \abs{\vec{q}}, \mu, \Lambda \right)^{[\op_\vec{A}]-D+\epsilon} = \sup\left( 1, \frac{\abs{\vec{q}}}{\sup(\mu, \Lambda)} \right)^{[\op_\vec{A}]-D+\epsilon} \mu^{[\op_\vec{A}]-D+\epsilon} \eqend{,}
\end{equation}
and estimate
\begin{equation}
- (2s-1) ([\op_\vec{A}]+s) + \abs{\vec{w}}-\abs{\vec{v}} + [\op_\vec{A}]-D+\epsilon \leq 0 \eqend{,}
\end{equation}
since $s \geq 2$, $\abs{\vec{w}} \leq [\op_\vec{A}]$ and $\epsilon < 1$. Estimating finally the $x$-dependent terms as
\begin{equation}
\prod_{k=1}^s \sum_{\vec{w}_1 + \cdots + \vec{w}_s = \vec{w}} \sum_{\vec{v}_k \leq \vec{w}_k} \left( \mu \abs{x_k} \right)^{\abs{\vec{w}_k}-\abs{\vec{v}_k}} \leq c \Xi^{(2)}_\abs{\vec{w}}(\vec{x}) \leq c \Xi^{(2)}_{\sup(\abs{\vec{w}},D)}(\vec{x}) \leq \Xi^{\Lambda, \Lambda_1}_{\sup(\abs{\vec{w}},D), [\op_\vec{A}]-D+\epsilon}(\vec{x})
\end{equation}
since $\Lambda \leq \Lambda_1$, the bound~\eqref{bound_gs} follows.

The proof of Corollary~\ref{thm_gs_taylor} can be taken over verbatim from the one of Theorem~\ref{thm_fs_taylor}, replacing only the bounds~\eqref{bound_fs} with the bounds~\eqref{bound_gs}.

\section{The operator product expansion}
\label{sec_ope}

This section defines the OPE coefficients, shows some relations between oversubtracted functionals with insertions of composite operators and shows the existence of the OPE in an asymptotic sense. All these statements are proven by using that two functionals or combinations thereof are equal if they satisfy the same flow equation and the same boundary conditions, since then their difference satisfies a linear flow equation with vanishing boundary condition, and thus vanishes.

To shorten notation, we define for $n \in \mathbb{N}$ the multivariate Taylor operator by
\begin{equation}
\mathcal{T}^n_{\vec{x} \to \vec{y}} f(\vec{x}) \equiv \sum_{\abs{\vec{w}} = n} \frac{(\vec{x}-\vec{y})^\vec{w}}{\vec{w}!} \partial^\vec{w} f(\vec{y}) \eqend{.}
\end{equation}
We first show the so-called Lowenstein rules~\cite{lowenstein1971,clarklowenstein1976}. For the case of a single insertion of a composite operator, we have
\begin{proposition}
\label{thm_lowenstein_1}
For all composite operators $\op_A$, we have
\begin{equation}
\label{lowenstein_1}
\partial^a_x L^{\Lambda, \Lambda_0}\left( \op_A(x) \right) = L^{\Lambda, \Lambda_0}\left( \partial^a_x \op_A(x) \right)
\end{equation}
which again should be understood as a shorthand for the hierarchy of identities obtained when we expand the above equations in external fields and antifields to an arbitrary loop order $l$.
\end{proposition}
The proof is obtained by comparing the flow equation and the boundary conditions for both sides, which is straightforward but lengthy, and we refer the reader to Ref.~\cite{froebhollandhollands2015} for details. For functionals with multiple insertions, we have
\begin{proposition}
\label{thm_lowenstein_2}
For all composite operators $\op_{A_i}$ and all multiindices $\vec{a}$ not involving the last coordinate $x_s$, we have
\begin{equation}
\label{lowenstein_2}
\partial^\vec{a}_\vec{x} F^{\Lambda, \Lambda_0}_D\left( \bigotimes_{k=1}^s \op_{A_k}(x_k) \right) = F^{\Lambda, \Lambda_0}_D\left( \bigotimes_{k=1}^s \partial^{a_k} \op_{A_k}(x_k) \right)
\end{equation}
for $x_s = 0$.
\end{proposition}
To prove this equality, we again compare flow equations and boundary conditions. The flow equations are seen to be the same by using the flow equation~\eqref{f_sop_flow} for the $F$ functionals and the first Lowenstein rule~\eqref{lowenstein_1}, and the boundary conditions are the same (vanishing) because both have the same amount of oversubtraction.

As a simple corollary of Propositions~\ref{thm_lowenstein_1} and~\ref{thm_lowenstein_2}, we obtain
\begin{corollary}
\label{thm_lowenstein_3}
For all composite operators $\op_{A_i}$ and all multiindices $\vec{a}$ not involving the last coordinate $x_s$, we have
\begin{equation}
\label{lowenstein_3}
\partial^\vec{a}_\vec{x} G^{\Lambda, \Lambda_0}_D\left( \bigotimes_{k=1}^s \op_{A_k}(x_k) \right) = G^{\Lambda, \Lambda_0}_D\left( \bigotimes_{k=1}^s \partial^{a_k} \op_{A_k}(x_k) \right) \eqend{,}
\end{equation}
since both sides only depend on the differences $x_i-x_j$, and we can thus set w.l.o.g. $x_s = 0$.
\end{corollary}

Then we define
\begin{definition}
\label{ope_def}
The (regulated) OPE coefficient $\mathcal{C}^{\mu, \Lambda_0; \unitmatrix}_{A_1 \cdots A_s}$ is defined as
\begin{equation}
\label{ope_def_unit}
\mathcal{C}^{\mu, \Lambda_0; \unitmatrix}_{A_1 \cdots A_s}(\vec{x}) \equiv G^{\mu, \Lambda_0}\left( \bigotimes_{k=1}^s \op_{A_k}(x_k) \right) \eqend{.}
\end{equation}
For
\begin{equation}
\op_B(x) \equiv \left( \prod_{i=1}^m \partial^{w_i} \phi_{K_i}(x) \right) \left( \prod_{j=1}^n \partial^{w^\ddag_j} \phi^\ddag_{L_j}(x) \right)
\end{equation}
with $[\op_B] = [\vec{K}] + [\vec{L}^\ddag] + \abs{\vec{w}} + \abs{\vec{w}^\ddag} > 0$, the (regulated) OPE coefficients $\mathcal{C}^{\mu, \Lambda_0; B}_{A_1 \cdots A_s}$ are recursively defined as
\begin{equation}
\label{ope_def_recursive}
\mathcal{C}^{\mu, \Lambda_0; B}_{A_1 \cdots A_s}(\vec{x}) \equiv \mathcal{D}^B_\vec{0} \left[ G^{\mu, \Lambda_0}\left( \bigotimes_{k=1}^s \op_{A_k}(x_k) \right) - \sum_{C\colon [\op_C] < [\op_B]} \mathcal{C}^{\mu, \Lambda_0; C}_{A_1 \cdots A_s}(\vec{x}) L^{\mu, \Lambda_0}\left( \op_C(x_s) \right) \right] \eqend{.}
\end{equation}
\end{definition}

We want to show that the OPE exists with these definitions (at least) as an asymptotic expansion. We thus define the remainder functional
\begin{equation}
\label{remainder_def_recursive}
R^{\Lambda, \Lambda_0}_D\left( \bigotimes_{k=1}^s \op_{A_k}(x_k) \right) \equiv G^{\Lambda, \Lambda_0}\left( \bigotimes_{k=1}^s \op_{A_k}(x_k) \right) - \sum_{C\colon [\op_C] < D} \mathcal{C}^{\mu, \Lambda_0; C}_{A_1 \cdots A_s}\left( \vec{x} \right) L^{\Lambda, \Lambda_0}\left( \op_C(x_s) \right) \eqend{,}
\end{equation}
and show that it has the correct asymptotic behaviour under rescalings $\vec{x} \to \tau \vec{x}$. Note that using the remainder functional, we have
\begin{equation}
\label{ope_from_remainder}
\mathcal{C}^{\mu, \Lambda_0; B}_{A_1 \cdots A_s}(\vec{x}) = \mathcal{D}^B_\vec{0} R^{\mu, \Lambda_0}_{[\op_B]}\left( \bigotimes_{k=1}^s \op_{A_k}(x_k) \right) \eqend{.}
\end{equation}

We first prove some equalities which will be useful in the following. The first one expresses the difference between two oversubtracted ACs,
\begin{proposition}
\label{oversubtracted_diff}
For the difference between two oversubtracted ACs, $s \geq 2$ and $D' > D \geq 0$, we have
\begin{equation}
G^{\Lambda, \Lambda_0}_D\left( \bigotimes_{k=1}^s \op_{A_k} \right) - G^{\Lambda, \Lambda_0}_{D'}\left( \bigotimes_{k=1}^s \op_{A_k} \right) = - \hbar \sum_{C\colon D \leq [\op_C] < D'} L^{\Lambda, \Lambda_0}\left( \op_C(0) \right) \mathcal{D}^C_\vec{0} F^{\mu, \Lambda_0}_{[\op_C]}\left( \bigotimes_{k=1}^s \op_{A_k} \right) \eqend{.}
\end{equation}
\end{proposition}
On the left-hand side, we first write the telescopic sum
\begin{equation}
G^{\Lambda, \Lambda_0}_D\left( \bigotimes_{k=1}^s \op_{A_k} \right) - G^{\Lambda, \Lambda_0}_{D'}\left( \bigotimes_{k=1}^s \op_{A_k} \right) = \sum_{n=0}^{(D'-D)/\Delta-1} \left[ G^{\Lambda, \Lambda_0}_{D+n\Delta}\left( \bigotimes_{k=1}^s \op_{A_k} \right) - G^{\Lambda, \Lambda_0}_{D+(n+1)\Delta}\left( \bigotimes_{k=1}^s \op_{A_k} \right) \right] \eqend{.}
\end{equation}
Since furthermore the completely disconnected part (the product $\prod_{k=1}^s L^{\Lambda, \Lambda_0}\left( \op_{A_k} \right)$) cancels out in each difference, the equality we need to prove reduces to
\begin{equation}
F^{\Lambda, \Lambda_0}_D\left( \bigotimes_{k=1}^s \op_{A_k} \right) - F^{\Lambda, \Lambda_0}_{D+\Delta}\left( \bigotimes_{k=1}^s \op_{A_k} \right) = \sum_{C\colon [\op_C] = D} L^{\Lambda, \Lambda_0}\left( \op_C \right) \mathcal{D}^C_\vec{0} F^{\mu, \Lambda_0}_D\left( \bigotimes_{k=1}^s \op_{A_k} \right) \eqend{.}
\end{equation}
Both sides fulfil the same linear flow equation, and we only need to check the boundary conditions. Applying the operator $\mathcal{D}^B_\vec{q}$ on both sides, we distinguish three cases:
\begin{enumerate}
\item $[\op_B] > D$

In this case, we take $\Lambda = \Lambda_0$ and both sides vanish by the boundary conditions~\eqref{f_sop_bdy_d} and~\eqref{l_1op_bdy_d}.
\item $[\op_B] = D$

In this case, we take $\Lambda = \mu$ and $\vec{q} = 0$. The boundary conditions~\eqref{f_sop_bdy_d} tell us that $\mathcal{D}^B_\vec{0} F^{\mu, \Lambda_0}_{D+\Delta} = 0$, while according to equation~\eqref{l_1op_bdy_d} we have $\mathcal{D}^B_\vec{0} L^{\mu, \Lambda_0, l'}\left( \op_C(0) \right) = \delta^B_C \delta_{l,0}$, and equality follows by expanding both sides in $\hbar$.
\item $[\op_B] < D$

In this case, we take $\Lambda = \mu$ and $\vec{q} = 0$, and both sides vanish by the boundary conditions~\eqref{f_sop_bdy_d} and~\eqref{l_1op_bdy_d}.
\end{enumerate}
The next proposition is about oversubtracted functionals where all operators are at $x_k = 0$, and reads
\begin{proposition}
\label{oversubtracted_x0}
For $s \geq 2$, we have
\begin{splitequation}
\hbar F^{\Lambda, \Lambda_0}_{[\op_\vec{A}]+\Delta}\left( \bigotimes_{k=1}^s \op_{A_k}(0) \right) = \prod_{k=1}^s L^{\Lambda, \Lambda_0}\left( \op_{A_k}(0) \right) - \sum_{C\colon [\op_C] = [\op_\vec{A}]} L^{\Lambda, \Lambda_0}\left( \op_C(0) \right) \mathcal{D}^C_\vec{0} \prod_{k=1}^s L^{\mu, \Lambda_0}\left( \op_{A_k}(0) \right) \eqend{.}
\end{splitequation}
and for $D > [\op_\vec{A}]$, we have
\begin{splitequation}
G^{\Lambda, \Lambda_0}_D\left( \bigotimes_{k=1}^s \op_{A_k}(0) \right) &= \sum_{C\colon [\op_C] = [\op_\vec{A}]} L^{\Lambda, \Lambda_0}\left( \op_C(0) \right) \mathcal{D}^C_\vec{0} \prod_{k=1}^s L^{\mu, \Lambda_0}\left( \op_{A_k}(0) \right) \\
&\quad+ \hbar \sum_{C\colon [\op_\vec{A}] < [\op_C] < D} L^{\Lambda, \Lambda_0}\left( \op_C(0) \right) \mathcal{D}^C_\vec{0} F^{\mu, \Lambda_0}_{[\op_C]}\left( \bigotimes_{k=1}^s \op_{A_k}(0) \right) \eqend{.}
\end{splitequation}
\end{proposition}
Note that both sides of these equations are well-defined for $x_k = 0$ because an appropriate number of oversubtractions was made. The second equation follows from the first, the definition of the $G$ functionals~\eqref{g_sop_def} and Proposition~\ref{oversubtracted_diff}, such that it remains to prove the first. Using the definition of the $G$ functionals~\eqref{g_sop_def}, the difference between left- and right-hand side is seen to fulfil a linear flow equation, and we only need to check the boundary conditions. We again apply $\mathcal{D}^B_\vec{q}$ and distinguish three cases:
\begin{enumerate}
\item $[\op_B] > [\op_\vec{A}]$

In this case, we take $\Lambda = \Lambda_0$. Distributing $\mathcal{D}^B_\vec{q}$ among the $L^{\Lambda, \Lambda_0}\left( \op_{A_k}(0) \right)$ (giving, say, $\mathcal{D}^{B_k}_\vec{q}$), we have $[\op_{B_k}] > [\op_{A_k}]$ for at least one $k$ and the functional vanishes according to the boundary conditions~\eqref{l_1op_bdy_d}, exactly as $\mathcal{D}^B_\vec{q} L^{\Lambda_0, \Lambda_0}\left( \op_C(0) \right)$. According to equation~\eqref{f_sop_bdy_d}, we also have $\mathcal{D}^B_\vec{q} F^{\Lambda_0, \Lambda_0}_{[\op_\vec{A}]+\Delta} = 0$.
\item $[\op_B] = [\op_\vec{A}]$

In this case, we take $\Lambda = \mu$ and $\vec{q} = 0$. By the boundary conditions~\eqref{f_sop_bdy_d}, $\mathcal{D}^B_\vec{0} F^{\mu, \Lambda_0}_{[\op_\vec{A}]+\Delta} = 0$. On the right-hand side, we get $\mathcal{D}^B_\vec{0} L^{\mu, \Lambda_0}\left( \op_C(0) \right) = \delta^B_C \delta_{l,0}$, and the right-hand side is seen to vanish by expanding in $\hbar$.
\item $[\op_B] < [\op_\vec{A}]$

In this case, we take $\Lambda = \mu$ and $\vec{q} = 0$. By the boundary conditions~\eqref{f_sop_bdy_d}, $\mathcal{D}^B_\vec{0} F^{\mu, \Lambda_0}_{[\op_\vec{A}]+\Delta} = 0$. Distributing $\mathcal{D}^B_\vec{q}$ among the $L^{\Lambda, \Lambda_0}\left( \op_{A_k}(0) \right)$ (giving, say, $\mathcal{D}^{B_k}_\vec{q}$), we have $[\op_{B_k}] < [\op_{A_k}]$ for at least one $k$ and the functional vanishes according to the boundary conditions~\eqref{l_1op_bdy_d}, exactly as $\mathcal{D}^B_\vec{q} L^{\Lambda_0, \Lambda_0}\left( \op_C(0) \right)$.
\end{enumerate}

The next proposition is about Taylor expansions, and we prove
\begin{proposition}
\label{taylor_g}
For all $n \geq 0$ we have
\begin{equation}
\mathcal{T}^n_{\vec{x} \to (x_s, \ldots, x_s)} G^{\Lambda, \Lambda_0}_{[\op_\vec{A}]+n+\Delta}\left( \bigotimes_{k=1}^s \op_{A_k} \right) = \sum_{C\colon [\op_C] = [\op_\vec{A}]+n} L^{\Lambda, \Lambda_0}\left( \op_C(0) \right) \mathcal{T}^n_{\vec{x} \to (x_s, \ldots, x_s)} \mathcal{D}^C_\vec{0} \prod_{k=1}^s L^{\mu, \Lambda_0}\left( \op_{A_k} \right) \eqend{.}
\end{equation}
\end{proposition}
Note first that the left-hand sides are well-defined because an appropriate number of oversubtractions has been made. W.l.o.g. we can again take $x_s = 0$, and using the Lowenstein rule~\eqref{lowenstein_3} and Proposition~\ref{oversubtracted_x0} we obtain
\begin{splitequation}
\mathcal{T}^n_{\vec{x} \to \vec{0}} G^{\Lambda, \Lambda_0}_{[\op_\vec{A}]+n+\Delta}\left( \bigotimes_{k=1}^s \op_{A_k} \right) &= \sum_{\abs{\vec{w}} = n} \frac{\vec{x}^\vec{w}}{\vec{w}!} G^{\Lambda, \Lambda_0}_{[\op_\vec{A}]+n+\Delta}\left( \bigotimes_{k=1}^s \left( \partial^{w_k} \op_{A_k} \right)(0) \right) \\
&= \sum_{\abs{\vec{w}} = n} \frac{\vec{x}^\vec{w}}{\vec{w}!} \Bigg[ \sum_{C\colon [\op_C] = [\op_\vec{A}]+n} L^{\Lambda, \Lambda_0}\left( \op_C(0) \right) \mathcal{D}^C_\vec{0} \prod_{k=1}^s L^{\mu, \Lambda_0}\left( \left( \partial^{w_k} \op_{A_k} \right)(0) \right) \\
&\qquad+ \hbar \sum_{C\colon n < [\op_C]-[\op_\vec{A}] < n+\Delta} L^{\Lambda, \Lambda_0}\left( \op_C(0) \right) \mathcal{D}^C_\vec{0} F^{\mu, \Lambda_0}_{[\op_C]}\left( \bigotimes_{k=1}^s \left( \partial^{w_k} \op_{A_k} \right)(0) \right) \Bigg] \eqend{.}
\end{splitequation}
However, since the smallest difference in composite operator dimensions is $\Delta$, the sum in the last line vanishes, and using the Lowenstein rule~\eqref{lowenstein_1} the proposition follows.

We need one further proposition about Taylor expansions, which reads
\begin{proposition}
\label{taylor_l}
For all $B$ we have
\begin{equation}
\mathcal{D}^B_\vec{0} \prod_{k=1}^s L^{\mu, \Lambda_0}\left( \op_{A_k} \right) = \sum_{n \leq [\op_B] - [\op_\vec{A}]} \mathcal{T}^n_{\vec{x} \to (x_s, \ldots, x_s)} \mathcal{D}^B_\vec{0} \prod_{k=1}^s L^{\mu, \Lambda_0}\left( \op_{A_k} \right) \eqend{,}
\end{equation}
where the sum over $n$ is not present if $[\op_B] < [\op_\vec{A}]$.
\end{proposition}
If $[\op_B] < [\op_\vec{A}]$, we distribute $\mathcal{D}^B_\vec{0}$ over the functionals on the left-hand side (giving, say, $\mathcal{D}^{B_k}_\vec{0}$) and have $[\op_{B_k}] < [\op_{A_k}]$ for at least one $k$, whence the functional vanishes according to the boundary conditions~\eqref{l_1op_bdy_d} (we first have to bring the operator insertion to $x_k = 0$, but then even less derivatives act on the functional). For $[\op_B] \geq [\op_\vec{A}]$, we perform a Taylor expansion with remainder of the functionals on the left-hand side (which is well-defined since the $L^{\Lambda, \Lambda_0}\left( \op_{A_k}(x_k) \right)$ are smooth in $x_k$). This gives
\begin{splitequation}
\prod_{k=1}^s L^{\mu, \Lambda_0}\left( \op_{A_k} \right) &= \sum_{n \leq [\op_B] - [\op_\vec{A}]} \mathcal{T}^n_{\vec{x} \to (x_s, \ldots, x_s)} \prod_{k=1}^s L^{\mu, \Lambda_0}\left( \op_{A_k} \right) + \int_0^1 \frac{r}{(1-t)} \mathcal{T}^r_{\vec{x} \to t \vec{x}} \prod_{k=1}^s L^{\mu, \Lambda_0}\left( \op_{A_k} \right) \total t \eqend{,}
\end{splitequation}
where $r$ is the smallest integer strictly greater than $[\op_B] - [\op_\vec{A}]$. The integral over $t$ is uniformly convergent since at least one factor of $(1-t)$ comes from the Taylor operator $\mathcal{T}^r_{\vec{x} \to t \vec{x}}$, such that applying $\mathcal{D}^B_\vec{0}$ we can pull it inside the integral. However, since $r > [\op_B] - [\op_\vec{A}]$ we have
\begin{equation}
\mathcal{D}^B_\vec{0} \mathcal{T}^r_{\vec{x} \to t \vec{x}} \prod_{k=1}^s L^{\mu, \Lambda_0}\left( \op_{A_k} \right) = 0
\end{equation}
because of the boundary conditions~\eqref{l_1op_bdy_d} by the same argumentation as before, and the proposition follows.


It is advantageous to rewrite the remainder functionals (and thus the OPE coefficients) in a different form, given by
\begin{proposition}
\label{ope_remainder_def}
For all $B$, we have
\begin{equation}
\label{remainder_def2}
R^{\Lambda, \Lambda_0}_D\left( \bigotimes_{k=1}^s \op_{A_k}(x_k) \right) = G^{\Lambda, \Lambda_0}_D\left( \bigotimes_{k=1}^s \op_{A_k}(x_k) \right) - \sum_{n < D - [\op_\vec{A}]} \mathcal{T}^n_{\vec{x} \to (x_s, \ldots, x_s)} G^{\Lambda, \Lambda_0}_D\left( \bigotimes_{k=1}^s \op_{A_k}(x_k) \right)
\end{equation}
and
\begin{equation}
\label{ope_def2}
\mathcal{C}^{\mu, \Lambda_0; B}_{A_1 \cdots A_s}(\vec{x}) = \mathcal{D}^B_\vec{0} \left[ G^{\mu, \Lambda_0}_{[\op_B]}\left( \bigotimes_{k=1}^s \op_{A_k}(x_k) \right) - \sum_{n < [\op_B] - [\op_\vec{A}]} \mathcal{T}^n_{\vec{x} \to (x_s, \ldots, x_s)} G^{\mu, \Lambda_0}_{[\op_B]}\left( \bigotimes_{k=1}^s \op_{A_k}(x_k) \right) \right] \eqend{,}
\end{equation}
where the sums are empty for $D \leq [\op_\vec{A}]$, resp. $[\op_B] \leq [\op_\vec{A}]$.
\end{proposition}
Using equation~\eqref{ope_from_remainder}, the second half of the proposition follows from the first. To prove the first half, we proceed by induction. Inspecting the recursive definition~\eqref{remainder_def_recursive}, equation~\eqref{remainder_def2} holds for $D = 0$ since $G^{\Lambda, \Lambda_0}_0 = G^{\Lambda, \Lambda_0}$. W.l.o.g. we set $x_s = 0$ in the following, since all functionals are translation invariant. Assume thus that equation~\eqref{remainder_def2} (and thus also equation~\eqref{ope_def2}) holds for all $D' \leq D$, and calculate using the definition~\eqref{remainder_def_recursive}
\begin{equation}
R^{\Lambda, \Lambda_0}_{D+\Delta}\left( \bigotimes_{k=1}^s \op_{A_k} \right) = R^{\Lambda, \Lambda_0}_D\left( \bigotimes_{k=1}^s \op_{A_k} \right) - \sum_{C\colon [\op_C] = D} \mathcal{C}^{\mu, \Lambda_0; C}_{A_1 \cdots A_s} L^{\Lambda, \Lambda_0}\left( \op_C(0) \right) \eqend{.}
\end{equation}
We now insert the induction hypothesis~\eqref{remainder_def2} and~\eqref{ope_def2} on the right-hand side, and use Proposition~\ref{oversubtracted_diff} with $D' = D+\Delta$ to obtain
\begin{splitequation}
R^{\Lambda, \Lambda_0}_{D+\Delta}\left( \bigotimes_{k=1}^s \op_{A_k} \right) &= G^{\Lambda, \Lambda_0}_{D+\Delta}\left( \bigotimes_{k=1}^s \op_{A_k} \right) - \sum_{n < D+\Delta - [\op_\vec{A}]} \mathcal{T}^n_{\vec{x} \to \vec{0}} G^{\Lambda, \Lambda_0}_{D+\Delta}\left( \bigotimes_{k=1}^s \op_{A_k} \right) \\
&+ \sum_{D - [\op_\vec{A}] \leq n < D+\Delta - [\op_\vec{A}]} \mathcal{T}^n_{\vec{x} \to \vec{0}} G^{\Lambda, \Lambda_0}_{D+\Delta}\left( \bigotimes_{k=1}^s \op_{A_k} \right) \\
&- \sum_{C\colon [\op_C] = D} L^{\Lambda, \Lambda_0}\left( \op_C(0) \right) \left( 1 - \sum_{n < D - [\op_\vec{A}]} \mathcal{T}^n_{\vec{x} \to \vec{0}} \right) \mathcal{D}^C_\vec{0} \prod_{k=1}^s L^{\mu, \Lambda_0}\left( \op_{A_k} \right) \eqend{.}
\end{splitequation}
The proposition follows if we can show that the terms in the second and third line cancel. If $D - [\op_\vec{A}]$ is an integer, the sum in the second line only involves $n = D - [\op_\vec{A}]$, and using Proposition~\ref{taylor_g} we can absorb this term by extending the sum in the third line to also include $n = D - [\op_\vec{A}]$. If $D - [\op_\vec{A}]$ is not an integer, the sum in the second line is empty, and we can trivially extend the sum in the third line. Using Proposition~\ref{taylor_l}, we conclude that the sum in the third line then vanishes, and the proposition follows.

We can then show that the OPE exists (at least) in the asymptotic sense:
\begin{proposition}
\label{thm_ope_asymptotic}
The OPE exists in the asymptotic sense, i.e., for all $\Lambda \leq \mu$ the remainder vanishes if we scale the points $x_k$ of the operator insertions together. Concretely, we have
\begin{equation}
\label{remainder_scaling}
\lim_{\tau \to 0} \tau^{[\op_\vec{A}]-D + \delta} R^{\Lambda, \Lambda_0}_D\left( \bigotimes_{k=1}^s \op_{A_k}(\tau x_k) \right) = 0
\end{equation}
for all $\delta > 0$ and all $D \geq 0$. The OPE coefficients scale in the same way, i.e.,
\begin{equation}
\label{ope_scaling}
\lim_{\tau \to 0} \tau^{[\op_\vec{A}]-[\op_B] + \delta} \, \mathcal{C}^{\mu, \Lambda_0; B}_{A_1 \cdots A_s}(\tau \vec{x}) = 0
\end{equation}
for all $\delta > 0$.
\end{proposition}
The scaling of the OPE coefficients follows from the scaling of the remainder by using equation~\eqref{ope_from_remainder}, such that we only have to prove the first part. For the proof we distinguish the cases $D \leq [\op_\vec{A}]$ and $D > [\op_\vec{A}]$. Since the remainder is by definition translation invariant, we may furthermore set $x_s = 0$. First note that for $\tau \leq 1$ we have
\begin{equation}
\label{xi_estimate}
\Xi^{\Lambda,\Lambda_1}_{p,p'}(\tau \vec{x}) \leq \tau^{-p'} \Xi^{\Lambda,\Lambda_1}_{p,p'}(\vec{x})
\end{equation}
if $p' \geq 0$, which follows directly from the definition~\eqref{xi_lambda_def} taking $\rho = 0$. Proposition~\ref{ope_remainder_def} then shows that in the case $D \leq [\op_\vec{A}]$, we simply have
\begin{equation}
R^{\Lambda, \Lambda_0}_D\left( \bigotimes_{k=1}^s \op_{A_k}(\tau x_k) \right) = G^{\Lambda, \Lambda_0}_D\left( \bigotimes_{k=1}^s \op_{A_k}(\tau x_k) \right) \eqend{,}
\end{equation}
and the bounds~\eqref{bound_gs} together with the estimate~\eqref{xi_estimate} show that for $\tau \leq 1$ we have
\begin{splitequation}
\abs{ \mathcal{G}^{\Lambda, \Lambda_0, l}_{\vec{K} \vec{L}^\ddag; D}\left( \bigotimes_{k=1}^s \op_{A_k}(\tau x_k); \vec{q} \right) } &\leq \tau^{-([\op_\vec{A}]-D+\epsilon)} \mu^{[\op_\vec{A}]-D+\epsilon} \, \Xi^{\Lambda,\Lambda_1}_{D, [\op_\vec{A}]-D+\epsilon}(\vec{x}) \\
&\quad\times \sup\left( 1, \frac{\abs{\vec{q}}}{\sup(\mu,\Lambda)} \right)^{g^{(s)}\left( [\op_\vec{A}], m+n+2l, 0 \right)} \\
&\quad\times \sum_{T^* \in \mathcal{T}^*_{m+n}} \mathsf{G}^{T^*,0}_{\vec{K} \vec{L}^\ddag; D-\epsilon}(\vec{q}; \mu, \Lambda) \, \mathcal{P}\left( \ln_+ \frac{\sup\left( \abs{\vec{q}}, \mu \right)}{\sup(\inf(\mu, \bar{\eta}(\vec{q})), \Lambda)}, \ln_+ \frac{\Lambda}{\mu} \right)
\end{splitequation}
since we have $p' = [\op_\vec{A}]-D+\epsilon > 0$ in this case. If $D > [\op_\vec{A}]$, Proposition~\ref{ope_remainder_def} shows that
\begin{equation}
R^{\Lambda, \Lambda_0}_D\left( \bigotimes_{k=1}^s \op_{A_k}(\tau x_k) \right) = \left( 1 - \sum_{n < D - [\op_\vec{A}]} \mathcal{T}^n_{\tau \vec{x} \to \vec{0}} \right) G^{\Lambda, \Lambda_0}_D\left( \bigotimes_{k=1}^s \op_{A_k}(\tau x_k) \right) \eqend{,}
\end{equation}
and the bounds~\eqref{bound_gs_taylor} together with the estimate~\eqref{xi_estimate} show that for $\tau \leq 1$ we have
\begin{splitequation}
&\abs{ \left( 1 - \sum_{n < D - [\op_\vec{A}]} \mathcal{T}^n_{\tau \vec{x} \to \vec{0}} \right) \mathcal{G}^{\Lambda, \Lambda_0, l}_{\vec{K} \vec{L}^\ddag; D}\left( \bigotimes_{k=1}^s \op_{A_k}(\tau x_k); \vec{q} \right) } \leq \tau^{-([\op_\vec{A}]-D+\epsilon)} \mu^{[\op_\vec{A}]+r-D+\epsilon} \sup_{i\in\{1,\ldots,s\}} \abs{x_i}^r \\
&\quad\times \Xi^{\Lambda,\Lambda_1}_{D, [\op_\vec{A}]+r-D+\epsilon}(\vec{x}) \ \sup\left( 1, \frac{\abs{\vec{q}}}{\sup(\mu,\Lambda)} \right)^{g^{(s)}\left( [\op_\vec{A}]+r, m+n+2l, 0 \right)} \\
&\quad\times \sum_{T^* \in \mathcal{T}^*_{m+n}} \mathsf{G}^{T^*,0}_{\vec{K} \vec{L}^\ddag; D-\epsilon}(\vec{q}; \mu, \Lambda) \, \mathcal{P}\left( \ln_+ \frac{\sup\left( \abs{\vec{q}}, \mu \right)}{\sup(\inf(\mu, \bar{\eta}(\vec{q})), \Lambda)}, \ln_+ \frac{\Lambda}{\mu} \right) \eqend{,} \raisetag{1.7\baselineskip}
\end{splitequation}
where $r$ is the smallest integer $\geq D - [\op_\vec{A}]$, since also in this case we have $p' = [\op_\vec{A}]+r-D+\epsilon > 0$. Taking then in both cases $\epsilon = \inf\left( \delta, \Delta \right)/2$ (fulfilling the restriction $0 < \epsilon < \Delta$ of Propositions~\ref{thm_gs} and~\ref{thm_gs_taylor}), the proposition follows.

For later use, we also need to rewrite the partially oversubtracted disconnected functionals in a form which involves OPE coefficients similar to the remainder functionals, and we prove
\begin{proposition}
\label{thm_g_ssop_ope}
For all $D \leq [\op_\vec{A}']$ we have
\begin{splitequation}
\label{g_ssop_as_ope}
&G^{\Lambda, \Lambda_0}_D\left( \bigotimes_{k=1}^{s'} \op_{A_k}(x_k); \bigotimes_{l=s'+1}^s \op_{A_l}(x_l) \right) = G^{\Lambda, \Lambda_0}\left( \bigotimes_{k=1}^s \op_{A_k}(x_k) \right) \\
&\qquad- \sum_{B\colon [\op_B] < D} \mathcal{C}^{\mu, \Lambda_0; B}_{A_1 \cdots A_{s'}}(x_1, \ldots, x_{s'}) G^{\Lambda, \Lambda_0}\left( \bigotimes_{l=s'+1}^s \op_{A_l}(x_l) \otimes \op_B(x_{s'}) \right) \eqend{.}
\end{splitequation}
\end{proposition}
To prove this equality, we compare flow equations and boundary conditions on both sides. First, one easily checks from the definition~\eqref{g_sop_def}, the flow equation for the functionals with one insertion~\eqref{l_sop_flow} (taking $s=1$ in that formula) and the flow equation~\eqref{f_sop_flow} for the oversubtracted almost disconnected functionals that the oversubtracted disconnected functionals fulfil the linear flow equation
\begin{splitequation}
\label{g_sop_flow}
\partial_\Lambda G^{\Lambda, \Lambda_0}_D\left( \bigotimes_{k=1}^s \op_{A_k}(x_k) \right) &= \frac{\hbar}{2} \left\langle \frac{\delta}{\delta \phi_K}, \left( \partial_\Lambda C^{\Lambda, \Lambda_0}_{KL} \right) \ast \frac{\delta}{\delta \phi_L} \right\rangle G^{\Lambda, \Lambda_0}_D\left( \bigotimes_{k=1}^s \op_{A_k}(x_k) \right) \\
&\quad- \left\langle \frac{\delta}{\delta \phi_K} L^{\Lambda, \Lambda_0}, \left( \partial_\Lambda C^{\Lambda, \Lambda_0}_{KL} \right) \ast \frac{\delta}{\delta \phi_L} G^{\Lambda, \Lambda_0}_D\left( \bigotimes_{k=1}^s \op_{A_k}(x_k) \right) \right\rangle \eqend{.}
\end{splitequation}
From the definition~\eqref{g_ssop_def}, the flow equation~\eqref{h_sop_flow} for the partially oversubtracted almost disconnected functionals and the linear flow equation~\eqref{g_sop_flow} it then follows that also the partially oversubtracted disconnected functionals fulfil a linear flow equation, given by
\begin{splitequation}
\label{g_ssop_flow}
&\partial_\Lambda G^{\Lambda, \Lambda_0}_D\left( \bigotimes_{k=1}^{s'} \op_{A_k}; \bigotimes_{l=s'+1}^s \op_{A_l}(x_l) \right) = \frac{\hbar}{2} \left\langle \frac{\delta}{\delta \phi_K}, \left( \partial_\Lambda C^{\Lambda, \Lambda_0}_{KL} \right) \ast \frac{\delta}{\delta \phi_L} \right\rangle G^{\Lambda, \Lambda_0}_D\left( \bigotimes_{k=1}^{s'} \op_{A_k}; \bigotimes_{l=s'+1}^s \op_{A_l} \right) \\
&\qquad- \left\langle \frac{\delta}{\delta \phi_K} L^{\Lambda, \Lambda_0}, \left( \partial_\Lambda C^{\Lambda, \Lambda_0}_{KL} \right) \ast \frac{\delta}{\delta \phi_L} G^{\Lambda, \Lambda_0}_D\left( \bigotimes_{k=1}^{s'} \op_{A_k}; \bigotimes_{l=s'+1}^s \op_{A_l} \right) \right\rangle \eqend{.} \raisetag{1.5\baselineskip}
\end{splitequation}
Thus, both sides of equation~\eqref{g_ssop_as_ope} fulfil the same linear flow equation, and it remains to check boundary conditions. For this, we evaluate equation~\eqref{g_ssop_as_ope} at $\Lambda = \Lambda_0$. Since the partially oversubtracted almost diconnected functionals have vanishing boundary conditions at $\Lambda = \Lambda_0$~\eqref{h_sop_bdy}, the definition~\eqref{g_ssop_def} shows that the left-hand side reduces to a product of two disconnected functionals, the first one being oversubtracted. We then insert the definition~\eqref{g_sop_def} of the oversubtracted disconnected functionals in terms of almost disconnected ones and a product of functionals with one insertion of a composite operator. Since the almost disconnected functionals have all vanishing boundary conditions at $\Lambda = \Lambda_0$ if they are not oversubtracted~\eqref{f_sop_bdy_d}, many terms vanish and equation~\eqref{g_ssop_as_ope} reduces to
\begin{equation}
\hbar F^{\Lambda_0, \Lambda_0}_D\left( \bigotimes_{k=1}^{s'} \op_{A_k}(x_k) \right) = \sum_{B\colon [\op_B] < D} \mathcal{C}^{\mu, \Lambda_0; B}_{A_1 \cdots A_{s'}}(x_1, \ldots, x_{s'}) L^{\Lambda_0, \Lambda_0}\left( \op_B(x_{s'}) \right) \eqend{.}
\end{equation}
We now use the definition of the remainder functional~\eqref{remainder_def_recursive} to replace the sum over OPE coefficients, and again the definition~\eqref{g_sop_def} together with the vanishing of the not-oversubtracted almost disconnected functionals at $\Lambda = \Lambda_0$~\eqref{f_sop_bdy_d} to obtain
\begin{equation}
G^{\Lambda_0, \Lambda_0}_D\left( \bigotimes_{k=1}^{s'} \op_{A_k}(x_k) \right) = R^{\Lambda_0, \Lambda_0}_D\left( \bigotimes_{k=1}^{s'} \op_{A_k}(x_k) \right) \eqend{.}
\end{equation}
However, Proposition~\ref{ope_remainder_def} shows that both sides are equal for $D \leq [\op_\vec{A}']$, and thus the proposition is proven.

\section{Parameter derivatives of functionals}
\label{sec_param}

In this section, we derive explicit formulas for the derivative of functionals with insertions of composite operators with respect to a coupling constant $g$ appearing in the interaction Lagrangian. Since the OPE coefficients are given by evaluating the disconnected functionals with insertions at a fixed point~\eqref{ope_def_recursive}, we thus also obtain a formula for the $g$-derivative of the OPE coefficients.

\begin{proposition}
\label{thm_l0_g}
For each parameter $g$ appearing in the interaction Lagrangian $L^{\Lambda_0}$ but not the covariance $C^{\Lambda, \Lambda_0}_{MN}$, there exist (possibly $g$-dependent) coefficients $\mathcal{I}^E$ which are nonvanishing only for $1 \leq [\op_E] \leq 4$, such that defining the ``interaction operator'' $\opint$ by~\eqref{op_g_def}
\begin{equation}
\opint \equiv \sum_E \mathcal{I}^E \op_E
\end{equation}
we have
\begin{equation}
\label{l0_g_deriv}
\partial_g L^{\Lambda, \Lambda_0} = \int L^{\Lambda, \Lambda_0}\left( \opint(y) \right) \total^4 y
\end{equation}
for all $\Lambda \geq 0$, which should be understood as a shorthand for the hierarchy of identities obtained when we expand the above equations in external fields and antifields to an arbitrary loop order $l$.
\end{proposition}
To show this identity, we first expand in external fields, antifields and $\hbar$, perform a Fourier transformation and use the shift property~\eqref{func_sop_shift} to isolate the $y$ dependence of the functional with one insertion and perform the $y$ integral. This leaves us with
\begin{splitequation}
\label{l0_g_deriv_2}
(2\pi)^4 \delta^4\left( \sum_{i=1}^{m+n} q_i \right) \partial_g \mathcal{L}^{\Lambda, \Lambda_0, l}_{\vec{K} \vec{L}^\ddag}\left( \vec{q} \right) &= \int \mathe^{- \mathi y \sum_{i=1}^{m+n} q_i} \mathcal{L}^{\Lambda, \Lambda_0, l}_{\vec{K} \vec{L}^\ddag}\left( \opint(0); \vec{q} \right) \total^4 y \\
&= (2\pi)^4 \delta^4\left( \sum_{i=1}^{m+n} q_i \right) \mathcal{L}^{\Lambda, \Lambda_0, l}_{\vec{K} \vec{L}^\ddag}\left( \opint(0); \vec{q} \right) \eqend{.}
\end{splitequation}
Removing the momentum-conserving $\delta$ distribution, the identity~\eqref{l0_g_deriv} follows if we can show that both sides of
\begin{equation}
\partial_g \mathcal{L}^{\Lambda, \Lambda_0, l}_{\vec{K} \vec{L}^\ddag}\left( \vec{q} \right) = \mathcal{L}^{\Lambda, \Lambda_0, l}_{\vec{K} \vec{L}^\ddag}\left( \opint(0); \vec{q} \right)
\end{equation}
(under the constraint that $\sum_{i=1}^{m+n} q_i = 0$) fulfil the same flow equation and have the same boundary conditions. This however is very easy: taking a derivative with respect to $g$ of the flow equation~\eqref{l_0op_flow} for $L^{\Lambda, \Lambda_0}$ and noting that we interchange the derivatives w.r.t. $\Lambda$ and $g$ in any finite order $l$ of perturbation theory (since then $L^{\Lambda, \Lambda_0}$ is a formal power series in $g$ if this is the case for $L^{\Lambda_0}$), we obtain the same flow equation as for the functionals with one operator insertion. For all irrelevant functionals with $[\vec{K}]+[\vec{L}^\ddag]+\abs{w} > 4$, the boundary conditions for the left-hand side are vanishing at $\Lambda = \Lambda_0$, which is also the case for the right-hand side if $[\opint] \leq 4$, \ie, if the coefficients $\mathcal{I}^E$ vanish for $[\op_E] > 4$. The boundary conditions for relevant and marginal functionals are then given by evaluating the left-hand side at $\Lambda = \mu$ and vanishing momenta, and the exact $g$-dependence of these conditions then determines the exact form of $\opint$, respectively the $\mathcal{I}^E$. Note that while the left-hand side vanishes for relevant functionals with $[\vec{K}]+[\vec{L}^\ddag]+\abs{w} < 4$ at $\Lambda = 0$ and vanishing momenta, functionals with one operator insertion generically diverge at this point, and we must give boundary conditions either at $\Lambda = \mu$ and vanishing momenta, or $\Lambda = 0$ but non-exceptional momenta. Thus, since the relevant functionals of the left-hand side generically do not vanish at $\Lambda = \mu$, we cannot assert that $[\opint] = 4$ exactly, but only that $[\opint] \leq 4$. Furthermore, the coefficient of the unit operator is determined by only one boundary condition for marginal functionals, namely when no external legs and no derivatives are present. However, by adding a field-independent constant to the bare action we can always arrange that these functionals vanish, such that no unit operator appears in $\opint$. Since all basic fields have dimension $\geq 1$, we thus must also have $[\opint] \geq 1$, \ie, the coefficients $\mathcal{I}^E$ also vanish for $[\op_E] < 1$.

\begin{proposition}
\label{thm_l1_g}
For each parameter $g$ appearing in the interaction Lagrangian $L^{\Lambda_0}$ but not the covariance $C^{\Lambda, \Lambda_0}_{MN}$ and all (monomial) composite operators $\op_A$, we have
\begin{equation}
\label{l1_g_deriv}
\partial_g L^{\Lambda, \Lambda_0}\left( \op_A(x) \right) = \int F^{\Lambda, \Lambda_0}_{[\op_A] + \Delta}\left( \opint(y) \otimes \op_A(x) \right) \total^4 y
\end{equation}
for all $\Lambda \geq \Lambda_1$ with an arbitrary $0 < \Lambda_1 \leq \mu$, with the composite operator $\opint$ of Proposition~\ref{thm_l0_g}. Again, this identity should be understood as a shorthand for the hierarchy of identities obtained when we expand the above equations in external fields and antifields to an arbitrary loop order $l$. Especially, since $\Lambda_1$ is arbitrary we obtain in the physical limit
\begin{equation}
\label{l1_g_deriv_phys}
\partial_g L^{0, \infty}\left( \op_A(x) \right) = \lim_{\Lambda \to 0} \int F^{\Lambda, \infty}_{[\op_A] + \Delta}\left( \opint(y) \otimes \op_A(x) \right) \total^4 y \eqend{.}
\end{equation}
\end{proposition}
Expanding in external fields, antifields and $\hbar$ and using the shift property~\eqref{func_sop_shift} to bring the insertion of $\op_A(x)$ to $x = 0$, we have to show that
\begin{equation}
\label{l1_g_deriv_expanded}
\partial_g \mathcal{L}^{\Lambda, \Lambda_0, l}_{\vec{K} \vec{L}^\ddag}\left( \op_A(0); \vec{q} \right) = \int \mathcal{F}^{\Lambda, \Lambda_0, l}_{\vec{K} \vec{L}^\ddag; [\op_A] + \Delta}\left( \opint(y-x) \otimes \op_A(0); \vec{q} \right) \total^4 y \eqend{.}
\end{equation}
For two composite operator insertions and $D = [\op_A]+\Delta$, the function $\Xi^{\Lambda,\Lambda_1}$ appearing in the bounds~\eqref{bound_fs} for $\mathcal{F}^{\Lambda, \Lambda_0, l}_{\vec{K} \vec{L}^\ddag; D}\left( \opint(y-x) \otimes \op_A(0); \vec{q} \right)$ which governs the $x$ dependence reduces to
\begin{equation}
\left( \mu \abs{y-x} \right)^{-([\opint]-\Delta+\epsilon+\rho)} \left( \frac{\mu}{\Lambda_1} \right)^{[\opint]-\Delta+\epsilon+\rho}
\end{equation}
for all $0 < \epsilon < \Delta$ and all $\rho \geq 0$, since $\Lambda \geq \Lambda_1$. Thus the $y$ integral converges absolutely: in the UV region where $\mu \abs{y-x} \leq 1$, we choose $\rho = 0$. Since $[\opint] \leq 4$, we have
\begin{equation}
[\opint]-\Delta+\epsilon+\rho \leq 4-\Delta+\epsilon < 4 \eqend{,}
\end{equation}
and the integral over this region is absolutely convergent. In the IR region where $\mu \abs{y-x} > 1$, we choose $\rho = 4$ such that
\begin{equation}
[\opint]-\Delta+\epsilon+\rho > 4 \eqend{,}
\end{equation}
since $\Delta \leq 1$ and $[\opint] \geq 1$, and the integral over this region is also absolutely convergent. We thus can also shift the integration variable $y \to y+x$. The identity~\eqref{l1_g_deriv_expanded} then follows if we can show that both sides fulfill the same flow equation and the same boundary conditions. For this, we have to exchange the $\Lambda$ derivative and the $y$ integral, which is allowed if $\partial_\Lambda \mathcal{F}^{\Lambda, \Lambda_0, l}_{\vec{K} \vec{L}^\ddag; [\op_A] + \Delta}\left( \opint(y) \otimes \op_A(0); \vec{q} \right)$ is absolutely integrable. However, the bounds~\eqref{bound_fs_lambdaderiv} for the $\Lambda$ derivative have the same $x$ dependence as the bounds~\eqref{bound_fs} for the functional itself, and thus the $\Lambda$ derivative is also absolutely integrable. Taking a derivative with respect to $g$ of the flow equation for $L^{\Lambda, \Lambda_0}\left( \op_A(0) \right)$ and using Proposition~\ref{thm_l0_g} [or, more directly, equation~\eqref{l0_g_deriv_2}], we then obtain the same flow equation for both sides of~\eqref{l1_g_deriv_expanded}. The boundary conditions for the left-hand side are then given by equation~\eqref{l_1op_bdy_d}, vanishing at $\Lambda = \Lambda_0$ for irrelevant functionals which have $[\vec{K}]+[\vec{L}^\ddag]+\abs{w} > [\op_A]$, and given by some $g$-independent terms at $\Lambda = \mu$ and vanishing momenta for marginal and relevant functionals with $[\vec{K}]+[\vec{L}^\ddag]+\abs{w} \leq [\op_A]$. Thus, the term on the left-hand side of the identity~\eqref{l1_g_deriv_expanded} has vanishing boundary conditions at $\Lambda = \mu$ and vanishing momenta for $[\vec{K}]+[\vec{L}^\ddag]+\abs{w} \leq [\op_A]$, which coincides with the boundary conditions of the term on the right-hand side if $D = [\op_A] + \Delta$, as claimed in equation~\eqref{l1_g_deriv_expanded}, such that the proposition follows.

\begin{corollary}
\label{thm_l1_g_ope}
Under the conditions and with the notation of Proposition~\ref{thm_l1_g}, we have
\begin{splitequation}
\label{l1_g_deriv_ope}
\hbar \partial_g L^{\Lambda, \Lambda_0}\left( \op_A(x) \right) = \int &\Bigg[ \hbar F^{\Lambda, \Lambda_0}\left( \opint(y) \otimes \op_A(x) \right) \\
&\quad+ \sum_E \mathcal{I}^E \sum_{C\colon [\op_C] \leq [\op_A]} \mathcal{C}^{\mu, \Lambda_0; C}_{E A}(y,x) L^{\Lambda, \Lambda_0}\left( \op_C(x) \right) \Bigg] \total^4 y \eqend{.}
\end{splitequation}
Especially, since $\Lambda_1$ is arbitrary we obtain in the physical limit
\begin{splitequation}
\label{l1_g_deriv_ope_phys}
\hbar \partial_g L^{0, \infty}\left( \op_A(x) \right) = \lim_{\Lambda \to 0} \int &\Bigg[ \hbar F^{\Lambda, \infty}\left( \opint(y) \otimes \op_A(x) \right) \\
&\quad+ \sum_E \mathcal{I}^E \sum_{C\colon [\op_C] \leq [\op_A]} \mathcal{C}^{\mu, \infty; C}_{E A}(y,x) L^{\Lambda, \infty}\left( \op_C(x) \right) \Bigg] \total^4 y \eqend{.}
\end{splitequation}
\end{corollary}
This just reexpresses the equality~\eqref{l1_g_deriv} in terms of non-oversubtracted functionals and partial OPE sums. We first express the oversubtracted functional of Proposition~\ref{thm_l1_g} as
\begin{equation}
\hbar F^{\Lambda, \Lambda_0}_{[\op_A] + \Delta}\left( \opint(y) \otimes \op_A(x) \right) = L^{\Lambda, \Lambda_0}\left( \opint(y) \right) L^{\Lambda, \Lambda_0}\left( \op_A(x) \right) - G^{\Lambda, \Lambda_0}_{[\op_A] + \Delta}\left( \opint(y) \otimes \op_A(x) \right) \eqend{.}
\end{equation}
Sincer $[\opint] \geq 1$ we have $[\op_A] + \Delta < [\opint] + [\op_A]$, and Proposition~\ref{ope_remainder_def} shows that thus
\begin{equation}
G^{\Lambda, \Lambda_0}_{[\op_A] + \Delta}\left( \opint(y) \otimes \op_A(x) \right) = R^{\Lambda, \Lambda_0}_{[\op_A] + \Delta}\left( \opint(y) \otimes \op_A(x) \right) \eqend{.}
\end{equation}
By the definition of the remainder functional~\eqref{remainder_def_recursive} and the interaction operator~\eqref{op_g_def}, we then obtain
\begin{splitequation}
\label{f_sop_oversub_coeffs}
&\hbar F^{\Lambda, \Lambda_0}_{[\op_A] + \Delta}\left( \opint(y) \otimes \op_A(x) \right) = L^{\Lambda, \Lambda_0}\left( \opint(y) \right) L^{\Lambda, \Lambda_0}\left( \op_A(x) \right) - G^{\Lambda, \Lambda_0}\left( \opint(y) \otimes \op_A(x) \right) \\
&\hspace{12em}+ \sum_E \mathcal{I}^E \sum_{C\colon [\op_C] < [\op_A] + \Delta} \mathcal{C}^{\mu, \Lambda_0; C}_{E A}(y,x) L^{\Lambda, \Lambda_0}\left( \op_C(x) \right) \\
&\quad= \hbar F^{\Lambda, \Lambda_0}\left( \opint(y) \otimes \op_A(x) \right) + \sum_E \mathcal{I}^E \sum_{C\colon [\op_C] \leq [\op_A]} \mathcal{C}^{\mu, \Lambda_0; C}_{E A}(y,x) L^{\Lambda, \Lambda_0}\left( \op_C(x) \right) \eqend{,}
\end{splitequation}
and the proposition is proven.

\begin{proposition}
\label{thm_fs_g}
For each parameter $g$ appearing in the interaction Lagrangian $L^{\Lambda_0}$ but not the covariance $C^{\Lambda, \Lambda_0}_{MN}$, all (monomial) composite operators $\op_{A_k}$ and $s \geq 2$, we have
\begin{splitequation}
\label{fs_g_deriv}
&\hbar \partial_g F^{\Lambda, \Lambda_0}\left( \bigotimes_{k=1}^s \op_{A_k}(x_k) \right) = \int \Bigg[ - F^{\Lambda, \Lambda_0}\left( \bigotimes_{k=1}^s \op_{A_k}(x_k) \otimes \opint(y) \right) \\
&\qquad\quad+ F^{\Lambda, \Lambda_0}\left( \bigotimes_{k=1}^s \op_{A_k}(x_k) \right) L^{\Lambda, \Lambda_0}\left( \opint(y) \right) \\
&\qquad\quad+ \sum_{k=1}^s F^{\Lambda, \Lambda_0}\left( \op_{A_k}(x_k) \otimes \opint(y) \right) \prod_{l \in \{1,\ldots,s\} \setminus \{k\}} L^{\Lambda, \Lambda_0}\left( \op_{A_l}(x_l) \right) \\
&\qquad\quad+ \sum_{l=1}^s \sum_E \mathcal{I}^E \sum_{C\colon [\op_C] \leq [\op_{A_l}]} \mathcal{C}^{\mu, \Lambda_0; C}_{E A_l}(y,x_l) F^{\Lambda, \Lambda_0}\left( \bigotimes_{k \in \{1,\ldots,s\} \setminus \{l\}} \op_{A_k}(x_k) \otimes \op_C(x_l) \right) \Bigg] \total^4 y \\
\end{splitequation}
for all $\Lambda \geq \Lambda_1$, with the composite operator $\opint$ of Proposition~\ref{thm_l0_g}. Again, this identity should be understood as a shorthand for the hierarchy of identities obtained when we expand the above equations in external fields and antifields to an arbitrary loop order $l$.
\end{proposition}
Before we present the proof, note that using the relation~\eqref{g_sop_def}, Proposition~\ref{thm_l1_g} and equation~\eqref{f_sop_oversub_coeffs} this proposition leads to
\begin{corollary}
\label{thm_gs_g}
For each parameter $g$ appearing in the interaction Lagrangian $L^{\Lambda_0}$ but not the covariance $C^{\Lambda, \Lambda_0}_{MN}$, all (monomial) composite operators $\op_{A_k}$ and $s \geq 2$, we have
\begin{splitequation}
\label{gs_g_deriv}
&\hbar \partial_g G^{\Lambda, \Lambda_0}\left( \bigotimes_{k=1}^s \op_{A_k}(x_k) \right) = \int \Bigg[ - G^{\Lambda, \Lambda_0}\left( \bigotimes_{k=1}^s \op_{A_k}(x_k) \otimes \opint(y) \right) \\
&\qquad\quad+ G^{\Lambda, \Lambda_0}\left( \bigotimes_{k=1}^s \op_{A_k}(x_k) \right) L^{\Lambda, \Lambda_0}\left( \opint(y) \right) \\
&\qquad\quad+ \sum_{k=1}^s \sum_E \mathcal{I}^E \sum_{C\colon [\op_C] \leq [\op_{A_k}]} \mathcal{C}^{\mu, \Lambda_0; C}_{E A_k}(y,x_k) G^{\Lambda, \Lambda_0}\left( \bigotimes_{l \in \{1,\ldots,s\} \setminus \{k\}} \op_{A_l}(x_l) \otimes \op_C(x_k) \right) \Bigg] \total^4 y \\
\end{splitequation}
for all $\Lambda \geq \Lambda_1$, with the composite operator $\opint$ of Proposition~\ref{thm_l0_g}. Again, this identity should be understood as a shorthand for the hierarchy of identities obtained when we expand the above equations in external fields and antifields to an arbitrary loop order $l$. Especially, since $\Lambda_1$ is arbitrary we obtain in the physical limit
\begin{splitequation}
\label{gs_g_deriv_phys}
&\hbar \partial_g G^{0, \infty}\left( \bigotimes_{k=1}^s \op_{A_k}(x_k) \right) = \lim_{\Lambda \to 0} \int \Bigg[ - G^{\Lambda, \infty}\left( \bigotimes_{k=1}^s \op_{A_k}(x_k) \otimes \opint(y) \right) \\
&\qquad\quad+ G^{\Lambda, \infty}\left( \bigotimes_{k=1}^s \op_{A_k}(x_k) \right) L^{\Lambda, \infty}\left( \opint(y) \right) \\
&\qquad\quad+ \sum_{k=1}^s \sum_E \mathcal{I}^E \sum_{C\colon [\op_C] \leq [\op_{A_k}]} \mathcal{C}^{\mu, \infty; C}_{E A_k}(y,x_k) G^{\Lambda, \infty}\left( \bigotimes_{l \in \{1,\ldots,s\} \setminus \{k\}} \op_{A_l}(x_l) \otimes \op_C(x_k) \right) \Bigg] \total^4 y \eqend{.}
\end{splitequation}
\end{corollary}
Again, we first have to show that the $y$ integral is absolutely convergent. To show equality, we then simply show that both sides fulfill the same flow equation and boundary conditions, for which we also need absolute convergence of its $\Lambda$ derivative. Since we have already shown the absolute convergence of the $y$ integral (and its $\Lambda$ derivative) appearing in equation~\eqref{l1_g_deriv} of Proposition~\ref{thm_l1_g}, it follows that the $y$ integral of equation~\eqref{fs_g_deriv} is absolutely convergent if the $y$ integral of equation~\eqref{gs_g_deriv} converges absolutely, and \vv, and the same is true for its $\Lambda$ derivative. In the same way, afterwards we can show equality of flow equations and boundary conditions for any of the two claimed identities, and equality follows for both. It turns out that absolute convergence is easier to show for equation~\eqref{gs_g_deriv}, which we do now. There are various UV regions $\Omega_k$ where $y$ lies close to some $x_k$,
\begin{equation}
\label{region_uv_def}
\Omega_k \equiv \left\{ y\colon \abs{x_k - y} \leq \inf_{1 \leq l < l' \leq s} \frac{\abs{x_l-x_{l'}}}{2} \right\} \eqend{,}
\end{equation}
which are all disjoint on account of the factor $1/2$, and the IR region
\begin{equation}
\label{region_ir_def}
\Omega_0 \equiv \mathbb{R}^4 \setminus \left( \Omega_1 \cup \cdots \cup \Omega_s \right) \eqend{.}
\end{equation}

For the IR region, we use the definition of the partially oversubtracted disconnected functionals~\eqref{g_ssop_def} to obtain
\begin{equation}
G^{\Lambda, \Lambda_0}_D\left( \bigotimes_{k=1}^s \op_{A_k}(x_k) \right) L^{\Lambda, \Lambda_0}\left( \opint(y) \right) = \hbar H^{\Lambda, \Lambda_0}_D\left( \bigotimes_{k=1}^s \op_{A_k}(x_k); \opint(y) \right) + G^{\Lambda, \Lambda_0}_D\left( \bigotimes_{k=1}^s \op_{A_k}(x_k); \opint(y) \right)
\end{equation}
for an arbitrary $D \geq 0$. Proposition~\ref{thm_g_ssop_ope} shows that for $D = 0$, the last functional reduces to $G^{\Lambda, \Lambda_0}\left( \bigotimes_{k=1}^s \op_{A_k}(x_k) \otimes \opint(y) \right)$, such that the $y$ integral over this region reduces to
\begin{splitequation}
&\int_{\Omega_0} \sum_E \mathcal{I}^E \Bigg[ \hbar H^{\Lambda, \Lambda_0}\left( \bigotimes_{k=1}^s \op_{A_k}(x_k); \op_E(y) \right) \\
&\qquad\quad+ \sum_{k=1}^s \sum_{C\colon [\op_C] \leq [\op_{A_k}]} \mathcal{C}^{\mu, \Lambda_0; C}_{E A_k}(y,x_k) G^{\Lambda, \Lambda_0}\left( \bigotimes_{l \in \{1,\ldots,s\} \setminus \{k\}} \op_{A_l}(x_l) \otimes \op_C(x_k) \right) \Bigg] \total^4 y \eqend{.}
\end{splitequation}
The bounds~\eqref{bound_hs} show that the $x$-dependence of the first term is given by
\begin{splitequation}
\left( \frac{\mu}{\Lambda_1} \right)^{[\op_\vec{A}]+[\op_E]+\epsilon+\rho} \sup_{1 \leq i \leq s} \left( \mu \abs{x_i-y} \right)^{-[\op_\vec{A}]-[\op_E]-\epsilon-\rho} \sup\left[ 1, \Xi^{\Lambda, \Lambda_1}_{r+\rho, [\op_\vec{A}]+\epsilon}(\vec{x}) \right] \eqend{,}
\end{splitequation}
since $\Lambda \geq \Lambda_1$, where $\Xi^{\Lambda, \Lambda_1}$ is defined by~\eqref{xi_lambda_def}, where $r$ is the smallest integer such that $r > [\op_\vec{A}]+[\op_E]$ and where $\rho \geq 0$ is arbitrary. Choosing thus $\rho = 4$, we see that the integral of this term is absolutely convergent. For the OPE coefficients in the second line, we use expression~\eqref{ope_def2}, noting that the sum is empty since $[\op_C] \leq [\op_{A_k}]$, and the definition~\eqref{g_sop_def} of the oversubtracted disconnected functionals to obtain
\begin{equation}
\label{ope_coeff_2_in_lf}
\mathcal{C}^{\mu, \Lambda_0; C}_{E A_k}(y,x_k) = \mathcal{D}^C_\vec{0} \left[ L^{\mu, \Lambda_0}\left( \op_{A_k}(x_k) \right) L^{\mu, \Lambda_0}\left( \op_E(y) \right) - \hbar F^{\mu, \Lambda_0}_{[\op_C]}\left( \op_{A_k}(x_k) \otimes \op_E(y) \right) \right] \eqend{.}
\end{equation}
Since $[\op_C] \leq [\op_{A_k}]$ and $[\op_E] \geq 1$ (since otherwise the coefficients $\mathcal{I}^E$ vanish), the boundary conditions~\eqref{l_1op_bdy_d} show that the first term vanishes. The bounds~\eqref{bound_fs} then show that the $x$-dependence of the second functional is given by
\begin{equation}
\label{ope_coeff_2_bounds}
\Xi^{\Lambda, \Lambda_1}_{[\op_C], [\op_{A_k}]+[\op_E]-[\op_C]+\epsilon}(x,y) = \left( \mu \abs{x_k-y} \right)^{-[\op_{A_k}]-[\op_E]+[\op_C]-\epsilon-\rho} \left( \frac{\mu}{\Lambda_1} \right)^{[\op_{A_k}]+[\op_E]-[\op_C]+\epsilon+\rho}
\end{equation}
for an arbitrary $\rho \geq 0$, since $\Lambda \geq \Lambda_1$, the number of derivatives $\abs{\vec{w}}$ contained in $\op_C$ is less than $D = [\op_C]$, and thus $\sup(\abs{\vec{w}},D) = [\op_C]$ and $D = [\op_C] < [\op_{A_k}]+[\op_E]$. Thus, choosing again $\rho = 4$ shows that also the integral of these terms is absolutely convergent, and thus the whole integral over the IR region.

We now turn to the UV regions, and choose some fixed $\Omega_k$. Using Proposition~\ref{thm_g_ssop_ope}, we can combine the first term of the integrand with one of the sums involving OPE coefficients, namely the one involving $x_k$, which results in a term
\begin{equation}
G^{\Lambda, \Lambda_0}_{[\op_{A_k}]+\Delta}\left( \opint(y) \otimes \op_{A_k}(x_k); \bigotimes_{l \in \{1,\ldots,s\} \setminus \{k\}} \op_{A_l}(x_l) \right)
\end{equation}
since the smallest difference between operator dimensions is $\Delta$, and thus $[\op_C] \leq [\op_{A_k}]$ is equivalent to $[\op_C] < [\op_{A_k}] + \Delta$. Using then also the definition of the partially oversubtracted disconnected functionals~\eqref{g_ssop_def} and the one of the oversubtracted disconnected functionals~\eqref{g_sop_def}, the $y$ integral over this UV region reduces to
\begin{splitequation}
&\int_{\Omega_k} \sum_E \mathcal{I}^E \left[ \hbar H^{\Lambda, \Lambda_0}_{[\op_{A_k}]+\Delta}\left( \op_E(y) \otimes \op_{A_k}(x_k); \bigotimes_{l \in \{1,\ldots,s\} \setminus \{k\}} \op_{A_l}(x_l) \right) \right. \\
&\qquad\quad+ \hbar F^{\Lambda, \Lambda_0}_{[\op_{A_k}]+\Delta}\left( \op_E(y) \otimes \op_{A_k}(x_k) \right) G^{\Lambda, \Lambda_0}\left( \bigotimes_{l \in \{1,\ldots,s\} \setminus \{k\}} \op_{A_l}(x_l) \right) \\
&\qquad\quad+ L^{\Lambda, \Lambda_0}\left( \op_E(y) \right) \left[ G^{\Lambda, \Lambda_0}\left( \bigotimes_{l=1}^s \op_{A_l}(x_l) \right) - L^{\Lambda, \Lambda_0}\left( \op_{A_k}(x_k) \right) G^{\Lambda, \Lambda_0}\left( \bigotimes_{l \in \{1,\ldots,s\} \setminus \{k\}} \op_{A_l}(x_l) \right) \right] \\
&\qquad\quad+ \left. \sum_{l \in \{1,\ldots,s\} \setminus \{k\}} \sum_{C\colon [\op_C] \leq [\op_{A_l}]} \mathcal{C}^{\mu, \Lambda_0; C}_{E A_l}(y,x_l) G^{\Lambda, \Lambda_0}\left( \bigotimes_{l' \in \{1,\ldots,s\} \setminus \{l\}} \op_{A_{l'}}(x_{l'}) \otimes \op_C(x_l) \right) \right] \total^4 y \eqend{.}
\end{splitequation}
The bounds~\eqref{bound_hs} show that the $x$-dependence of the first line is given by
\begin{splitequation}
\label{ope_coeff_2_boundsuv}
&\left( \frac{\mu}{\Lambda_1} \right)^{[\op_\vec{A}]+[\op_E]+\epsilon+\rho} \left[ \mu \inf_{l \in \{1,\ldots,s\} \setminus \{k\}} \inf \left( \abs{y-x_l}, \abs{x_k-x_l} \right) \right]^{-[\op_\vec{A}]-[\op_E]-\epsilon-\rho} \\
&\quad\times \sup\left[ 1, \left( \mu \abs{y-x_k} \right)^{-([\op_E]-\Delta+\epsilon)-\rho'} \left( \frac{\mu}{\Lambda_1} \right)^{[\op_E]-\Delta+\epsilon+\rho'} \right] \\
&\quad\times \sup\left[ 1, \Xi^{\Lambda, \Lambda_1}_{r+\rho, [\op_\vec{A}]-[\op_{A_k}]+\epsilon}(x_1,\ldots,x_{k-1},x_{k+1},\ldots,x_s) \right]
\end{splitequation}
since $\Lambda \geq \Lambda_1$, where $\Xi^{\Lambda, \Lambda_1}$ is defined by~\eqref{xi_lambda_def}, where $r$ is the smallest integer such that $r > [\op_\vec{A}]+[\op_E]$ and where $\rho,\rho' \geq 0$ are arbitrary. From the definition of the UV region $\Omega_k$~\eqref{region_uv_def} we obtain
\begin{equation}
\abs{y - x_l} \geq \inf_{1 \leq l < l' \leq s} \frac{\abs{x_l-x_{l'}}}{2}
\end{equation}
for all $l \neq k$, such that the first term in equation~\eqref{ope_coeff_2_boundsuv} can be estimated independently of $y$. For the second term, since $1 \leq [\op_E] \leq 4$ (because otherwise the coefficients $\mathcal{I}^E$ vanish), $0 < \Delta \leq 1$ and $0 < \epsilon < \Delta$ taking $\rho' = 0$ we obtain $[\op_E]-\Delta+\epsilon < 4$, and thus the $y$ integral over the first line is absolutely convergent. For the second line, the bounds~\eqref{bound_fs} show that the $x$-dependence is given by
\begin{equation}
\left( \mu \abs{y-x_k} \right)^{-([\op_E]-\Delta+\epsilon)-\rho} \left( \frac{\mu}{\Lambda_1} \right)^{[\op_E]-\Delta+\epsilon+\rho}
\end{equation}
for an arbitrary $\rho \geq 0$, and taking $\rho = 0$ the $y$ integral of the second line is also absolutely convergent. Using the shift property~\eqref{func_sop_shift}, one sees that also the integral of the third line is absolutely convergent. For the OPE coefficients in the last line, we again use the expression~\eqref{ope_coeff_2_in_lf} and obtain the bounds~\eqref{ope_coeff_2_bounds}. Since by definition the UV regions are disjoint, these bounds are finite for $y \in \Omega_k$ for all $x_l \neq x_k$, and thus also the integral of these terms is absolutely convergent.

Furthermore, since the bounds~\eqref{bound_hs_lambdaderiv} for the $\Lambda$ derivative of the partially oversubtracted almost disconnected functionals have the same $x$-dependence as the bounds~\eqref{bound_hs} for the functionals themselves, and the bounds~\eqref{bound_fs_lambdaderiv} for the $\Lambda$ derivative of the oversubtracted almost disconnected functionals have the same $x$-dependence as the bounds~\eqref{bound_fs} for the functionals themselves, also the integral over the $\Lambda$ derivative is absolutely convergent, and thus we can interchange the $y$ integration and the derivative with respect to $\Lambda$.

It thus remains to compare flow equations and boundary conditions, which is easier for equation~\eqref{fs_g_deriv}. Since $F^{\Lambda_0, \Lambda_0}\left( \bigotimes_{k=1}^s \op_{A_k}(x_k) \right) = 0$, both sides of the equation vanish for $\Lambda = \Lambda_0$. We thus only have to check equality of flow equations. We start with the term on the left-hand side and the first term on the right-hand side. Applying $\partial_g$ on the flow equation~\eqref{f_sop_flow} and exchanging both derivatives (since the functionals are formal power series in $g$), we use equation~\eqref{l0_g_deriv} to replace $\partial_g L^{\Lambda, \Lambda_0}$ and equation~\eqref{l1_g_deriv} to replace all occurrences of $\partial_g L^{\Lambda, \Lambda_0}\left( \op_A \right)$ and obtain
\begin{splitequation}
&\partial_\Lambda \partial_g F^{\Lambda, \Lambda_0}\left( \bigotimes_{k=1}^s \op_{A_k}(x_k) \right) = \frac{\hbar}{2} \left\langle \frac{\delta}{\delta \phi_K}, \left( \partial_\Lambda C^{\Lambda, \Lambda_0}_{KL} \right) \ast \frac{\delta}{\delta \phi_L} \right\rangle \partial_g F^{\Lambda, \Lambda_0}\left( \bigotimes_{k=1}^s \op_{A_k}(x_k) \right) \\
&\qquad- \left\langle \frac{\delta}{\delta \phi_K} L^{\Lambda, \Lambda_0}, \left( \partial_\Lambda C^{\Lambda, \Lambda_0}_{KL} \right) \ast \frac{\delta}{\delta \phi_L} \partial_g F^{\Lambda, \Lambda_0}\left( \bigotimes_{k=1}^s \op_{A_k}(x_k) \right) \right\rangle \\
&\qquad- \left\langle \frac{\delta}{\delta \phi_K} \int L^{\Lambda, \Lambda_0}\left( \opint(y) \right) \total^4 y, \left( \partial_\Lambda C^{\Lambda, \Lambda_0}_{KL} \right) \ast \frac{\delta}{\delta \phi_L} F^{\Lambda, \Lambda_0}\left( \bigotimes_{k=1}^s \op_{A_k}(x_k) \right) \right\rangle \\
&\qquad- \sum_{l=1}^s \sum_{\subline{1 \leq k < k' \leq s \\ k \neq l \neq k'}} \left\langle \frac{\delta}{\delta \phi_K} L^{\Lambda, \Lambda_0}\left( \op_{A_k}(x_k) \right) , \left( \partial_\Lambda C^{\Lambda, \Lambda_0}_{KL} \right) \ast \frac{\delta}{\delta \phi_L} L^{\Lambda, \Lambda_0}\left( \op_{A_{k'}}(x_{k'}) \right) \right\rangle \\
&\hspace{8em}\times \int F^{\Lambda, \Lambda_0}_{[\op_{A_l}]+\Delta}\left( \opint(y) \otimes \op_{A_l}(x_l) \right) \total^4 y \prod_{k'' \in \{1, \ldots, s\} \setminus \{k,k',l\}} L^{\Lambda, \Lambda_0}\left( \op_{A_{k''}}(x_{k''}) \right) \\
&\qquad- \sum_{l=1}^s \sum_{k \in \{1, \ldots, s\} \setminus \{l\}} \left\langle \frac{\delta}{\delta \phi_K} \int F^{\Lambda, \Lambda_0}_{[\op_{A_l}+\Delta]}\left( \opint(y) \otimes \op_{A_l}(x_l) \right) \total^4 y, \left( \partial_\Lambda C^{\Lambda, \Lambda_0}_{KL} \right) \ast \frac{\delta}{\delta \phi_L} L^{\Lambda, \Lambda_0}\left( \op_{A_k}(x_k) \right) \right\rangle \\
&\hspace{8em}\times \prod_{k' \in \{1, \ldots, s\} \setminus \{k,l\}} L^{\Lambda, \Lambda_0}\left( \op_{A_{k'}}(x_{k'}) \right) \eqend{.} \raisetag{1.1\baselineskip}
\end{splitequation}
Some terms of this equation are not present for a low number of insertions: if $s = 2$, both the last line and the product in the penultimate one are absent, and for $s = 3$ the product in the last line is absent. We can then perform the $y$ integrals last: in the third line, this follows because of the shift property~\eqref{func_sop_shift} which implies that the integral over $y$ only gives a $\delta$ distribution as in equation~\eqref{l0_g_deriv_2}; in the fourth line nothing needs to be done, and in the fifth line because the bound~\eqref{bound_fs} for the partially connected functionals and the bound~\eqref{prop_abl} for the covariance imply absolute convergence of the $y$ integral and the momentum integral hidden in the scalar product $\left\langle \cdot, \cdot \right\rangle$ when they are done in any order. To derive the flow equation for the remaining terms on the right-hand side, we exchange the derivative with respect to $\Lambda$ with the $y$ integral, which is allowed because of the absolute convergence as proven earlier. To shorten notation, we denote the integrand by $J^{\Lambda, \Lambda_0}\left( \bigotimes_{k=1}^s \op_{A_k}(x_k) \otimes \opint(y) \right)$. Using again the flow equation~\eqref{f_sop_flow}, we derive
\begin{splitequation}
&\partial_\Lambda J^{\Lambda, \Lambda_0}\left( \bigotimes_{k=1}^s \op_{A_k}(x_k) \otimes \opint(y) \right) = \frac{\hbar}{2} \left\langle \frac{\delta}{\delta \phi_K}, \left( \partial_\Lambda C^{\Lambda, \Lambda_0}_{KL} \right) \ast \frac{\delta}{\delta \phi_L} \right\rangle J^{\Lambda, \Lambda_0}\left( \bigotimes_{k=1}^s \op_{A_k}(x_k) \otimes \opint(y) \right) \\
&\qquad- \left\langle \frac{\delta}{\delta \phi_K} L^{\Lambda, \Lambda_0}, \left( \partial_\Lambda C^{\Lambda, \Lambda_0}_{KL} \right) \ast \frac{\delta}{\delta \phi_L} J^{\Lambda, \Lambda_0}\left( \bigotimes_{k=1}^s \op_{A_k}(x_k) \right) \right\rangle \\
&\qquad- \hbar \left\langle \frac{\delta}{\delta \phi_K} L^{\Lambda, \Lambda_0}\left( \opint(y) \right), \left( \partial_\Lambda C^{\Lambda, \Lambda_0}_{KL} \right) \ast \frac{\delta}{\delta \phi_L} F^{\Lambda, \Lambda_0}\left( \bigotimes_{k=1}^s \op_{A_k}(x_k) \right) \right\rangle \\
&\qquad- \sum_{l=1}^s \sum_{\subline{1 \leq k < k' \leq s \\ k \neq l \neq k'}} \left\langle \frac{\delta}{\delta \phi_K} L^{\Lambda, \Lambda_0}\left( \op_{A_k}(x_k) \right), \left( \partial_\Lambda C^{\Lambda, \Lambda_0}_{KL} \right) \ast \frac{\delta}{\delta \phi_L} L^{\Lambda, \Lambda_0}\left( \op_{A_{k'}}(x_{k'}) \right) \right\rangle \\
&\qquad\qquad\times \prod_{k'' \in \{1,\ldots,s\} \setminus \{l,k,k'\}} L^{\Lambda, \Lambda_0}\left( \op_{A_{k''}}(x_{k''}) \right) \\
&\qquad\qquad\times \left[ \hbar F^{\Lambda, \Lambda_0}\left( \op_{A_l}(x_l) \otimes \opint(y) \right) + \sum_{\smash{C\colon [\op_C] \leq [\op_{A_l}]}} \mathcal{C}^{\mu, \Lambda_0; C}_{g A_l}(y,x_l) L^{\Lambda, \Lambda_0}\left( \op_C(x_l) \right) \right] \\
&\qquad- \sum_{l=1}^s \sum_{k \in \{1,\ldots,s\} \setminus \{l\}} \left\langle \frac{\delta}{\delta \phi_K} \left[ \hbar F^{\Lambda, \Lambda_0}\left( \op_{A_l}(x_l) \otimes \opint(y) \right) + \sum_E \mathcal{I}^E \sum_{\smash{C\colon [\op_C] \leq [\op_{A_l}]}} \mathcal{C}^{\mu, \Lambda_0; C}_{E A_l}(y,x_l) L^{\Lambda, \Lambda_0}\left( \op_C(x_l) \right) \right], \right. \\
&\hspace{12em}\left. \left( \partial_\Lambda C^{\Lambda, \Lambda_0}_{KL} \right) \ast \frac{\delta}{\delta \phi_L} L^{\Lambda, \Lambda_0}\left( \op_{A_k}(x_k) \right) \right\rangle \prod_{k' \in \{1,\ldots,s\} \setminus \{k,l\}} L^{\Lambda, \Lambda_0}\left( \op_{A_{k'}}(x_{k'}) \right) \eqend{.}
\end{splitequation}
Using equation~\eqref{f_sop_oversub_coeffs} to convert the sums in the last two lines into oversubtracted partially connected functionals, it follows that $\int J^{\Lambda, \Lambda_0}\left( \bigotimes_{k=1}^s \op_{A_k}(x_k) \otimes \opint(y) \right) \total^4 y$ satisfies the same flow equation as $\hbar \partial_g F^{\Lambda, \Lambda_0}\left( \bigotimes_{k=1}^s \op_{A_k}(x_k) \right)$, and the proposition is proven.

With all this preparation, we are ready to state the main result of this section:
\begin{proposition}
\label{thm_coeffs_g}
For each parameter $g$ appearing in the interaction Lagrangian $L^{\Lambda_0}$ but not the covariance $C^{\Lambda, \Lambda_0}_{MN}$, all (monomial) composite operators $\op_{A_k}$ and $s \geq 2$, the derivative of the OPE coefficients with respect to $g$ can be expressed as
\begin{splitequation}
\label{ope_coeffs_g_recursive_1}
\hbar \partial_g \mathcal{C}^{\mu, \Lambda_0; B}_{A_1 \cdots A_s}(\vec{x}) = \int \sum_E \mathcal{I}^E &\Bigg[ - \mathcal{C}^{\mu, \Lambda_0; B}_{E A_1 \cdots A_s}(y, \vec{x}) + \sum_{C\colon [\op_C] < [\op_B]} \mathcal{C}^{\mu, \Lambda_0; C}_{A_1 \cdots A_s}(\vec{x}) \mathcal{C}^{\mu, \Lambda_0; B}_{E C}(y, x_s) \\
&\quad+ \sum_{k=1}^s \sum_{C\colon [\op_C] \leq [\op_{A_k}]} \mathcal{C}^{\mu, \Lambda_0; C}_{E A_k}(y,x_k) \mathcal{C}^{\mu, \Lambda_0; B}_{A_1 \cdots A_{k-1} C A_{k+1} \cdots A_s}(\vec{x}) \Bigg] \total^4 y \eqend{,}
\end{splitequation}
with the coefficients $\mathcal{I}^E$ of Proposition~\ref{thm_l0_g}.
\end{proposition}
To prove the proposition, we apply a derivative with respect to $g$ on the recursive definition of the OPE coefficients~\eqref{ope_def_recursive}, use Propositions~\ref{thm_l1_g_ope} and~\ref{thm_gs_g} to replace the $g$ derivatives, the relation~\eqref{g_sop_def} to replace all partially connected functionals and the definition~\eqref{op_g_def} of the interaction operator $\opint$, and obtain
\begin{splitequation}
\label{ope_coeffs_g_1}
&\hbar \partial_g \mathcal{C}^{\mu, \Lambda_0; B}_{A_1 \cdots A_s}(\vec{x}) = \mathcal{D}^B_\vec{0} \int \sum_F \mathcal{I}^F \Bigg[ - G^{\mu, \Lambda_0}\left( \bigotimes_{k=1}^s \op_{A_k}(x_k) \otimes \op_F(y) \right) \\
&\qquad\quad+ \left[ G^{\mu, \Lambda_0}\left( \bigotimes_{k=1}^s \op_{A_k}(x_k) \right) - \sum_{C\colon [\op_C] < [\op_B]} \mathcal{C}^{\mu, \Lambda_0; C}_{A_1 \cdots A_s}(\vec{x}) L^{\mu, \Lambda_0}\left( \op_C(x_s) \right) \right] L^{\mu, \Lambda_0}\left( \op_F(y) \right) \\
&\qquad\quad+ \sum_{C\colon [\op_C] < [\op_B]} \mathcal{C}^{\mu, \Lambda_0; C}_{A_1 \cdots A_s}(\vec{x}) \left[ G^{\mu, \Lambda_0}\left( \op_F(y) \otimes \op_C(x_s) \right) - \sum_{E\colon [\op_E] \leq [\op_C]} \mathcal{C}^{\mu, \Lambda_0; E}_{F C}(y,x_s) L^{\mu, \Lambda_0}\left( \op_E(x_s) \right) \right] \\
&\qquad\quad+ \sum_{k=1}^s \sum_{C\colon [\op_C] \leq [\op_{A_k}]} \mathcal{C}^{\mu, \Lambda_0; C}_{F A_k}(y,x_k) G^{\mu, \Lambda_0}\left( \bigotimes_{l \in \{1,\ldots,s\} \setminus \{k\}} \op_{A_l}(x_l) \otimes \op_C(x_k) \right) \Bigg] \total^4 y \\
&\quad- \hbar \mathcal{D}^B_\vec{0} \sum_{C\colon [\op_C] < [\op_B]} \partial_g \mathcal{C}^{\mu, \Lambda_0; C}_{A_1 \cdots A_s}(\vec{x}) L^{\mu, \Lambda_0}\left( \op_C(x_s) \right) \eqend{.}
\end{splitequation}
From the recursive definition of the OPE coefficients~\eqref{ope_def_recursive} we obtain by adding and subtracting the same term and using the boundary conditions for functionals with one insertion of a composite operator~\eqref{l_1op_bdy_d} that
\begin{splitequation}
\mathcal{C}^{\mu, \Lambda_0; B}_{A_1 \cdots A_s}(\vec{x}) &= \mathcal{D}^B_\vec{0} \left[ G^{\mu, \Lambda_0}\left( \bigotimes_{k=1}^s \op_{A_k}(x_k) \right) - \sum_{C\colon [\op_C] \leq [\op_B]} \mathcal{C}^{\mu, \Lambda_0; C}_{A_1 \cdots A_s}(\vec{x}) L^{\mu, \Lambda_0}\left( \op_C(x_s) \right) \right] \\
&\quad+ \sum_{C\colon [\op_C] = [\op_B]} \mathcal{C}^{\mu, \Lambda_0; C}_{A_1 \cdots A_s}(\vec{x}) \delta^B_C
\end{splitequation}
(since the functionals with one insertion of a composite operator are translation invariant at zero momentum according to the shift property~\eqref{func_sop_shift}, and thus do not depend on $x_s$), and thus
\begin{equation}
\mathcal{D}^B_\vec{0} G^{\mu, \Lambda_0}\left( \bigotimes_{k=1}^s \op_{A_k}(x_k) \right) = \mathcal{D}^B_\vec{0} \sum_{C\colon [\op_C] \leq [\op_B]} \mathcal{C}^{\mu, \Lambda_0; C}_{A_1 \cdots A_s}(\vec{x}) L^{\mu, \Lambda_0}\left( \op_C(x_s) \right) \eqend{.}
\end{equation}
which allows us to replace the terms involving the disconnected functionals of equation~\eqref{ope_coeffs_g_1} by finite sums over OPE coefficients and functionals with one insertion. Because of the boundary conditions~\eqref{l_1op_bdy_d}, we can actually extend the sum on the right-hand side, including an arbitrary number of operators $\op_C$ with $[\op_C] > [\op_B]$. From this, we conclude that
\begin{splitequation}
&\mathcal{D}^B_\vec{0} \left[ \left( G^{\mu, \Lambda_0}\left( \bigotimes_{k=1}^s \op_{A_k}(x_k) \right) - \sum_{C\colon [\op_C] < [\op_B]} \mathcal{C}^{\mu, \Lambda_0; C}_{A_1 \cdots A_s}(\vec{x}) L^{\mu, \Lambda_0}\left( \op_C(x_s) \right) \right) L^{\mu, \Lambda_0}\left( \op_F(y) \right) \right] \\
&= \sum_{B' \leq B} c_{BB'} \mathcal{D}^{B'}_\vec{0} \left[ G^{\mu, \Lambda_0}\left( \bigotimes_{k=1}^s \op_{A_k}(x_k) \right) - \sum_{C\colon [\op_C] < [\op_B]} \mathcal{C}^{\mu, \Lambda_0; C}_{A_1 \cdots A_s}(\vec{x}) L^{\mu, \Lambda_0}\left( \op_C(x_s) \right) \right] \mathcal{D}^{B-B'}_\vec{0} L^{\mu, \Lambda_0}\left( \op_F(y) \right) \\
&= 0 \eqend{,} \raisetag{1.1\baselineskip}
\end{splitequation}
since the first term vanishes whenever $[\op_{B'}] < [\op_B]$ according to the above, while the second term vanishes whenever $[\op_B]-[\op_{B'}] < [\op_F]$ according to the boundary conditions~\eqref{l_1op_bdy_d}, but we have $[\op_F] \geq 1$ since otherwise the coefficients $\mathcal{I}^F$ vanish. Replacing the disconnected functionals of equation~\eqref{ope_coeffs_g_1} according to these formulas, we obtain (renaming summation indices)
\begin{splitequation}
\label{ope_coeffs_g_2}
&\hbar \partial_g \mathcal{C}^{\mu, \Lambda_0; B}_{A_1 \cdots A_s}(\vec{x}) = \mathcal{D}^B_\vec{0} \sum_{C\colon [\op_C] \leq [\op_B]} L^{\mu, \Lambda_0}\left( \op_C(x_s) \right) \int \sum_F \mathcal{I}^F \Bigg[ \sum_{E\colon [\op_E] < [\op_C]} \mathcal{C}^{\mu, \Lambda_0; E}_{A_1 \cdots A_s}(\vec{x}) \mathcal{C}^{\mu, \Lambda_0; C}_{F E}(y, x_s) \\
&\hspace{8em}- \mathcal{C}^{\mu, \Lambda_0; C}_{F A_1 \cdots A_s}(y, \vec{x}) + \sum_{k=1}^s \sum_{E\colon [\op_E] \leq [\op_{A_k}]} \mathcal{C}^{\mu, \Lambda_0; E}_{F A_k}(y,x_k) \mathcal{C}^{\mu, \Lambda_0; C}_{A_1 \cdots A_{k-1} E A_{k+1} \cdots A_s}(\vec{x}) \Bigg] \total^4 y \\
&\quad- \hbar \mathcal{D}^B_\vec{0} \sum_{C\colon [\op_C] < [\op_B]} \partial_g \mathcal{C}^{\mu, \Lambda_0; C}_{A_1 \cdots A_s}(\vec{x}) L^{\mu, \Lambda_0}\left( \op_C(x_s) \right) \eqend{.} \raisetag{1.1\baselineskip}
\end{splitequation}
Using that because of the boundary conditions~\eqref{l_1op_bdy_d} we have
\begin{equation}
\mathcal{D}^B_\vec{0} \sum_{C\colon [\op_C] = [\op_B]} \partial_g \mathcal{C}^{\mu, \Lambda_0; C}_{A_1 \cdots A_s}(\vec{x}) L^{\mu, \Lambda_0}\left( \op_C(x_s) \right) = \partial_g \mathcal{C}^{\mu, \Lambda_0; B}_{A_1 \cdots A_s}(\vec{x})
\end{equation}
to combine the terms in the last line with the one on the left-hand side of equation~\eqref{ope_coeffs_g_2}, we finally arrive at
\begin{splitequation}
\label{ope_coeffs_g_3}
&\hbar \sum_{C\colon [\op_C] \leq [\op_B]} \partial_g \mathcal{C}^{\mu, \Lambda_0; C}_{A_1 \cdots A_s}(\vec{x}) \mathcal{D}^B_\vec{0} L^{\mu, \Lambda_0}\left( \op_C(x_s) \right) \\
&= \sum_{C\colon [\op_C] \leq [\op_B]} \mathcal{D}^B_\vec{0} L^{\mu, \Lambda_0}\left( \op_C(x_s) \right) \int \sum_F \mathcal{I}^F \Bigg[ - \mathcal{C}^{\mu, \Lambda_0; C}_{F A_1 \cdots A_s}(y, \vec{x}) + \sum_{E\colon [\op_E] < [\op_C]} \mathcal{C}^{\mu, \Lambda_0; E}_{A_1 \cdots A_s}(\vec{x}) \mathcal{C}^{\mu, \Lambda_0; C}_{F E}(y, x_s) \\
&\hspace{8em}+ \sum_{k=1}^s \sum_{E\colon [\op_E] \leq [\op_{A_k}]} \mathcal{C}^{\mu, \Lambda_0; E}_{F A_k}(y,x_k) \mathcal{C}^{\mu, \Lambda_0; C}_{A_1 \cdots A_{k-1} E A_{k+1} \cdots A_s}(\vec{x}) \Bigg] \total^4 y \eqend{.} \raisetag{1.1\baselineskip}
\end{splitequation}
We now show that~\eqref{ope_coeffs_g_recursive_1} follows from equation~\eqref{ope_coeffs_g_3} by induction in $[\op_B]$. We start with $[\op_B] = 0$, which means $B = \unitmatrix$. Thus, the sums over $C$ in equation~\eqref{ope_coeffs_g_3} reduce to the single term $C = \unitmatrix$, and the boundary conditions~\eqref{l_1op_bdy_d} show that
\begin{equation}
\mathcal{D}^B_\vec{0} L^{\mu, \Lambda_0}\left( \op_C(x_s) \right) = 1
\end{equation}
such that equation~\eqref{ope_coeffs_g_recursive_1} follows in this case. Assume thus that equation~\eqref{ope_coeffs_g_recursive_1} holds for all $B$ with $[\op_B] < D$, and take an arbitrary $B'$ with $[\op_{B'}] = D$. Splitting the sums in equation~\eqref{ope_coeffs_g_3} into terms of dimension smaller than and equal to $D$, the terms of dimension smaller than $D$ cancel by the induction hypothesis~\eqref{ope_coeffs_g_recursive_1}, while for the terms with dimension equal to $D$ we use that
\begin{equation}
\mathcal{D}^{B'}_\vec{0} L^{\mu, \Lambda_0}\left( \op_C(x_s) \right) = \delta^{B'}_C
\end{equation}
according to the boundary conditions~\eqref{l_1op_bdy_d} and the fact that at zero momentum, the functionals with one insertion are independent of $x_s$ according to the shift property~\eqref{func_sop_shift}. Thus equation~\eqref{ope_coeffs_g_recursive_1} follows also for $B'$, and since $B'$ was arbitrary for all terms of dimension equal to $D$.

Finally, we can show also associativity of the OPE coefficients:
\begin{proposition}
\label{thm_ope_associative}
Under the conditions and with the notations of Proposition~\ref{thm_coeffs_g}, we have
\begin{equation}
\mathcal{C}^{\mu, \Lambda_0; B}_{A_1 \cdots A_s}(\vec{x}) = \sum_C \mathcal{C}^{\mu, \Lambda_0; C}_{A_1 \cdots A_k}(x_1, \cdots, x_k) \mathcal{C}^{\mu, \Lambda_0; B}_{C A_{k+1} \cdots A_s}(x_k, \cdots, x_s)
\end{equation}
for all insertion points $x_i$ such that
\begin{equation}
\max_{1 \leq i < k} \abs{x_i-x_k} < \min_{k < j \leq s} \abs{x_j - x_k} \eqend{.}
\end{equation}
\end{proposition}
The proof can be taken over from the work of Holland and Hollands~\cite{hollandhollands2015b}, and involves two ingredients: the recursion formula for the OPE coefficients~\eqref{ope_coeffs_g_recursive_1}, which is exactly the same as for the scalar field theory considered in Ref.~\cite{hollandhollands2015b}, and the OPE coefficients of the free theory, which is different since they consider massive scalar theory while our OPE coefficients are defined in a massless theory at fixed IR cutoff $\mu$. However, the only relevant estimate is [Ref.~\cite{hollandhollands2015b}, Eq. (A8)]
\begin{equation}
\abs{ \partial^u C^{0,\infty}_{m^2}(x) } \leq \left( \frac{4}{x^2} \right)^{(\abs{u}+\delta)/2+1} \frac{\Gamma(\abs{u} + \delta+1)}{4 \pi^2 m^\delta}
\end{equation}
for any $\delta \in [0,1]$ and any multiindex $u$. In our case, the covariance with fixed IR cutoff can be written as the heat kernel integral
\begin{equation}
C^{\mu,\infty}_0(x) = \frac{1}{16 \pi^2} \int_0^{\mu^{-2}} t^{-2} \mathe^{- \frac{x^2}{4t}} \total t \eqend{,}
\end{equation}
which together with the inequality [Ref.~\cite{hollandskopper2012}, Eq. (56)]
\begin{equation}
\abs{ \partial^u \mathe^{- \frac{x^2}{4t}} } \leq 2 (2t)^{-\frac{\abs{u}}{2}} \sqrt{\abs{u}!} \mathe^{- \frac{x^2}{8t}}
\end{equation}
leads to
\begin{splitequation}
\abs{ \partial^u C^{\mu,\infty}_0(x) } &\leq \frac{\sqrt{\abs{u}!}}{8 \pi^2} 2^{-\frac{\abs{u}}{2}} \int_0^{\mu^{-2}} t^{-2-\frac{\abs{u}}{2}} \mathe^{- \frac{x^2}{8t}} \total t \leq \frac{\sqrt{\abs{u}!}}{8 \pi^2} 2^{-\frac{\abs{u}}{2}} \mu^{-\delta} \int_0^\infty t^{-2-\frac{\abs{u}+\delta}{2}} \mathe^{- \frac{x^2}{8t}} \total t \\
&\leq \left( \frac{4}{x^2} \right)^{(\abs{u}+\delta)/2+1} \frac{\Gamma(\abs{u} + \delta+1)}{4 \pi^2 \mu^\delta} \eqend{,}
\end{splitequation}
such that we obtain exactly the same estimate with $\mu$ instead of $m$, and the proposition is proven.

\section{Identities for gauge theories}
\label{sec_gauge}

We now consider the interplay between the OPE and gauge theories. First, we want to show gauge invariance for the OPE coefficients, at least asymptotically. Since the functionals with insertions fulfill the appropriate Ward identities only in the physical limit $\Lambda_0 \to \infty$, we also except that also the OPE coefficients are gauge invariant only in this limit. The proper statement is given by
\begin{proposition}
\label{thm_ope_gaugeinv}
The OPE coefficients are gauge invariant to all asymptotic orders, \ie, for all composite operators $\op_{A_k}$, all $D \geq 0$ and all composite operators $\op_B$ with $[\op_B] < D+1$, they fulfil
\begin{equation}
\label{ope_gaugeinv_scaling}
\lim_{\tau \to 0} \tau^{[\op_\vec{A}]-D+\delta} \mathcal{K}^B_{A_1 \cdots A_s;D}\left( \tau \vec{x} \right) = 0
\end{equation}
for all $\delta > 0$, where
\begin{splitequation}
\label{ope_coeffs_gaugeinv_k_def}
\mathcal{K}^B_{A_1 \cdots A_s;D}\left( \vec{x} \right) &\equiv \sum_{k=1}^s \sum_{C\colon [\op_C] \leq [\op_{A_k}]+1} \stQ_{A_k}{}^C \mathcal{C}^{\mu, \infty; B}_{A_1 \cdots A_{k-1} C A_{k+1} \cdots A_s}\left( \vec{x} \right) - \sum_{C\colon [\op_C] < D} \stQ_C{}^B \mathcal{C}^{\mu, \infty; C}_{A_1 \cdots A_s}\left( \vec{x} \right) \\
&\quad- \hbar \sum_{1 \leq k < l \leq s} \sum_{E\colon [\op_E] \leq [\op_{A_k}]+[\op_{A_l}]-3} \mathcal{C}^{\mu, \infty; B}_{A_1 \cdots A_{k-1} E A_{k+1} \cdots A_{l-1} A_{l+1} \cdots A_s}\left( \vec{x} \right) \\
&\qquad\quad\times \sum_{w\colon \abs{w} = [\op_{A_k}]+[\op_{A_l}]-[\op_E]-3} \bvq^{E,w}_{A_k A_l} \partial^w_{x_k} \delta^4(x_k-x_l) \eqend{.}
\end{splitequation}
The elements of the nilpotent ``quantum BRST matrix'' $\stQ_C{}^B$ are defined as the expansion coefficients of the quantum BRST differential $\stq$ in monomials, given by equation~\eqref{stq_matrix_def}. Since $\stq$ increases the dimension by one, and thus $\stQ_A{}^B = 0$ if $[\op_B] > [\op_A]+1$, the sum over $B$ can also be extended to all $B$. Similarly, the elements of the ``quantum BV bracket matrix'' $\bvq^{C,w}_{AB}$ are defined as the expansion coefficients of the quantum BV bracket $\left( \cdot, \cdot \right)_\hbar$~\eqref{bvq_matrix_def}, and since the coefficients $\bvq^{C,w}_{AB}$ vanish for $\abs{w} \neq [\op_A]+[\op_B]-[\op_C]-3$ and thus for $[\op_C] > [\op_A]+[\op_B]-3$, such that also the restrictions in the sum over $C$ and $w$ can be removed.
\end{proposition}
To prove this invariance, we again choose w.l.o.g. $x_s = 0$ and apply $\st_0$ on the definition of the remainder functional~\eqref{remainder_def_recursive} for $\Lambda = 0$ and $\Lambda_0 = \infty$ (which we know exists for non-exceptional momenta). Using that the action of $\st_0$ on the disconnected functionals is given by the Ward identity~\eqref{ward_gs}, we can apply the OPE again to the functionals on the right-hand side of that equation, in the form of the definition of the remainder functional~\eqref{remainder_def_recursive} but this time with $D' = D + 1$, and use the definition~\eqref{stq_matrix_def} to obtain
\begin{splitequation}
&\sum_{l=1}^s \sum_{C\colon [\op_C] < D+1} \sum_{E\colon [\op_E] \leq [\op_{A_l}]+1} \stQ_{A_l}{}^E \mathcal{C}^{\mu, \infty; C}_{A_1 \cdots A_{l-1} E A_{l+1} \cdots A_s}\left( \vec{x} \right) L^{0, \infty}\left( \op_C(0) \right) \\
&\quad= \sum_{C\colon [\op_C] < D} \sum_{E\colon [\op_E] \leq [\op_C]+1} \stQ_C{}^E \mathcal{C}^{\mu, \infty; C}_{A_1 \cdots A_s}\left( \vec{x} \right) L^{0, \infty}\left( \op_E(0) \right) \\
&\qquad+ \hbar \sum_{1 \leq l < l' \leq s} \sum_{C\colon [\op_C] < {D+1}} \sum_{E\colon [\op_E] \leq [\op_{A_l}]+[\op_{A_{l'}}]-3} \mathcal{C}^{\mu, \infty; C}_{A_1 \cdots A_{l-1} E A_{l+1} \cdots A_{l'-1} A_{l'+1} \cdots A_s}\left( \vec{x} \right) \\
&\hspace{12em}\times \sum_w \bvq^{E,w}_{A_l A_{l'}} \partial^w_{x_l} \delta^4(x_l-x_{l'}) L^{0, \infty}\left( \op_C(0) \right) \\
&\qquad+ \st_0 R^{0, \infty}_D\left( \bigotimes_{k=1}^s \op_{A_k}(x_k) \right) - \sum_{l=1}^s R^{0, \infty}_{D+1}\left( \bigotimes_{k \in \{1,\ldots,s\} \setminus \{l\}} \op_{A_k}(x_k) \otimes (\stq \op_{A_l})(x_l) \right) \\
&\qquad+ \hbar \sum_{1 \leq l < l' \leq s} R^{0, \infty}_{D+1}\left( \bigotimes_{k \in \{1,\ldots,s\} \setminus \{l,l'\}} \op_{A_k}(x_k) \otimes \left( \op_{A_l}(x_l), \op_{A_{l'}}(x_{l'}) \right)_\hbar \right) \eqend{.}
\end{splitequation}
Since $\stQ_C{}^E = 0$ if $[\op_E] > [\op_C]+1$, we can extend the sum over $E$ in the second line up to the maximum value of $[\op_C]$ without changing it and sum over all $E$ with $[\op_E] < D+1$, and then exchange the sums over $C$ and $E$ and rename indices to simplify the result. We now recall the reader that the composite operators $\op_A$ are defined by boundary conditions at $\Lambda = \mu$ and vanishing momenta, i.e., by the boundary conditions~\eqref{l_1op_bdy_d}. However, the above equation is given for $\Lambda = 0$ where Ward identities were proven in Ref.~\cite{froebhollandhollands2015}. We thus introduce a different set of composite operators $\tilde{\op}_A$, for which the functionals with one insertion fulfill the same linear flow equation, but satisfy the boundary conditions
\begin{equations}[l_1op_bdy_tilde]
\mathcal{D}^B_{\vec{p}_\text{ne}} L^{0, \Lambda_0}\left( \tilde{\op}_A(0) \right) &= \delta^B_A \delta_{l,0} \quad\text{ for } [\tilde{\op}_B] \leq [\tilde{\op}_A] \eqend{,} \\
\mathcal{D}^B_{\vec{q}} L^{\Lambda_0, \Lambda_0}\left( \tilde{\op}_A(0) \right) &= \rlap{0}\phantom{\delta^B_A \delta_{l,0}} \quad\text{ for } [\tilde{\op}_B] > [\tilde{\op}_A]
\end{equations}
for some fixed non-exceptional momenta $\vec{p}_\text{ne}$ (depending on $B$). It has been shown in Ref.~\cite{froebhollandhollands2015} and argued in section~\ref{sec_framework} that there is a one-to-one correspondence between these two sets of composite operators, so that there exists a ``mixing matrix'' $\mathcal{Z}_A{}^B$ such that
\begin{equation}
\tilde{\op}_A = \sum_B \mathcal{Z}_A{}^B \op_B \eqend{.}
\end{equation}
Furthermore, since this is only a change in the boundary conditions for marginal and relevant functionals, we have $\mathcal{Z}_A{}^B = 0$ if $[\op_B] > [\tilde{\op}_A]$ and the sum is finite. Because the correspondence is one-to-one, the inverse of $\mathcal{Z}_A{}^B$ exists, and we have
\begin{equation}
\op_A = \sum_B (\mathcal{Z}^{-1})_A{}^B \tilde{\op}_B \eqend{,}
\end{equation}
where the sum again only extends over $B$ with $[\tilde{\op}_B] \leq [\op_A]$ since otherwise $(\mathcal{Z}^{-1})_A{}^B = 0$. Because of the linearity of the functionals with one operator insertion, we can then write
\begin{equation}
\label{l_1op_inoptilde}
L^{0, \infty}\left( \op_C(0) \right) = \sum_{F\colon [\tilde{\op}_F] \leq [\op_C]} (\mathcal{Z}^{-1})_C{}^F L^{0, \infty}\left( \tilde{\op}_F(0) \right) \eqend{.}
\end{equation}
Applying $\mathcal{D}^B_{\vec{p}_\text{ne}}$ with some $B$ of dimension $[\tilde{\op}_B] < D+1$ and using the boundary conditions~\eqref{l_1op_bdy_tilde}, we finally arrive at
\begin{splitequation}
&\sum_{C,F\colon [\tilde{\op}_F] \leq [\op_C] < D+1} (\mathcal{Z}^{-1})_C{}^F \mathcal{K}^C_{A_1 \cdots A_s; D}\left( \vec{x} \right) \mathcal{D}^B_{\vec{p}_\text{ne}} L^{0, \infty}\left( \tilde{\op}_F(0) \right) \\
&\quad= \mathcal{D}^B_{\vec{p}_\text{ne}} \Bigg[ \st_0 R^{0, \infty}_D\left( \bigotimes_{k=1}^s \op_{A_k}(x_k) \right) - \sum_{l=1}^s R^{0, \infty}_{D+1}\left( \bigotimes_{k \in \{1,\ldots,s\} \setminus \{l\}} \op_{A_k}(x_k) \otimes (\stq \op_{A_l})(x_l) \right) \\
&\qquad\qquad+ \hbar \sum_{1 \leq l < l' \leq s} R^{0, \infty}_{D+1}\left( \bigotimes_{k \in \{1,\ldots,s\} \setminus \{l,l'\}} \op_{A_k}(x_k) \otimes \left( \op_{A_l}(x_l), \op_{A_{l'}}(x_{l'}) \right)_\hbar \right) \Bigg] \eqend{.} \raisetag{1.1\baselineskip}
\end{splitequation}
Since $\hat{s}_0$ only exchanges some external fields, and $\mathcal{D}^B_{\vec{p}_\text{ne}}$ evaluates the functionals at some specific value of the momenta and external fields, the asymptotic scaling in the $x_k$ of the remainder terms on the right-hand side is unchanged by these operations according to Proposition~\ref{thm_ope_asymptotic}. Replacing the $x_k$ by $\tau x_k$, multiplying by $\tau^{[\op_\vec{A}]-D + \delta}$ for some arbitrary $\delta > 0$ and taking the limit $\tau \to 0$, the right-hand side vanishes according to equation~\eqref{remainder_scaling}, since $[\stq \op_A] = [\op_A] + 1$ and $\left[ \left( \op_{A_k}, \op_{A_l} \right)_\hbar \right] = [\op_{A_k}] + [\op_{A_l}] + 1$. We therefore obtain
\begin{splitequation}
&\lim_{\tau \to 0} \tau^{[\op_\vec{A}]-D + \delta} \sum_{C,F\colon [\tilde{\op}_F] \leq [\op_C] < D+1} (\mathcal{Z}^{-1})_C{}^F \mathcal{K}^C_{A_1 \cdots A_s; D}\left( \tau \vec{x} \right) \mathcal{D}^B_{\vec{p}_\text{ne}} L^{0, \infty}\left( \tilde{\op}_F(0) \right) = 0 \eqend{.}
\end{splitequation}
We choose now an arbitrary $G$ with $[\op_G] < D+1$, multiply this equation with $\mathcal{Z}_B{}^G$ and sum, for each $F$, over all $B$ with $[\tilde{\op}_B] \leq [\tilde{\op}_F]$. The boundary conditions~\eqref{l_1op_bdy_tilde} show that the last term reduces to $\delta^B_F$, such that the sum over $B$ collapses to a single term and we obtain
\begin{splitequation}
&\lim_{\tau \to 0} \tau^{[\op_\vec{A}]-D + \delta} \sum_{C,F\colon [\tilde{\op}_F] \leq [\op_C] < D+1} (\mathcal{Z}^{-1})_C{}^F \mathcal{Z}_F{}^G \mathcal{K}^C_{A_1 \cdots A_s; D}\left( \tau \vec{x} \right) = 0 \eqend{.}
\end{splitequation}
Since $(\mathcal{Z}^{-1})_C{}^F = 0$ whenever $[\tilde{\op}_F] > [\op_C]$, we can remove the restriction on the sum over $F$, and then use that
\begin{equation}
\label{mixing_z_zinv}
\sum_F (\mathcal{Z}^{-1})_C{}^F \mathcal{Z}_F{}^G = \delta_C^G \eqend{.}
\end{equation}
Because $[\op_G] < D+1$ and we sum over all $C$ with $[\op_C] < D+1$, the sum over $C$ does not vanish but reduces just to equation~\eqref{ope_gaugeinv_scaling}, and the proposition is proven.

The question is then how exactly this gauge invariance is related with the recursive definition of the OPE coefficients according to Proposition~\ref{thm_coeffs_g}. We first prove
\begin{proposition}
\label{thm_opg_stq}
The integrals in Propositions~\ref{thm_l0_g}--\ref{thm_coeffs_g} are unchanged under the substitution $\opint \to \opint + \partial^a \op$ for an arbitrary multiindex $a$ and composite operator $\op$, provided $[\op] \leq 4-\abs{a}$. Furthermore, $\stq \opint = \total \op$ for some composite operator $\op$.
\end{proposition}
To prove the first part, we simply note that the restriction on $[\op]$ is such that the estimates for the integrals appearing in these propositions are unchanged, such that they are still absolutely convergent. Thus, adding a total derivative to $\opint$ and remembering the definition of the OPE coefficients~\eqref{ope_def2}, we can use the second and third Lowenstein rules~\eqref{lowenstein_2} and~\eqref{lowenstein_3} to extract this derivative from the correlation functions and thus the OPE coefficients appearing in the integrals of Propositions~\ref{thm_l0_g}--\ref{thm_coeffs_g}, since $\opint$ is never the last composite operator appearing in these coefficients. The estimates for the integrals in Propositions~\ref{thm_l0_g}--\ref{thm_coeffs_g} are then such that the $y$ integral of a total derivative vanishes, and the first part of the proposition is proven. For the second part, we apply $\st_0$ on equation~\eqref{l0_g_deriv} in the physical limit $\Lambda = 0$ and $\Lambda_0 = \infty$. Since $\st_0$ is $g$-independent, using the Ward identity~\eqref{ward_l0} it follows that
\begin{equation}
0 = \int L^{0, \infty}\left( (\stq \opint)(y) \right) \total^4 y \eqend{.}
\end{equation}
Thus, using the first Lowenstein rule~\eqref{lowenstein_1}, there exists a multiindex $a$ with $\abs{a} = 1$ and a composite operator $\opint{}_{,a}$ such that
\begin{equation}
\stq \opint = \partial^a \opint{}_{,a} \eqend{,}
\end{equation}
which proves the second part.

As a next step, we consider the $g$-dependence of the quantum BRST matrix, which is given by
\begin{proposition}
\label{thm_stq_g}
For each parameter $g$ appearing in the interaction Lagrangian $L^{\Lambda_0}$ but not the covariance $C^{\Lambda, \Lambda_0}_{MN}$ and all (monomial) composite operators $\op_A$, $\op_B$, the derivative of the quantum BRST matrix with respect to $g$ can be expressed as
\begin{splitequation}
\label{stq_g}
\hbar \partial_g \stQ_A{}^B &= \hbar \sum_E \mathcal{I}^E \sum_{C,w} (-1)^\abs{w} \bvq^{C,w}_{EA} \delta\left( \op_B, \partial^w \op_C \right) \\
&\quad+ \int \sum_E \mathcal{I}^E \Bigg[ \sum_{C\colon [\op_C] \leq [\op_A]} \mathcal{C}^{\mu, \infty; C}_{E A}(y,0) \stQ_C{}^B - \sum_{C\colon [\op_B] \leq [\op_C] \leq [\op_A]+1} \stQ_A{}^C \mathcal{C}^{\mu, \infty; B}_{E C}(y,0) \Bigg] \total^4 y
\end{splitequation}
with the coefficients $\mathcal{I}^E$ of Proposition~\ref{thm_l0_g}, assuming that the physical limit $\Lambda \to 0$ can be taken inside the integral in Proposition~\ref{thm_l1_g}. Here, the expression $\delta\left( \op_B, \partial^w \op_C \right)$ is equal to the number of times $\op_B$ appears in the expansion of $\partial^w \op_C$ in monomials.
\end{proposition}
While we haven't been able to prove that the limit $\Lambda \to 0$ can be taken inside the integral in Proposition~\ref{thm_l1_g}, this should hold on physical grounds. In massless theories, correlation functions only decay as a power law at large distances, instead of exponentially as in massive theories. A finite IR cutoff behaves as a mass in this sense, but the form of the bounds that we have derived only show that the correlation functions decay faster than any polynomial in this case; thus, it seems to be that the bounds are not (yet) sharp enough to pinpoint the power-law decay in the massless case without IR cutoff. Since the Ward identities only hold in the unregulated theory, taking $\Lambda = 0$ is inevitable, but one could imagine putting the theory in finite volume. Then, the integral in Proposition~\ref{thm_l1_g} has no convergence properties at all in the IR (taking $\Lambda_1 = \mu$ in the appropriate bounds), and one can straightforwardly take the limit $\Lambda \to 0$ inside the integral. Since the Ward identities will hold also in finite volume, one would arrive at a formula similar to~\eqref{stq_g} in finite volume, and could then take the infinite-volume limit afterwards. Since the OPE coefficients are defined at the scale $\mu$, there are no IR convergence problems for the integrals in~\eqref{stq_g}, and thus one should arrive exactly at formula~\eqref{stq_g} in this way. Thus, we are convinced that the assumption we are making in the proposition is only necessary because of technical reasons, and one could make it completely rigorous with some additional work.

To prove the proposition, we apply $\st_0$ to equation~\eqref{l1_g_deriv_ope_phys} for $x = 0$ and pulling the limit $\Lambda \to 0$ inside the integral, taking into account relation~\eqref{g_sop_def} between the partially connected and disconnected functionals and the action of $\st_0$ on the disconnected functionals~\eqref{ward_gs}. Since $\st_0$ is $g$-independent, this results in
\begin{splitequation}
\hbar \partial_g L^{0, \infty}\left( \stq \op_A(0) \right) &= \int \left[ \hbar F^{0, \infty}\left( \opint(y) \otimes \stq \op_A(0) \right) + \sum_E \mathcal{I}^E \sum_{C\colon [\op_C] \leq [\op_A]} \mathcal{C}^{\mu, \infty; C}_{E A}(y,0) L^{0, \infty}\left( \stq \op_C(0) \right) \right] \total^4 y \\
&\quad+ \hbar \int G^{0, \infty}\left( (\opint(y), \op_A(0))_\hbar \right) \total^4 y \eqend{,}
\end{splitequation}
where we also used Proposition~\ref{thm_opg_stq}, which shows the vanishing of some terms. We now insert the definition of the quantum BRST matrix~\eqref{stq_matrix_def} and the quantum BV bracket matrix~\eqref{bvq_matrix_def} and use the multilinearity of the functionals with composite operator insertions, which results in
\begin{splitequation}
&\hbar \sum_B \left( \partial_g \stQ_A{}^B \right) L^{0, \infty}\left( \op_B(0) \right) = \hbar \sum_{C,w} \int L^{0, \infty}\left( \op_C(y) \right) \bvq^{C,w}_{gA} \partial^w_y \delta^4(y) \total^4 y - \hbar \sum_B \stQ_A{}^B \partial_g L^{0, \infty}\left( \op_B(0) \right) \\
&\quad+ \int \sum_E \mathcal{I}^E \sum_B \left[ \stQ_A{}^B \hbar F^{0, \infty}\left( \op_E(y) \otimes \op_B(0) \right) + \sum_{C\colon [\op_C] \leq [\op_A]} \mathcal{C}^{\mu, \infty; C}_{E A}(y,0) \stQ_C{}^B L^{0, \infty}\left( \op_B(0) \right) \right] \total^4 y \eqend{.}
\end{splitequation}
For the first term on the right-hand side, we integrate the $\abs{w}$ derivatives with respect to $y$ by parts and perform the $y$ integral, while for the second term we use equation~\eqref{l1_g_deriv_ope} again. Some terms then cancel, and after renaming some summation indices and reordering we obtain in total
\begin{splitequation}
0 &= \int \sum_{B\colon [\op_B] \leq [\op_A]+1} \Bigg[ \hbar \partial_g \stQ_A{}^B \delta^4(y) - \sum_E \mathcal{I}^E \Bigg( \hbar \sum_{C,w} (-1)^\abs{w} \bvq^{C,w}_{EA} \delta\left( \op_B, \partial^w \op_C \right) \delta^4(y) \\
&\qquad+ \sum_{C\colon [\op_C] \leq [\op_A]} \mathcal{C}^{\mu, \infty; C}_{E A}(y,0) \stQ_C{}^B - \sum_{C\colon [\op_B] \leq [\op_C] \leq [\op_A]+1} \stQ_A{}^C \mathcal{C}^{\mu, \infty; B}_{E C}(y,0) \Bigg) \Bigg] L^{0, \infty}\left( \op_B(0) \right) \total^4 y \eqend{,}
\end{splitequation}
where we could restrict the sum over $B$ since otherwise the terms in the sum vanish anyway.

We then again switch to the basis $\tilde{\op}_A$ using equation~\eqref{l_1op_inoptilde}, and apply $\mathcal{D}^E_{\vec{p}_\text{ne}}$ with some $E$, which results in
\begin{splitequation}
0 &= \int \sum_{B\colon [\op_B] \leq [\op_A]+1} \Bigg[ \hbar \partial_g \stQ_A{}^B \delta^4(y) - \sum_E \mathcal{I}^E \Bigg( \hbar \sum_{C,w} (-1)^\abs{w} \bvq^{C,w}_{gA} \delta\left( \op_B, \partial^w \op_C \right) \delta^4(y) \\
&\qquad\quad+ \sum_{C\colon [\op_C] \leq [\op_A]} \mathcal{C}^{\mu, \infty; C}_{E A}(y,0) \stQ_C{}^B - \sum_{C\colon [\op_B] \leq [\op_C] \leq [\op_A]+1} \stQ_A{}^C \mathcal{C}^{\mu, \infty; B}_{E C}(y,0) \Bigg) \Bigg] \\
&\quad\times \sum_{F\colon [\tilde{\op}_F] \leq [\op_B]} (\mathcal{Z}^{-1})_B{}^F \mathcal{D}^E_{\vec{p}_\text{ne}} L^{0, \infty}\left( \tilde{\op}_F(0) \right) \total^4 y \eqend{.}
\end{splitequation}
We choose now an arbitrary $G$ with $[\op_G] \leq [\op_A]+1$, multiply this equation with $\mathcal{Z}_E{}^G$ and sum, for each $F$, over all $E$ with $[\tilde{\op}_E] \leq [\tilde{\op}_F]$. The boundary conditions~\eqref{l_1op_bdy_tilde} show that the last term reduces to $\delta^E_F$, such that the sum over $E$ collapses to a single term. Since $(\mathcal{Z}^{-1})_B{}^F = 0$ if $[\tilde{\op}_F] > [\op_B]$, we can remove the restriction on the sum over $F$ and use equation~\eqref{mixing_z_zinv} to obtain $\delta^G_B$. Since $[\op_G] \leq [\op_A]+1$ and we sum over all $B$ with $[\op_B] \leq [\op_A]+1$, the sum over $B$ collapses to a single term, which is equation~\eqref{stq_g} upon noting that $\stQ_A{}^C$ vanishes whenever $[\op_C] > [\op_A]+1$.

We also obtain a formula for the $g$-dependence of the quantum BV bracket matrix, which is given by
\begin{proposition}
\label{thm_bvq_g}
For each parameter $g$ appearing in the interaction Lagrangian $L^{\Lambda_0}$ but not the covariance $C^{\Lambda, \Lambda_0}_{MN}$ and all (monomial) composite operators $\op_{A_1}$, $\op_{A_2}$ and $\op_B$, the derivative of the quantum BV bracket matrix with respect to $g$ can be expressed as
\begin{splitequation}
\label{bvq_g}
&\sum_{E,w} (-1)^\abs{w} \hbar \partial_g \bvq^{E,w}_{A_1 A_2} \delta\left( \op_B, \partial^w \op_E \right) = \int \sum_F \mathcal{I}^F \sum_{E,w} (-1)^\abs{w} \Bigg[ - \sum_C \delta\left( \op_C, \partial^w \op_E \right) \bvq^{E,w}_{A_1 A_2} \mathcal{C}^{\mu, \infty; B}_{F C}(y,0) \\
&\qquad+ \left[ \sum_{C\colon [\op_C] \leq [\op_{A_1}]} \bvq^{E,w}_{C A_2} \mathcal{C}^{\mu, \infty; C}_{F A_1}(y,0) + \sum_{C\colon [\op_C] \leq [\op_{A_2}]} \bvq^{E,w}_{A_1 C} \mathcal{C}^{\mu, \infty; C}_{F A_2}(y,0) \right] \delta\left( \op_B, \partial^w \op_E \right) \Bigg] \total^4 y \eqend{,}
\end{splitequation}
with the coefficients $\mathcal{I}^F$ of Proposition~\ref{thm_l0_g}, assuming that the physical limit $\Lambda \to 0$ can be taken inside the integrals in Propositions~\ref{thm_l1_g} and~\ref{thm_gs_g}.
\end{proposition}
The same remarks we made after Proposition~\ref{thm_stq_g} about the assumption also apply in this case. To prove the proposition, we take the equation for the action of $\st_0$ on the disconnected functionals~\eqref{ward_gs} for $s = 2$, isolate the last term which is supported on the diagonal $x = 0$ and integrate over $x$ (where the integral is thus well-defined). Removing the constraints on the sums over $E$ (since otherwise the quantum BRST matrix $\stQ_{A_k}{}^E$ or the quantum BV bracket matrix $\bvq^{E,w}_{A_1 A_2}$ vanish anyway), this gives
\begin{splitequation}
&\hbar \sum_{E,w} \int L^{0, \infty}\left( \op_E(x) \right) \bvq^{E,w}_{A_1 A_2} \partial^w_x \delta^4(x) \total^4 x = \int \Bigg[ - \st_0 G^{0, \infty}\left( \op_{A_1}(x) \otimes \op_{A_2}(0) \right) \\
&\qquad\qquad+ \sum_E \stQ_{A_1}{}^E G^{0, \infty}\left( \op_E(x) \otimes \op_{A_2}(0) \right) + \sum_E \stQ_{A_2}{}^E G^{0, \infty}\left( \op_{A_1}(x) \otimes \op_E(0) \right) \Bigg] \total^4 x \eqend{.}
\end{splitequation}
On the left-hand side, we now integrate by parts, use the Lowenstein rule~\eqref{lowenstein_1} to take the derivatives inside the functional and then perform a derivative with respect to $g$, using that $\st_0$ is independent of $g$. We then use equations~\eqref{gs_g_deriv_phys} and~\eqref{l1_g_deriv_ope_phys}, pulling the limit $\Lambda \to 0$ inside the integral, to replace the $g$ derivative of disconnected functionals, and equation~\eqref{ward_gs} for the action of $\st_0$ on the disconnected functionals, which results after some partial integrations, use of the Lowenstein rules~\eqref{lowenstein_1}--\eqref{lowenstein_3} and renaming of indices in
\begin{splitequation}
&\hbar^2 \sum_{C,E,w} (-1)^\abs{w} \partial_g \bvq^{E,w}_{A_1 A_2} \delta\left( \op_C, \partial^w \op_E \right) L^{0, \infty}\left( \op_C(0) \right) \\
&\quad= \hbar \int \sum_H \mathcal{I}^H \sum_{E,w} (-1)^\abs{w} \Bigg[ \sum_{C\colon [\op_C] \leq [\op_{A_2}]} \mathcal{C}^{\mu, \infty; C}_{H A_2}(y,0) \bvq^{E,w}_{A_1 C} L^{0, \infty}\left( \partial^w \op_E(0) \right) \\
&\qquad\qquad+ \sum_{u \leq w} \frac{w!}{u! (w-u)!} \sum_{C\colon [\op_C] \leq [\op_{A_1}]} \partial_x^u \mathcal{C}^{\mu, \infty; C}_{H A_1}(y,x) \bvq^{E,w}_{C A_2} L^{0, \infty}\left( \partial^{w-u} \op_E(0) \right) \Big\rvert_{x=0} \\
&\qquad\qquad- \sum_C \delta\left( \op_C, \partial^w \op_E \right) \bvq^{E,w}_{A_1 A_2} \sum_{F\colon [\op_F] \leq [\op_C]} \mathcal{C}^{\mu, \infty; F}_{H C}(y,0) L^{0, \infty}\left( \op_F(0) \right) \Bigg] \total^4 y \\
&\quad+ \int \sum_C G^{0, \infty}\left( \op_C(x) \otimes \op_{A_2}(0) \right) \Bigg[ \hbar \partial_g \stQ_{A_1}{}^C - \hbar \sum_H \mathcal{I}^H \sum_{E,w} (-1)^{\abs{w}} \bvq^{E,w}_{H A_1} \delta\left( \op_C, \partial^w \op_E \right) \\
&\qquad\qquad- \int \sum_H \mathcal{I}^H \left( \sum_{E\colon [\op_E] \leq [\op_{A_1}]} \stQ_E{}^C \mathcal{C}^{\mu, \infty; E}_{H A_1}(y,x) - \sum_{E\colon [\op_E] \geq [\op_C]} \stQ_{A_1}{}^E \mathcal{C}^{\mu, \infty; C}_{H E}(y,x) \right) \total^4 y \Bigg] \total^4 x \\
&\qquad+ \int \sum_C G^{0, \infty}\left( \op_{A_1}(x) \otimes \op_C(0) \right) \Bigg[ \hbar \partial_g \stQ_{A_2}{}^C - \hbar \sum_H \mathcal{I}^H \sum_{E,w} (-1)^{\abs{w}} \bvq^{E,w}_{H A_2} \delta\left( \op_C, \partial^w \op_E \right) \\
&\qquad\qquad- \int \sum_H \mathcal{I}^H \left( \sum_{E\colon [\op_E] \leq [\op_{A_2}]} \stQ_E{}^C \mathcal{C}^{\mu, \infty; E}_{H A_2}(y,0) - \sum_{E\colon [\op_E] \geq [\op_C]} \stQ_{A_2}{}^E \mathcal{C}^{\mu, \infty; C}_{H E}(y,0) \right) \total^4 y \Bigg] \total^4 x \eqend{,}
\end{splitequation}
where many terms have cancelled. Proposition~\ref{thm_stq_g} shows that the last term vanishes. Since the integral in equation~\eqref{stq_g} is absolutely convergent, we can shift the integration variable $y \to y-x$, and since the OPE coefficients are translation invariant and thus only depend on the difference $y-x$, also the second-to-last expression vanishes. Furthermore, because of translation invariance we can convert the $\partial_x^u$ derivatives in the second line of the right-hand side into derivatives with respect to $y$, but since no other term in this line depends on $y$ the corresponding expression vanishes except if $u = 0$. Renaming some summation indices, we thus obtain altogether
\begin{splitequation}
0 &= \int \sum_{E,w} (-1)^\abs{w} \sum_{B\colon [\op_B] \leq [\op_E]+\abs{w}} \Bigg[ \hbar \partial_g \bvq^{E,w}_{A_1 A_2} \delta\left( \op_B, \partial^w \op_E \right) \delta^4(y) \\
&\qquad- \sum_H \mathcal{I}^H \left[ \sum_{C\colon [\op_C] \leq [\op_{A_1}]} \bvq^{E,w}_{C A_2} \mathcal{C}^{\mu, \infty; C}_{H A_1}(y,0) + \sum_{C\colon [\op_C] \leq [\op_{A_2}]} \bvq^{E,w}_{A_1 C} \mathcal{C}^{\mu, \infty; C}_{H A_2}(y,0) \right] \delta\left( \op_B, \partial^w \op_E \right) \\
&\qquad+ \sum_H \mathcal{I}^H \sum_C \delta\left( \op_C, \partial^w \op_E \right) \bvq^{E,w}_{A_1 A_2} \mathcal{C}^{\mu, \infty; B}_{H C}(y,0) \Bigg] L^{0, \infty}\left( \op_B(0) \right) \total^4 y \eqend{.} \raisetag{1.5\baselineskip}
\end{splitequation}
The same operator redefinitions as in the proof of Proposition~\ref{thm_stq_g} and some simplifications then lead to equation~\eqref{bvq_g}, and the proposition is proven. Taking equation~\eqref{ward_gs} for $s > 2$ and applying a derivative with respect to $g$ does not give any more conditions.

It is now seen that we can achieve gauge invariance for the OPE coefficients in a stronger sense than just asymptotic vanishing (as asserted by Proposition~\ref{thm_ope_gaugeinv}), namely we obtain
\begin{proposition}
\label{thm_ope_gaugeinv_strong}
Assuming that the physical limit $\Lambda \to 0$ can be taken inside the integrals in Propositions~\ref{thm_l1_g} and~\ref{thm_gs_g}, the OPE coefficients are gauge invariant; \ie, for all composite operators $\op_{A_k}$ and all composite operators $\op_B$, they fulfil
\begin{equation}
\label{ope_gaugeinv}
\mathcal{K}^B_{A_1 \cdots A_s;D}(\vec{x}) = 0
\end{equation}
for all $D > [\op_B]-1$ with the $\mathcal{K}^B_{A_1 \cdots A_s;D}(\vec{x})$ defined by equation~\eqref{ope_coeffs_gaugeinv_k_def}. Especially, the limit $D \to \infty$ exists.
\end{proposition}
Using Propositions~\ref{thm_stq_g},~\ref{thm_bvq_g} and~\ref{thm_coeffs_g}, one obtains after a long but straightforward calculation the formula
\begin{splitequation}
\label{ope_gaugeinv_recursion}
\hbar \partial_g \mathcal{K}^B_{A_1 \cdots A_s;D}\left( \vec{x} \right) &= \int \sum_E \mathcal{I}^E \Bigg[ - \mathcal{K}^B_{E A_1 \cdots A_s;D}(y, \vec{x}) + \sum_{C\colon [\op_C] < D} \mathcal{C}^{\mu, \infty; C}_{A_1 \cdots A_s}(\vec{x}) \mathcal{K}^B_{E C;D}(y,0) \\
&\quad+ \sum_{C\colon [\op_C] < [\op_B]} \mathcal{C}^{\mu, \infty; B}_{E C}(y,0) \mathcal{K}^C_{A_1 \cdots A_s;D}(\vec{x}) \\
&\quad+ \sum_{k=1}^s \sum_{C\colon [\op_C] \leq [\op_{A_k}]} \mathcal{C}^{\mu, \infty; C}_{E A_k}(y,x_k) \mathcal{K}^B_{A_1 \cdots A_{k-1} C A_{k+1} \cdots A_s;D}(\vec{x}) \Bigg] \total^4 y \eqend{.}
\end{splitequation}
On the right-hand side, all terms but one involve a factor $\mathcal{K}^B_{\cdots;D}$, and the remaining term has a factor $\mathcal{K}^C_{\cdots;D}$ with $[\op_C] < [\op_B]$, such that $D$ is always larger than $[\op_B]-1$ (resp. $[\op_C]-1$). If we thus can show that equation~\eqref{ope_gaugeinv} holds in the free theory for $g = 0$, it will hold, via the recursion formula~\eqref{ope_gaugeinv_recursion}, to all orders in $g$. Note that since in the free theory $\stQ_A{}^B = 0$ for all $B$ with $[\op_B] \neq [\op_A]+1$, the $\mathcal{K}^B_{A_1 \cdots A_s;D}(\vec{x})$ in the free theory are independent of $D$ as long as $D > [\op_B]-1$, as can be directly seen from the definition~\eqref{ope_coeffs_gaugeinv_k_def}. In the free theory, the OPE coefficients are obtained by a sort of Wick expansion, as was proven in Ref.~\cite{holland2013}. While the concrete proof is given for scalar fields, it is straightforward to extend it to other types of fields, and the only substantial difference is that in our case the propagator with finite IR cutoff $\Lambda = \mu$ must be used according to Definition~\ref{ope_def}. Thus, the OPE coefficients in the free theory are of the form
\begin{equation}
\mathcal{C}^{\mu, \infty; B}_{A_1 \cdots A_s}(\vec{x}) = \sum_{P \in \mathcal{P}\left( A_1, \ldots, A_s, B \right)} \prod_{\subline{(v_{A_k},v_{A_l}) \in \mathcal{E}(P) \\ v_{A_k} \not\in V_B \not\ni v_{A_l}}} f_{v_{A_k},v_{A_l}}(x_k,x_l) \prod_{\subline{(v_{A_k},w) \in \mathcal{E}(P) \\ w \in V_B}} g_{v_{A_k},w}(x_k,0) \eqend{,}
\end{equation}
where the sum ranges over all graphs $P$ in the set $\mathcal{P}\left( A_1, \ldots, A_s, B \right)$ of graphs with $s+1$ labeled vertices, which are decorated with multiindices $v$, and where $V_B$ is the set of multiindices belonging to a line ending in the vertex labeled $B$ (see Ref.~\cite{holland2013}, section 3.6.3 for more details). The functions $f_{v,w}(x,y)$ and $g_{v,w}(x,y)$ are given by
\begin{equations}
f_{v,w}(x,y) &\equiv \partial^v_x \partial^w_y C^{\mu,\infty}(x,y) \eqend{,} \\
g_{v,w}(x,y) &\equiv \partial^v_x \frac{(x-y)^w}{w!}
\end{equations}
for scalar fields, and by a similar formula (involving Lorentz and colour indices) for other types of fields. For the cutoff covariance one easily calculates
\begin{equation}
C^{\mu, \infty}(x,y) = \int \frac{1 - \mathe^{-\frac{p^2}{\mu^2}}}{p^2} \mathe^{- \mathi p (x-y)} \frac{\total^4 p}{(2\pi)^4} = \frac{\mathe^{- \frac{1}{4} \mu^2 (x-y)^2}}{4 \pi^2 (x-y)^2} \eqend{,}
\end{equation}
which is a meromorphic function of the difference $x-y$. Thus, the OPE coefficients of the free theory are meromorphic functions of the $x_i$ with poles of finite order at $x_i - x_j = 0$ for any pair $(i,j)$, where the order is bounded by $[\op_\vec{A}]-[\op_B]$. From this, it follows that if
\begin{equation}
\lim_{\tau \to 0} \tau^{[\op_\vec{A}]-[\op_B]-\delta} \mathcal{C}^{\mu, \infty; B}_{A_1 \cdots A_s}(\tau \vec{x}) = 0
\end{equation}
for some $\delta > 0$, we have
\begin{equation}
\mathcal{C}^{\mu, \infty; B}_{A_1 \cdots A_s}(\vec{x}) = 0
\end{equation}
identically. From the definition~\eqref{ope_coeffs_gaugeinv_k_def} of the $\mathcal{K}^B_{A_1 \cdots A_s;D}(\vec{x})$ it follows that in the free theory they are also meromorphic functions of the $x_i$, with the order of poles bounded by $[\op_\vec{A}]+1-[\op_B]$, since the $\delta$ distribution contact type terms are absent in the free theory where $g = \hbar = 0$. Theorem~\ref{thm_ope_gaugeinv} shows that
\begin{equation}
\lim_{\tau \to 0} \tau^{[\op_\vec{A}]-D+\delta} \mathcal{K}^B_{A_1 \cdots A_s;D}\left( \tau \vec{x} \right) = \lim_{\tau \to 0} \tau^{[\op_\vec{A}]+1-[\op_B]-(D+1-[\op_B]-\delta)} \mathcal{K}^B_{A_1 \cdots A_s;D}\left( \tau \vec{x} \right) = 0
\end{equation}
for all $\delta > 0$ and all $[\op_B] < D+1$, and since thus $(D+1-[\op_B]-\delta) > 0$ we conclude that
\begin{equation}
\mathcal{K}^B_{A_1 \cdots A_s;D}(\vec{x}) = 0
\end{equation}
in the free theory, such that the proposition is proven.

\section{Discussion}

In this article, we have shown that Yang-Mills gauge theories based on compact semisimple Lie algebras admit an (asymptotic) operator product expansion, valid in any well-behaved state, and have given explicit recursive formulas to calculate the OPE coefficients in perturbation theory. Moreover, we have shown that the OPE respects gauge invariance, in the sense that if only gauge-invariant operators appear in the correlation function on the left-hand side, also the right-hand side only involves gauge-invariant operators. This invariance is expressed in the form of ``Ward identities'' for the OPE coefficients, which we derive using the Batalin-Vilkovisky formalism for gauge-fixed theories. These identities are given in terms of a ``quantum Slavnov-Taylor differential'' $\stq$ and a ``quantum antibracket'' $( \cdot, \cdot )_\hbar$, which in general differ from the naive classical expressions by terms of order $\bigo{\hbar}$, and for which we also derived explicit recursion formulas in perturbation theory. These recursion formulas are self-consistent in the sense that the only input are the OPE coefficients of the free theory, and the interaction operator $\opint$ (which is constrained by gauge invariance), and they are finite, \ie, no further renormalisation is necessary. In turn, the free theory coefficients are easily obtained using a sort of Wick expansion as shown in Ref.~\cite{holland2013}, where in our case one only has to take the covariance at finite IR cutoff $\mu$ as required by our definition of the OPE coefficients. This is necessary for the convergence of the integrals in the recursion formulas, but does not affect the short-distance expansion other than providing a scale for the logarithmic corrections which appear in perturbation theory. We note that a similar formula to~\eqref{thm3_ope_recursive} has been derived independently by Bochicchio and Becchetti~\cite{bochicchio2017,becchettibochicchio2018}, who have also calculated examples up to two loops. While their formula in principle should be valid also outside of perturbation theory, it only holds for bare (unrenormalised) composite operators and renormalisation still has to be performed.

Our proof extends straightforwardly to other gauge theories where the BV-extended action is linear in the antifields, provided one can prove suitable Ward identities for the correlation functions with composite operator insertions. In turn, this follows from the results of Ref.~\cite{froebhollandhollands2015}, on which this article is based, if one can remove the potential anomaly appearing in the Ward identity for the functionals without insertions. In the flow equation framework as presented in Ref.~\cite{froebhollandhollands2015}, as in other previous frameworks, this question is decided by the relevant equivariant cohomology of the corresponding classical BV differential $\st$ at dimension $4$ and ghost number $1$. If this cohomology is empty, then any anomaly can automatically be removed, but otherwise, one would have to show explicitly that the numerical coefficient in front of the anomaly cancels (\eg, as for the chiral anomaly in the Standard Model~\cite{gengmarshak1989,minahanetal1990}). If the theory has a quadratic (or higher) dependence on antifields such as supergravity~\cite{vannieuwenhuizen1981}, one would have to modify the proof accordingly, but we believe that this does not pose any major problem since our proofs (and the ones of Ref.~\cite{froebhollandhollands2015}) work as long as one has the correct (naive) power-counting.

Finally, we would like to illustrate our formulas with some explicit examples, as was done for scalar field theory in~\cite{hollandhollands2015a} (an explicit study of the OPE for scalar operators of low scaling dimension was also done in~\cite{paganisonoda2018}). However, in contrast to the scalar case, the simplest non-trivial example in Yang-Mills theory concerns the OPE of $\tr F^2$ with itself, which already involves a large amount of computational work because of the Lorentz and Lie algebra indices, even to first order in perturbation theory. Since one also would like to work out the quantum Slavnov-Taylor differential and then check gauge invariance, at least to first order, which involves calculations of the same order of computational complexity, we leave examples for future work. Up to finite redefinitions (\ie, a change of renormalisation scheme), the coefficients of the singular parts in the results of course should coincide with the work of Zoller and Chetyrkin~\cite{zollerchetyrkin2012,zoller2014,zoller2016} for the OPE of the stress tensor $T^{\mu\nu}$ and $\tr F^2$ among each other and with themselves, but our formula also computes the regular part of the OPE.

\acknowledgments

This work was supported by ERC starting grant QC\&C 259562. The authors thank S.~Hollands for discussions, and M.~Bochicchio for discussions on Ref.~\cite{becchettibochicchio2018}.

\bibliography{literature}

\end{document}